\documentclass[longauth]{aa}

\usepackage{graphicx}
\usepackage{txfonts}
\usepackage{lipsum}
\usepackage{subcaption}         
\usepackage{lscape}             
\usepackage{placeins}           
                                
\usepackage[dvipsnames]{xcolor}
\usepackage{url}
\usepackage{mathtools}
\usepackage{siunitx}
\usepackage{empheq}
\usepackage{bm}

\usepackage{array}

\newcommand{\Hii}{\ion{H}{II}}
\newcommand{\Ha}{H${\alpha}$}
\newcommand{\Hb}{H${\beta}$}
\newcommand{\nii}{[\ion{N}{II}]}

\newcommand{\oiii}{[\ion{O}{III}]}
\newcommand{\oii}{[\ion{O}{II}]}
\newcommand{\cii}{[\ion{C}{II}]}

\newcommand{\neiii}{[\ion{Ne}{III}]}

\newcommand{\sii}{[\ion{S}{II}]}
\newcommand{\mgii}{[\ion{Mg}{II}]}
\newcommand{\heii}{\ion{He}{II}}

\newcommand{\Rthree}{\oiii\,$\lambda$5007/\Hb} 
\newcommand{\Ntwo}{\nii\,$\lambda$6583/\Ha}

\newcommand{\mstar}{$M_{\star}$}

\newcommand{\zcristal}{$4$\,$<$\,$z$\,$<$\,$6$}

\newcommand{\kms}{${\rm km\,s^{-1}}$}
\newcommand{\dexkpc}{dex\,kpc\,$^{-1}$}

\newcommand{\micron}{$\mu$m}
\newcommand{\znoon}{$z$\,$\sim$\,1$-$3}
\newcommand{\lOH}{$12+\log({\rm O/H})$}
\newcommand{\dlOH}{$\nabla_r \log({Z})$}

\newcommand{\gradimage}[1]{%
  \begin{minipage}[t]{0.5\textwidth}
    \centering
    \includegraphics[width=\linewidth]{figs/#1}%
  \end{minipage}%
}

\newcommand{\hc}{R. Herrera-Camus et al., in prep.}

\newcommand{\gomezespinoza}{D. G\'omez-Espinoza et al. (in prep.)}

\newcommand{\mediandlOH}{$+0.039$\,$\pm$\,$0.010$}
\newcommand{\sdlOH}{0.053}
\newcommand{\meandlOH}{$+0.043$\,$\pm$\,$0.019$} 
\newcommand{\meandlOHpix}{$+0.027$\,$\pm$\,$0.007$}

\newcommand{\dlOHdisk}{$+0.038$\,$\pm$\,$0.013$} 
\newcommand{\dlOHnondisk}{$+0.051$\,$\pm$\,$0.026$}

\newcommand{\dlOHdiskCRILIT}{$+0.040$\,$\pm$\,$0.011$} 
\newcommand{\dlOHnondiskCRILIT}{$+0.034$\,$\pm$\,$0.020$}

\newcommand{\dlOHdiskCRILITlowz}{$0.000$\,$\pm$\,$0.003$} 
\newcommand{\dlOHnondiskCRILITlowz}{$0.0195$\,$\pm$\,$0.009$}

\newcommand{\dlOHdisknondiskKS}{0.32}
\newcommand{\dlOHdisknondiskKSp}{0.64}
\newcommand{\Ngal}{20}
\newcommand{\Ngallowz}{206}
\newcommand{\acj}{JWST-ALPINE-CRISTAL}

\usepackage{txfonts}
\usepackage{hyperref}

\hypersetup{
    pdfnewwindow=true,  
    colorlinks=true,    
    linkcolor=blue,  
    citecolor=blue,   
    filecolor=cyan,    
    urlcolor=blue,     
    allcolors=blue
}

\begin{document} 

   \title{The ALPINE-CRISTAL-JWST Survey: Gas-phase abundance gradients of main-sequence star-forming galaxies and their kinematics at $4<z<6$}
   \subtitle{}
   
      \author{
      Lilian~L.~Lee\inst{1}
      \fnmsep\thanks{Corresponding author; lilian@mpe.mpg.de; mail@lilianlylee.com}
         \and N.~M.~Förster~Schreiber\inst{1}
        \and S.~Fujimoto\inst{2,3}
        \and A.~L.~Faisst\inst{4}
        \and R.~Herrera-Camus\inst{5,6}
        \and R.~Genzel\inst{1,7}
        \and L.~J.~Tacconi\inst{1}
        \and D.~Lutz\inst{1}
        \and A.~Renzini\inst{8}
        \and R.~Sanders\inst{9}
        \and E.~Wisnioski\inst{10}
        \and S.~Wuyts\inst{11}
        \and E.~Parlanti\inst{12,1}
        \and G.~Jones\inst{13,14}
        \and H.~Übler\inst{1}
        \and D.~Liu\inst{15}
        \and J.~Chen\inst{1}
        \and R.~I.~Davies\inst{1}
        \and G.~Tozzi\inst{1}
        \and A.~Burkert\inst{16,1}
        \and S.~H.~Price\inst{17}
        \and M.~Aravena\inst{18,19}
        \and M.~Boquien\inst{20}
        \and M.~Béthermin\inst{21}
        \and E.~da~Cunha\inst{22}
        \and R.~L.~Davies\inst{23}
        \and I.~De~Looze\inst{24}
        \and M.~Dessauges-Zavadsky\inst{25}
        \and A.~Ferrara\inst{12}
        \and D.~B.~Fisher\inst{26}
        \and S.~Gillman\inst{27,28}
        \and M.~Ginolfi\inst{29}
        \and E.~Ibar\inst{30,31}
        \and A.~M.~Koekemoer\inst{32}
        \and J.~Molina\inst{30,31}
        \and T.~Naab\inst{33}
        \and M.~Relaño\inst{34,35}
        \and D.~A.~Riechers\inst{36}
        \and D.~B.~Sanders\inst{37}
        \and J.~S.~Spilker\inst{38}
        \and L.~Vallini\inst{39}
        \and G.~Zamorani\inst{39}
        \and A.~Nanni\inst{40,41}
        \and P.~Dam\inst{42}
        \and T.~Diaz-Santos\inst{43,44}
        \and D.~Gómez-Espinoza\inst{30,31}
        \and A.~Hadi\inst{45}
        \and R.~Ikeda\inst{46,47}
        \and A.~Posses\inst{38}
        \and M.~Romano\inst{48,49}
        \and A.~Sternberg\inst{50,51}
        \and V.~Villanueva\inst{5}
        \and W.~Wang\inst{4}
        }

        \institute{
        Max-Planck-Institut für extraterrestrische Physik (MPE), Gießenbachstraße 1, 85748 Garching, Germany 
        \and David A. Dunlap Department of Astronomy and Astrophysics, University of Toronto, 50 St. George Street, Toronto, Ontario, M5S 3H4, Canada 
        \and Dunlap Institute for Astronomy and Astrophysics, 50 St. George Street, Toronto, Ontario, M5S 3H4, Canada 
        \and IPAC, California Institute of Technology, 1200 E. California Blvd. Pasadena, CA 91125, USA 
        \and Departamento de Astronom{\'i}a, Universidad de Concepci{\'o}n, Barrio Universitario, Concepci{\'o}n, Chile 
        \and Millenium Nucleus for Galaxies (MINGAL), Concepción, Chile 
        \and Departments of Physics and Astronomy, University of California, Berkeley, CA 94720, USA 
        \and Osservatorio Astronomico di Padova, Vicolo dell’Osservatorio 5, Padova, I-35122, Italy 
        \and Department of Physics and Astronomy, University of Kentucky, 505 Rose Street, Lexington, KY 40506, USA 
        \and Research School of Astronomy and Astrophysics, Australian National University, Canberra, ACT 2611, Australia 
        \and Department of Physics, University of Bath, Claverton Down, Bath BA2 7AY, UK 
        \and Scuola Normale Superiore, Piazza dei Cavalieri 7, I-56126 Pisa, Italy 
        \and Kavli Institute for Cosmology, University of Cambridge, Madingley Road, Cambridge CB3 0HA, UK 
        \and Cavendish Laboratory, University of Cambridge, 19 JJ Thomson Avenue, Cambridge CB3 0HE, UK 
        \and Purple Mountain Observatory, Chinese Academy of Sciences, 10 Yuanhua Road, Nanjing 210023, China 
        \and Universitäts-Sternwarte Ludwig-Maximilians-Universität (USM), Scheinerstr. 1, München, D-81679, Germany 
        \and Space Telescope Science Institute, 3700 San Martin Drive, MD 21218, USA 
        \and Instituto de Estudios Astrofísicos, Facultad de Ingeniería y Ciencias, Universidad Diego Portales, Av. Ejército 441, Santiago, Chile 
        \and Millenium Nucleus for Galaxies (MINGAL), Santiago, Chile 
        \and Université Côte d'Azur, Observatoire de la Côte d'Azur, CNRS, Laboratoire Lagrange, 06000, Nice, France 
        \and Observatoire Astronomique de Strasbourg, UMR 7550, CNRS, Université de Strasbourg, F-67000 Strasbourg, France 
        \and International Centre for Radio Astronomy Research (ICRAR), The University of Western Australia, M468, 35 Stirling Highway, Crawley, WA 6009, Australia 
        \and Swinburne University, John St, Hawthorn, VIC, 3122, Australia 
        \and Sterrenkundig Observatorium, Ghent University, Krijgslaan 281 S9, B-9000 Ghent, Belgium 
        \and Department of Astronomy, University of Geneva, Chemin Pegasi 51, 1290 Versoix, Switzerland 
        \and Swinburne University, John St, Hawthorn, VIC, 3183, Australia 
        \and Cosmic Dawn Center (DAWN), Copenhagen, Denmark 
        \and DTU-Space, Elektrovej, Building 328, 2800 Kgs. Lyngby, Denmark 
        \and Università di Firenze, Dipartimento di Fisica e Astronomia, via G. Sansone 1, 50019 Sesto F.no, Firenze, Italy 
        \and Instituto de F\'{i}sica y Astronom\'{i}a, Universidad de Valpara\'{i}so, Avda. Gran Breta\~{n}a 1111, Valpara\'{i}so, Chile 
        \and Millenium Nucleus for Galaxies (MINGAL), Avda. Gran Breta\~{n}a 1111, Valpara\'{i}so, Chile 
        \and Space Telescope Science Institute, 3700 San Martin Dr., Baltimore, MD 21218, USA 
        \and Max-Planck Institut f\"ur Astrophysik, Karl-Schwarzschild Straße 1, D-85748 Garching, Germany 
        \and Dept. Física Te\'{o}rica y del Cosmos, Campus de Fuentenueva, Edificio Mecenas, Universidad de Granada, E-18071, Granada, Spain 
        \and Instituto Universitario Carlos I de Física Te\'{o}rica y Computacional, Universidad de Granada, 18071, Granada, Spain 
        \and Institut für Astrophysik, Universität zu Köln, Zülpicher Straße 77, D-50937 Köln, Germany 
        \and Institute for Astronomy, University of Hawaii, 2680 Woodlawn Drive, Honolulu, HI 96822, USA 
        \and Department of Physics and Astronomy and George P. and Cynthia Woods Mitchell Institute for Fundamental Physics and Astronomy, Texas A\&M University, 4242 TAMU, College Station, TX 77843-4242, US 
        \and INAF – Osservatorio di Astrofisica e Scienza dello Spazio di Bologna, Via Gobetti 93/3, 40129 Bologna, Italy 
        \and National Centre for Nuclear Research, ul. Pasteura 7, 02-093 Warsaw, Poland 
        \and INAF – Osservatorio Astronomico d’Abruzzo, Via Maggini SNC, 64100 Teramo, Italy 
        \and Dipartimento di Fisica e Astronomia Galileo Galilei Universit\'a degli Studi di Padova, Vicolo dell’Osservatorio 3, 35122 Padova, Italy 
        \and Institute of Astrophysics, Foundation for Research and Technology-Hellas (FORTH), Heraklion, 70013, Greece 
        \and School of Sciences, European University Cyprus, Diogenes Street, Engomi 1516, Nicosia, Cyprus 
        \and Department of Physics \& Astronomy, University of California, Riverside, 900 University Ave., Riverside, CA 92521, USA 
        \and Department of Astronomy, School of Science, SOKENDAI (The Graduate University for Advanced Studies), 2-21-1 Osawa, Mitaka, Tokyo 181-8588, Japan 
        \and National Astronomical Observatory of Japan, 2-21-1 Osawa, Mitaka, Tokyo 181-8588, Japan 
        \and Max-Planck-Institut für Radioastronomie, Auf dem Hügel 69, 53121 Bonn, Germany 
        \and INAF – Osservatorio Astronomico di Padova, Vicolo dell’Osservatorio 5, I-35122 Padova, Italy 
        \and School of Physics and Astronomy, Tel Aviv University, Tel Aviv 69978, Israel 
        \and Centre for Computational Astrophysics, Flatiron Institute, 162 5th Avenue, New York, NY 10010, USA 
        }

   \date{Received XXX; accepted YYY}

  \abstract
   {
  We present gas-phase radial metallicity profiles for \Ngal\ main-sequence galaxies at $4<z<6$ primarily based on JWST NIRSpec IFU observations obtained as part of the \acj\ programme. Our study aims to connect the metallicity gradients of these galaxies with their kinematic properties from \cii\,158\micron\ ALMA observations.
  We mapped the radial profiles of oxygen abundance using the strong-line method, leveraging the rich set of rest-frame optical emission lines.
Linear fits to the annular-binned radial profiles show that, on average, the metallicity gradients are slightly positive with a median of \mediandlOH\,\dexkpc. However, only three galaxies show a gradient $>$\,$0.05\,$\dexkpc\ at $1\sigma$, and none have a significant negative gradient. There are no substantial systematic offsets in gradients when using different line diagnostics. We investigated the correlation between the metallicity gradients and the intrinsic gas velocity dispersion $\sigma_0$ as well as the $V_{\rm rot}/\sigma_0$ ratio of the disks. Combining our sample with mass-matched literature samples at $3\lesssim z\lesssim7$, we found a negative shallow correlation between $V_{\rm rot}/\sigma_0$ and the metallicity gradients, but no strong relationships with $\sigma_0$. As $V_{\rm rot}/\sigma_0$ increases towards later cosmic times, the observed negative trend with $V_{\rm rot}/\sigma_0$ is consistent with the overall cosmic evolution of metallicity gradients from high to low redshifts. This suggests that disk maturity plays a crucial role in shaping the radial metallicity gradients. We do not find the metallicity gradients of disk galaxies to be significantly different from non-disk galaxies, which could be attributed to the frequent accretion events that took place in these gas-rich systems. Additionally, we find no strong dependence of metallicity gradients on stellar mass and only a marginal positive dependence on specific star-formation rate. Our study extends the efforts to connect the internal kinematics of galaxies with their gas-phase chemical enrichment at kiloparsec scales from cosmic noon to $z$\,$>$\,$4$.
 }


\titlerunning{ }
\maketitle
\nolinenumbers
\section{Introduction}\label{sec:intro}
The distribution of heavy elements (metals) at high redshifts ($z\gtrsim1$) places important constraints on the effects of gas flows in the early phase of disk growth. 
The radial gradient of oxygen abundance (metallicity gradient henceforth), as well as that of other $\alpha$ elements in the gas phase, is commonly used as a sensitive probe of baryonic assembly and the complex gas flows driven by both galactic feedback and tidal interactions that can transport the stellar nucleosynthesis yields from where they are produced.
In the local Universe, observations from various large surveys indicate a typical negative gas-phase metallicity gradient of oxygen ranging from $\sim$\,$-0.05$ to $0$\,\dexkpc\ \citep[e.g.][]{Zaritsky1994,Sanchez2014,Ho2015,Grasha2022}, indicating a chemically richer central region compared to the outskirts.
Similarly for the Milky Way, typical radial metallicity gradients inferred from different tracers range from $-0.06$ to $-0.01$\,\dexkpc\ \citep[][and references therein]{MaiolinoMannucci2019}.
The observed negative metallicity gradient in the local Universe could arise from an inside-out growth scenario, 
where the inner disk forms stars first from the high density of gas and has more time to enrich the interstellar medium (ISM) than the outer regions \citep[e.g.][]{MatteucciFrancois1989}.

Metallicity measurements of galaxies at cosmic noon (\znoon), obtained with sensitive integral field units (IFUs) mounted on 8--10\,m class ground-based telescopes, have revealed a wide range of metallicity gradients, with both negative, positive, and flat gradients observed 
\citep{Cresci2010,Yuan2011,Queyrel2012,Swinbank2012b,Jones2013,Stott2014,Troncoso2014,Leethochawalit2016,EWuyts2016,XWang2017,XWang2022,Carton2018,nmfs2018,Curti2020a}.
Some studies suggest that more massive galaxies tend to have steeper metallicity gradients, while galaxies with higher  specific star formation rates (sSFRs) tend to have flatter gradients \citep[e.g.][]{Gillman2021,Cheng2024}.
The majority of $z$\,$\lesssim$\,$3$ results are consistent with flat gradients, within errors, or are only moderately negative.

In the case of the Milky Way, 
its metallicity gradient (of oxygen) 
has evolved from $\sim$\,$-0.01$\,\dexkpc\ at $z$\,$=$\,$1$ to the current steepness of $\sim$\,$-0.04$\,\dexkpc\ 
based on, for example, planetary nebulae that probe longer look-back times and chemical evolution models \citep{Chiappini2001,Stanghellini2014,Molla2019}.
Therefore, based on observations, one expects metallicity gradients to become increasingly negative with cosmic time, eventually matching the steeper gradients seen in nearby disks. This evolutionary trend is reproduced only in a subset of simulations \citep[e.g. MaGICC,][]{Gibson2013}. 
Most simulations, however, predict the opposite behaviour, namely that gradients start steeper at high redshift and subsequently flatten \citep{TaylorKobayashi2017,Acharyya2025,Garcia2025}, while others find very little evolution between $z$\,$\lesssim$\,$3$ and the present epoch \citep{Ma2017,Hemler2021,Tissera2022,Sun2025}. The difference in behaviour 
is largely because of the influence of feedback, where models or simulations with strong feedback lead to flatter gradients at higher redshifts when feedback and outflows were more important.

The advent of the \textit{James Webb} Space Telescope (JWST) has enabled access to multiple rest-frame strong optical lines at $z$\,$>$\,$4$.
Recent JWST-based metallicity measurements suggest flat to positive gradients for a large fraction of galaxies at 
$z$\,$>$\,$6$, but this is based on a heterogeneous sample
\citep[][see also \citealt{Birkin2023}]{Arribas2024,Venturi2024,Barisic2025,ZLi2025,Ju2025}. 
Notably, there have been no systematic studies of metallicity gradients at \zcristal\ to date.

Several internal secular processes can contribute to the flattening or the even inversion of the gas-phase metallicity gradients in disk galaxies and can even prevent the formation of gradients altogether.
Radial flows of gas and the migration of young massive stars, for instance, can dilute the central regions and flatten the metallicity gradients \citep{SpitoniMatteucci2011,Mott2013,Spitoni2015}.
Additionally, strong feedback in the form of galactic outflows can redistribute metals produced by the central starburst region towards the external regions of galactic disks, potentially inverting the gradient to become positive if the outflows have a strong metal loading \citep{Tissera2022}. 
The presence of a bar may also play a role in flattening the abundance gradient, likely due to the non-circular motions induced in the gas \citep[e.g.][]{Alloin1981,Zaritsky1994} and 
with its impact correlating with bar strength \citep[][but see 
\citealt{SM18} and \citealt{Wisz2025} for a contrasting view]{Martin1994}.

Positive gradients could also result from external stochastic processes such as galaxy interactions \citep{Rupke2010b,TaylorKobayashi2017} or the dilution of metallicity in the central regions by cold flows of pristine gas from the intergalactic medium \citep{Dekel2009}. 
The complex interplay between these various mechanisms could contribute to the observed diversity of metallicity gradients in galaxies \citep{Sharda2021a}.

Nevertheless, simulation work on metallicity gradients have largely been
focussed at $z$\,$\lesssim$\,$3$, with only a handful of studies covering earlier cosmic epochs \citep[e.g.][]{Garcia2025,IbrahimKobayashi2025}. 
Most simulations predict strong negative metallicity gradients at early cosmic times, likely due to their similar treatments of relatively smooth stellar feedback allowing for sustained inside-out growth and weak subgrid metal diffusion \citep[][see the discussion in \citealt{Garcia2025}]{Hemler2021,Tissera2022,Sun2025}.

From a dynamical perspective, kinematics and metal distribution are tightly connected.
At low-redshift and up to \znoon, an emerging trend suggests that more turbulent galaxies, 
characterised by their low rotational support relative to velocity dispersion, tend to have flatter metallicity gradients \citep[e.g.][]{Queyrel2012,Jones2013,Leethochawalit2016,Lyu2025}, 
a tendency also seen in simulations \citep{Ma2017,Hemler2021} and reproduced well with analytical models with efficient metal mixing \citep[e.g.][]{Sharda2021}. 
In general, the flat gradient can be attributed to the unstable internal dynamics of turbulent disks, 
making them more susceptible to disturbances caused by strong feedback-driven gas flows, in addition to mixing from advection and accretion.
As a result, large-scale metal redistribution over several kiloparsecs within the galaxy can occur, leading to the flat to positive gradients observed.

At cosmic noon and up to $z$\,$\sim$\,$4$, there is also growing evidence suggesting the prevalence of non-circular motion and radial gas flows in gas-rich disk galaxies, characterised by velocities of the order of a few tens of \kms, which could efficiently transport gas inwards \citep[e.g.][]{Price2021,Genzel2023,Uebler2024b,Huang2025,Pastras2025,Tsukui2024}. 
If these secular mixing processes occur on a timescale comparable to that of metal production by stars, 
the ISM could remain well mixed from the outset and stay in such way until the disk settles into a dynamically cold and thin disk, at which point the classical inside-out formation scenario may dominate, leading to a negative metallicity gradient.

While measurements of metallicity gradients in galactic disks at $z$\,$>$\,$4$ remain scarce due to rest-frame optical lines being shifted out of the atmospheric window, 
kinematics studies benefit from sub-millimetre interferometers such as ALMA and NOEMA, where
the bright \cii\,158\micron\ (\cii\ henceforth) line emission is a commonly used tracer.
The \cii\ line is a primary radiative line coolant of the cold ISM associated with photo-dissociation regions (PDRs) and \Hii\ regions \citep[e.g.][]{Wolfire2003,Vallini2015,Clark2019}. The high spectral and spatial resolution of the ALMA and NOEMA facilities have allowed for high-resolution resolved kinematics at scales of $\sim$\,1\,kpc \citep[e.g.][see also \citealt{Martinez-Cuadra2026}]{Tsukui2021,Rizzo2021,RomanOliveira2023,Rowland2024,Lee2025b}. 

In this work, we exploit the synergy between ALMA and JWST to investigate the connection between the kinematic nature and metallicity gradients of star-forming galaxies (SFGs) at \zcristal, a previously under-explored area. Our analysis combines ALMA-CRISTAL survey data \citep{HerreraCamus2025}, which provides high-resolution kinematic information via the \cii\ line emission, with JWST NIRSpec IFU data from primarily the \acj\ survey \citep{Faisst2026}, which enables the measurement of chemical abundance gradients using strong rest-optical nebular lines that are commonly used at lower redshift.

Our sample consists of a homogeneously selected set of main-sequence (MS) SFGs representative of the galaxy population at \zcristal. Building on the kinematic analysis presented in \citet{Lee2025b}, we aim to explore the relationship between the internal dynamics and chemical abundance distribution in these galaxies and investigate the potential differences between disk and non-disk galaxies classified in \citet{Lee2025b}. The differences between the kinematic classifications using the ALPINE and CRISTAL data are discussed in more detail in the work by \citet{Lee2025b}.

A separate work by \citet{Faisst2025} presents a more detailed analysis of the integrated metallicity and mass-metallicity relation. 
\citet{Fujimoto2025} discusses the cosmic evolution of metallicity gradients, and in conjunction with \citet{Lopez2026}, they present the metallicity maps for the full \acj\ sample.
This work builds upon the analysis of \citet{Fujimoto2025} by incorporating a broader range of strong-line diagnostics available from the data. Additionally, we investigate the impact of different diagnostics and beam smearing on the resulting metallicity gradients, thereby complementing the analysis of \citet{Fujimoto2025}.

This work is structured as follows.
In Sect.~\ref{sec:data_sample} we describe the sample selection and data reduction process. 
Section~\ref{sec:emission_line_fitting} presents the method of emission line fitting. In Sect.~\ref{sec:strong_line_method} we describe the methods used to infer metallicity and examine the systematic effects of different strong-line diagnostics on the inferred gradients. 
Section~\ref{sec:gradients_fitting} explores the relationship between the measured metallicity gradients and their kinematic nature as well as other physical properties of the samples. 
In Sect.~\ref{sec:effects_of_beam_smearing} we investigate the impacts of beam smearing on the inferred metallicity gradients.
Finally, in Sect.~\ref{sec:summary} we summarise the key findings.

Throughout this work we adopt a flat $\Lambda$ cold dark matter (CDM) cosmology with $H_0$\,$=$\,$70{\rm\,km\,s^{-1}\,Mpc^{-1}}$ 
and $\Omega_m$\,$=$\,$0.3$. 
Solar metallicity was taken to be \lOH\,$=$\,$8.69$ \citep{Asplund2009}.
Where relevant, the \citet{Chabrier2003} initial mass function (IMF) is used.
Physical size is always reported in physical kiloparsecs, but we simply state `kiloparsecs' henceforth for brevity.

\begin{table*}
\caption{Physical properties of the \Ngal\ galaxies in the sample.}
\label{tab:main_table}
\centering
\begin{tabular}{>{\small}l >{\small}l >{\small}l >{\small}c >{\small}c >{\small}c >{\small}c >{\small}c}
\hline\hline  
  CRISTAL ID & Full name & $z_{\rm [\ion{C}{II}]}$ & R.A. & Decl. & $\log{(M_\star/\rm{M_\odot})}$\tablefootmark{a} & $\log{[{\rm SFR}/(\rm{M_\odot}\,{\rm yr}^{-1}})]$\tablefootmark{a}& Class.\tablefootmark{b} \\
  &&&\small{($^{\circ}$)}&\small{($^{\circ}$)}&\small{(dex)}&\small{(dex)}
\\
\hline  
01 & DC\_842313 & 4.554 & 150.2271 & 2.5762 & 10.65 & 2.31 & Non-Disk \\
02$^{\dagger}$ & DC\_848185, HZ6, LBG-1 & 5.294 & 150.0896 & 2.5864 & 10.30 & 2.25  & Disk \\
03$^{\dagger}$ & DC\_536534, HZ1 & 5.689 & 149.9719 & 2.1182 & 10.40 & 1.79  & Best-Disk\tablefootmark{c} \\
04a & vc\_5100822662 & 4.520 & 149.7413 & 2.0809 & 10.15 & 1.89  & Non-Disk \\
05$^{\dagger}$ & DC\_683613, HZ3 & 5.541 & 150.0393 & 2.3372 & 10.16 & 1.83 & Disk \\
06a & vc\_5100541407 & 4.562 & 150.2538 & 1.8094 & 10.09 & 1.62& Disk \\
06b & \ldots & 4.562 & 150.2542 & 1.8097 & 9.19 & 1.07 & Non-Disk  \\
07a$^{\dagger}$ & DC\_873321, HZ8 & 5.154 & 150.0169 & 2.6266 & 10.00 & 1.89 & Disk  \\
07b & \ldots & 5.154 & 150.0166 & 2.6268 & \ldots & \ldots & Non-Disk  \\
09$^{\dagger}$ & DC\_519281 & 5.575 & 149.7537   & 2.0910 & 9.84 & 1.51 & Disk\tablefootmark{c}  \\
10a-E$^{\dagger}$ & DC\_417567, HZ2 & 5.671 & 150.5172 & 1.9290 & 9.99 & 1.86  & Disk \\
11 & DC\_630594 & 4.439 & 150.1358 & 2.2579 & 9.68 & 1.57 & Best Disk\tablefootmark{c} \\
13 & vc\_5100994794 & 4.579 & 150.1715 & 2.2873 & 9.65 & 1.51 & Non-Disk  \\
14 & DC\_709575 & 4.411 & 149.9461 & 2.3758 & 9.53 & 1.45& Non-Disk \tablefootmark{d}  \\
15 & vc\_5101244930 & 4.580 & 150.1986 & 2.3006 & 9.69 & 1.44 & Best Disk \\
19 & DC\_494763 & 5.233 & 150.0213 & 2.0534 & 9.51 & 1.45  & Best Disk\tablefootmark{c} \\
20 & DC\_494057, HZ4 & 5.545 & 149.6188 & 2.0518 & 10.11\textbf{} & 1.82 &Best Disk\tablefootmark{e} \\ 
22a & HZ10-E & 5.653 &  150.2471 & 1.5554  & 10.35 &2.13 & Disk \\ 
25 & vc\_5101218326 & 4.573 & 150.3021 & 2.3146 & 10.90 & 2.75  & Non-Disk \\
VC-7875 & vc\_5110377875 & 4.551 & 150.3848 & 2.4084 & 10.17 & 2.00 & Disk\tablefootmark{f}  \\ 
\hline
\end{tabular}
\tablefoot{
\tablefoottext{a}{Derived from SED modellings in \citet{JLi2024} and \cite{Mitsuhashi2024}. The typical uncertainty for $\log(M_\star)$ and $\log$(SFR) are about 0.3\,dex.} 
\tablefoottext{b}{Classified in \citet{Lee2025b}.} 
\tablefoottext{c}{Classified as rotating disks also in \gomezespinoza\ with the \Ha\ data.} 
\tablefoottext{d}{Classified as compact in \gomezespinoza.}
\tablefoottext{$^{\dagger}$}{Candidate Type-1 AGNs identified in \citet{Ren2025} based on the tentative detection of \Ha\ broad component.}
\tablefoottext{e}{See \citealt{Parlanti2025} for a contrasting view.}
\tablefoottext{f}{Classified as `Rotator' in \citet{Jones2021} but with signatures of interaction with an extended halo or outflow \citep{Fujimoto2020}, so it will fall closest to the `Disk' classification in \citet{Lee2025b}.} 
}
\end{table*}

\begin{figure*}
\centering
    \includegraphics[width=0.7\textwidth]{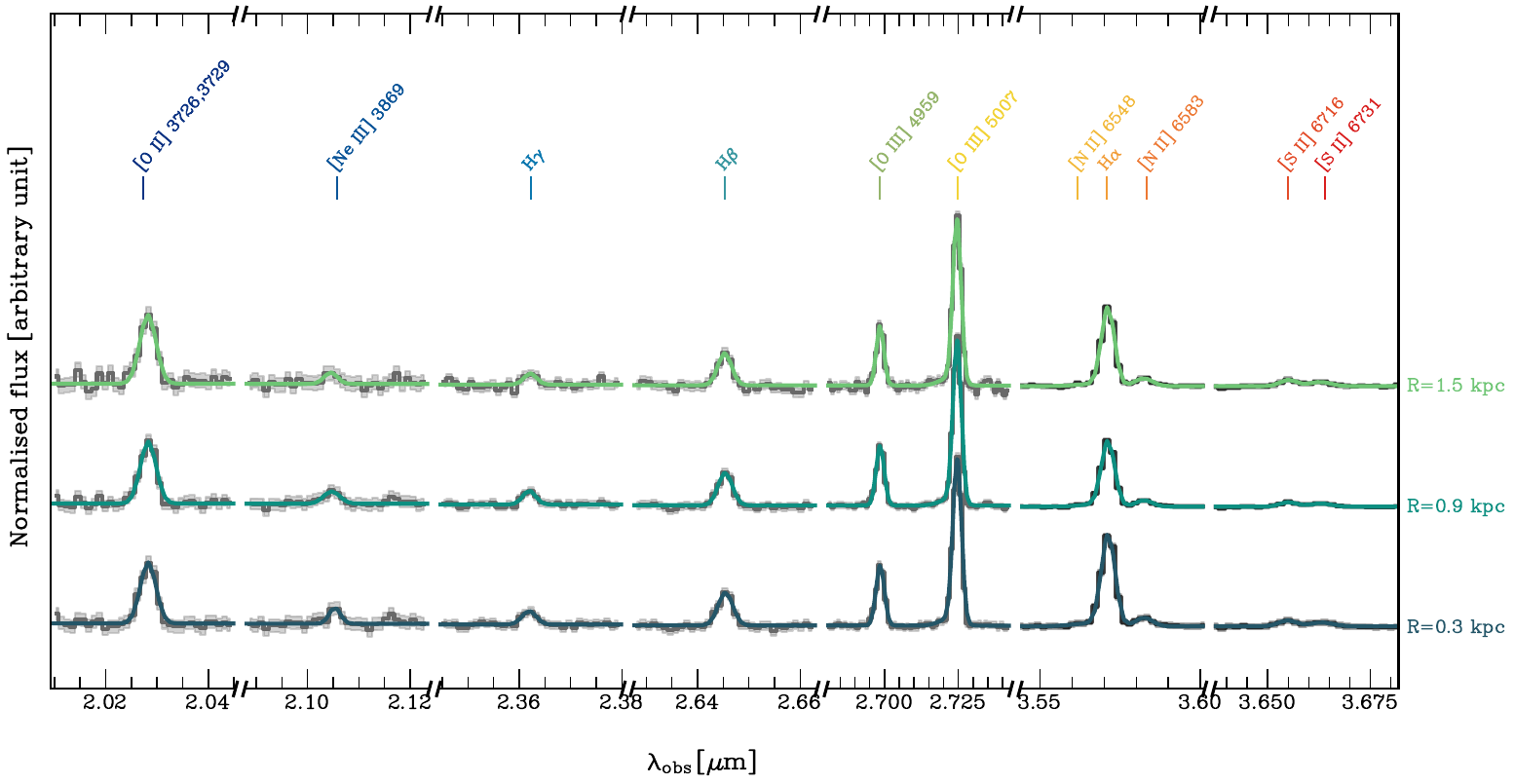}
    \raisebox{0.92cm}{\includegraphics[width=0.25\textwidth]{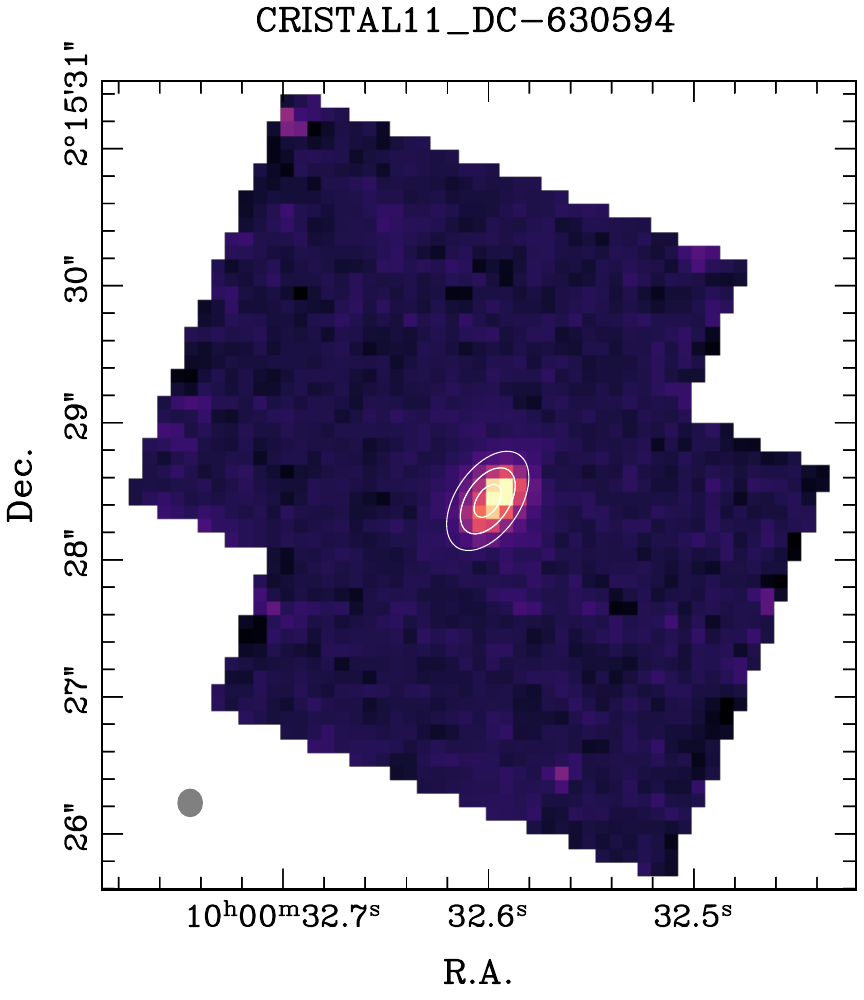}}

      \caption{\textit{Left}: Example spectral fit of CRISTAL-11 ($z=4.439$) for each annulus. 
      The fitting procedure of the spectra is described in Sect.~\ref{sec:emission_line_fitting}.
      The radius of the corresponding annulus is indicated in physical units on the right.
      The lines used for inferring metallicity in Sect.~\ref{sec:strong_line_method} are annotated at the top.
      \textit{Right}: Shape of the annuli  shown overlaid on the line flux map of \Ha. The PSF after PSF homogenisation is shown at the bottom left.
      }
      \label{fig:spectral_fit}
\end{figure*}

\section{Samples}\label{sec:data_sample}
In this study, we investigate the metals abundance distribution of the sample presented in \citet{Lee2025b}, which examines the kinematic properties of 32 galaxies at \zcristal\ using \cii\ emission, as part of the ALMA-CRISTAL survey \citep{HerreraCamus2025}
and samples supplemented from the ALMA archive. 
The NIRSpec/IFU data used for metallicity measurements are drawn from the \acj\ survey (PID: 3045, PI: A. Faisst), a GO programme (PID: 4265, PI: J. Gonzalez Lopez) and the Galaxy Assembly with NIRSpec Integral Field Spectroscopy (GA-NIFS) Guaranteed Time Observations (GTO) programme (PID: 1217, PI: N. Luetzgendorf).
The NIRSpec data from \acj\ survey additionally covers vc\_5110377875 (VC-7875 henceforth), for which the kinematics results are presented in \citet{Fujimoto2020} and \citet{Jones2021}.
The key properties of our sample are listed in Table~\ref{tab:main_table}. The NIRSpec/IFU data cover a wavelength range of $1.66$ to $5.10$\,\micron\ (rest-frame $0.28$--$0.85$\,\micron\ at $z=5$) at a medium spectral resolution of $R\sim1000$ ($\sigma$\,$\approx$\,$125\,$\kms\ at 3\,\micron) with two gratings: G235M and G395M. 

For a detailed description of the samples, we refer the reader to the survey papers: \citet{HerreraCamus2025} for the original ALMA-CRISTAL sample and \citet{Faisst2026} for the NIRSpec/IFU data and data reduction in \citet{Fujimoto2025}. In brief, ALMA-CRISTAL \citep{HerreraCamus2025} is a high-resolution follow-up of a subset of galaxies from the ALMA-ALPINE survey \citep{Bethermin2020,LeFevre2020,Faisst2020}, with stellar masses $\log(M_\star/{\rm M_\odot})$\,$\geqslant$\,$9.5$ and sSFRs within a factor of three of the MS, with higher angular resolution, resolving the galaxies at $\sim$\,$1$\,kpc scale.\footnote{The ALMA data of VC-7875 and CRISTAL-24 (DEIMOS\_COSMOS\_873756) also come from two other ALMA observations (ALMA project 2019.1.00226.S, 2022.1.01118.S, PIs: E. Ibar, M. B\'ethermin).}
Both programmes carried out ALMA observations in Band 7, targeting the \cii\ and dust continuum emission of MS SFGs at \zcristal. 
The \acj\ survey is a NIRSpec IFU follow-up of 18 galaxies from the two ALMA large programmes.

Since the \acj\ sample does not include all galaxies in the kinematics sample presented in \citet{Lee2025b}, we supplement our analysis by incorporating NIRSpec/IFU data for CRISTAL-01, 20, and 22a from two separate programmes. Specifically, we utilise data from a GO programme (PID: 4265, PI: J. Gonzalez Lopez, \citealt{Solimano2025}) for CRISTAL-01, and from the GA-NIFS survey for CRISTAL-20 and 22a.
The GO programme employed a nine-point (`SMALL') dither pattern, while the GA-NIFS GTO programme used an eight-point (`MEDIUM') dither pattern. Both datasets were obtained at high spectral resolution using the G395H/F290LP configuration, which provides a spectral resolution of $R$\,$\sim$\,$2700$.
For CRISTAL-20 and 22a, we adopt the published metallicity maps presented in \citet{Parlanti2025} and \citet{Jones2025}, respectively, which have a pixel scale of 0\farcs05\,pix$^{-1}$.
We examine their different adopted calibrations, and any differences they may introduce to the metallicity gradients in Sect.~\ref{sec:strong_line_method}.

The important rest-frame optical emission lines for inferring metallicity, listed in order of increasing wavelength, are: \oii\,$\lambda\lambda$3726,3729, \neiii\,$\lambda$3869, \Hb, \oiii\,$\lambda$4959,5007, \nii\,$\lambda\lambda$6548,6583, \Ha, and \sii\,$\lambda\lambda$6716,6731. Higher Balmer transition lines up to H$\epsilon$ are also detected in some cases \citep{Fujimoto2025,Faisst2026}.
The auroral line \oiii\,$\lambda$4363 is also detected in five galaxies, and for these cases, the non-$T_e$-based metallicities are in good agreement with the $T_e$-based metallicities, albeit with larger uncertainties \citep{Faisst2025}.

Six objects in our sample were classified as candidate Type-1 active galactic nuclei (AGNs) by \citet[see our Table~\ref{tab:main_table}]{Ren2025} 
due to the tentative presence of broad H$\alpha$ emission. 
We retain these objects in the sample because their \nii/\Ha\ ratios are low ($f_{\nii}/f_{\rm H\alpha}$\,$\lesssim$\,$0.1$) and they lie very close to the star-forming locus in the classical \citet*[][BPT]{BPT1981} diagrams. 
The same holds when using higher-ionisation diagnostics: the \heii\,$\lambda4686$-based criterion of \citet{Shirazi2012} and the \oiii\,$\lambda4363$/H$\gamma$-based demarcation of \citet{Mazzolari2024}; they would also fall well below the revised AGN boundary of \citet{Scholtz2025}.
The \heii\ and \oiii\,$\lambda4363$ lines are, however, too faint to construct spatially resolved maps that would separate AGN- and star-formation-dominated regions, leaving the less suitable \nii-based BPT diagram as the only feasible option (as in \citealt{Ren2025}). 
Hence, although an AGN contribution cannot be entirely ruled out, the available diagnostics provide no compelling evidence of a strong one; any putative AGN should therefore be weak, and contamination from AGN-driven shocks or excess hard ionising radiation is expected to be minor.

There is nonetheless one galaxy, CRISTAL-24, that has a very high \nii/\Ha\ \citep{Ren2025} hinting at strong AGN activity or the presence of shocks \citep[e.g.][]{Dopita1995}. We therefore excluded this galaxy from our analysis.
The stacking of \cii\ data indicates that the sample as a whole exhibits weak outflow signatures \citep{Birkin2025}, and for the majority of the sample, outflows are expected to have a negligible impact on the line fluxes. 
CRISTAL-17 (DEIMOS\_COSMOS\_742174) is also not included in this work since it is too faint in the ALMA data for kinematic classification and modelling in \citet{Lee2025b}.
The metallicity gradients of CRISTAL-17 and 24 are investigated in \citet{Fujimoto2025}. The final sample size of our study is \Ngal.

\section{Emission line measurements}\label{sec:emission_line_fitting}
To avoid introducing artificial metallicity gradients caused by variations in angular resolution as a function of wavelength, 
we homogenise the point spread function (PSF) of all channels to the size of the PSF at the wavelength of the reddest line used in our metallicity diagnostics (see Sect.~\ref{sec:strong_line_method}).
This line is chosen as either \sii\,$\lambda6731$ or \nii\,$\lambda6583$, depending on whether \sii\ is detected with sufficient S/N.
Since no stars are detected in our data cubes, 
we characterise the PSF using a data cube from another GO program (PID: 2957, PIs: H. \"Ubler and R. Maiolino). 
In each channel, we fit a 2D Gaussian to the star to determine the orientation and size of the PSF both across and along the slicer. We observe that the PSF is more elongated in the direction along the IFU slicer, consistent with \citet{DEugenio2024}.
The minor and major axes of the PSF range from $[0.16,0.18]\arcsec$ and $[0.19,0.20]\arcsec$, respectively, equivalent to $1$--$1.3$\,kpc at $z=5$.

We followed the same background subtraction method as in \citet{Fujimoto2025}.
After subtraction of the background from the datacube, and PSF homogenisation with the PSF models constructed above, by convolving each image with a kernel whose $\sigma$ equals the quadrature difference between its native PSF and the chosen target PSF, we sum spectra from spaxels located within concentric elliptical annuli.

The annuli had a width equal to one PSF HWHM (half-width at half-maximum, taken as the geometric mean of the along- and across-slicer directions), 
with typical width of $\lesssim 0\farcs2$, $\approx$\,$0.6\,$kpc at $z$\,$=$\,$5$.
 The galactocentric radius assigned to every spaxel is taken to be $(a_{\mathrm{in}}+a_{\mathrm{out}})/2$, where $a_{\mathrm{in}}$ and $a_{\mathrm{out}}$ 
are the inner and outer semi-major-axis lengths of the elliptical annulus that contains the spaxel.
Because an intrinsically thin disk is not compressed along its major axis, this definition already gives the true (de-projected) radius for such systems. 
However, for our galaxies that are likely not well described by thin disks, the same quantity is only a projected radius, with de-projection requiring an assumption on the intrinsic disk thickness. 
These geometric ambiguities introduce an additional systematic uncertainty when our gradients are compared to literature measurements that adopt different assumptions about galaxy geometry, as we discuss in Sect.~\ref{sec:gradients_fitting}.

We adjust the number of rings based on 
the outermost radius set by the S/N of the integrated line maps of \oiii\,$\lambda5007$ or \Ha. 
However, when calculating line flux ratios, we consider the S/N of the individual lines involved in the extracted spectra, the outermost radii for each line diagnostic therefore vary.
At this stage we do not apply any pixel-by-pixel rejection.
The spectra are extracted using all pixels in an annulus, and the selection relies solely on the azimuthally integrated S/N.
Rejection of pixels that have low fluxes to give robust line ratios is introduced later, in Sect.~\ref{subsec:azimuthal_variation}, where we investigate azimuthal variations across radii.
The apparent centre, axial ratio and position angle of the galaxy are derived from the emission line maps of the brightest line, either \oiii\,$\lambda5007$ or \Ha, by fitting a 2D Gaussian profile. 
In most cases, the centres coincide with those of the \cii\ moment-0 maps to within one NIRSpec pixel ($0\farcs1$), although the ALMA \cii\ data have a spatial resolution that is $2$--$4\times$ worse than that of NIRSpec.
The error spectrum is calculated as the root mean square (RMS) error using a moving window of $160$ channels, with emission lines and any spurious noise spikes masked out.

We corrected for dust reddening by comparing the observed Balmer line ratios to the theoretical values on an annulus-to-annulus basis. 
We assumed the theoretical \Ha/\Hb\ ratio equal to $2.86$ as expected for case-B recombination (i.e. ionising photons are absorbed as soon as they are emitted) 
and an electron temperature of $T_e$\,$\sim$\,$10^4$ K \citep[][Table~4.2]{Osterbrock2006}.
We then assume the nebular extinction that follows the \citet{Calzetti2000} curve.
For CRISTAL-06a and 25, the \Hb\ emission is too weak in the individual annuli, so we applied the correction using a global $A_{V}$, 
which is inferred from the spectrum integrated over a larger aperture with size covering the entire galaxy.
For other galaxies, the mean offset between the radially resolved and integrated $A_V$ is $0.05$ with a standard deviation of $0.34$, so while individual annuli can deviate, the typical variation in $A_V$ is small compared to the uncertainties and would not significantly change the measured gradients of CRISTAL-06a and 25. CRISTAL-06b has no \Hb\ detection even in the integrated spectrum, so we adopt only diagnostics that are close in wavelength for inferring metallicity.

For some galaxies, higher transition Balmer lines H$\gamma$ and H$\delta$ are also detected at lower S/N.\footnote{H$\epsilon$ is detected in some cases; however, due to its blending with \neiii\,$\lambda$3967, we refrain from using it to infer $A_{V}$.} They tend to give more uncertain $A_{V}$ than that inferred from \Ha/\Hb. We therefore adopted the $A_{V}$ from \Ha/\Hb\ only.
We do not correct for Galactic extinction explicitly, since for our sources, the foreground Galactic $A_{V}$\,$\lesssim$\,$0.4\,$mag \citep{SchlaflyFinkbeiner2011}, and becomes negligible at IR wavelengths following \citet{FitzpatrickMassa2007} extinction curve.

To extract the flux for each emission line used in Sect.~\ref{sec:strong_line_method}, 
we used \texttt{CapFit} \citep{Cappellari2023} to fit a single Gaussian profile to each detected line. We included an underlying first-order polynomial to model the baseline. 
If present in the spectra, we tie the line ratios of \oiii\ and \nii\ doublets to the theoretical values of $2.98$ and $3.071$ \citep{Storey2000}, respectively. 
All emission lines within the same annulus are fixed to the same redshift, 
but their line centroids are allowed to vary within two channels (0.002\,\micron\ or $\approx$\,$160$--$300$\,\kms\ at the wavelengths of interest)  
from the theoretical wavelength position. 
This margin accommodates bulk gas motions that can induce velocity shifts of at most $\sim$\,$150\,$\kms, as observed for \cii\ (see Table 2 of \citealt{Lee2025b}), as well as the wavelength-dependent line spread function (LSF) across the cube.
Their (observed) line widths are tied to the same value but can vary within five channels as a whole.

For five galaxies that are classified as candidate Type-1 AGNs in \citet[see our Table~\ref{tab:main_table}]{Ren2025}, we additionally include a broad Gaussian component 
to model the emission lines, which otherwise would result in residuals around the narrow component. 
However, except for the stronger lines such as \oiii\,$\lambda$5007, \Ha, and in some cases \Hb, the broad component flux converges to zero for other lines, suggesting the S/N for these fainter lines is insufficient to recover the broad component.
In cases where the broad component is non-zero, it contributes a relatively small fraction of the total line flux, ranging from a few to $<$\,$20\%$.
For our subsequent analysis, we only consider the flux from the narrow component. We find that the gradients are consistent with those obtained when using a single-component fit, only the normalisation differs. 
This agreement is expected, given that the broad component contributes only a minor fraction of the total line flux.
An example best-fit is shown in Fig.~\ref{fig:spectral_fit}.

\section{Strong line metallicity indicators and metallicity gradients}\label{sec:strong_line_method}

In this section we leverage the rich set of lines observed, giving access to multiple strong-line diagnostics.
As discussed below, different diagnostics have different strengths and combining them would give a tightened constraint on the metallicity. 
Building on the work of \citet{Fujimoto2025}, who presented metallicity gradients inferred from two diagnostics, we expand our analysis to include a broader range of strong-line diagnostics available from the data. We assess the impact of different diagnostics on the resulting metallicity gradients, complementing the analysis of \citet{Fujimoto2025}.

We derived the gas-phase metallicity by jointly analysing the measured emission line ratios using the empirical calibrations summarised in Table~\ref{tab:line_ratios}. The diagnostic line ratios often involve collisionally excited lines from singly and/or doubly ionised oxygen (O$^+$ and O$^{2+}$) or singly ionised nitrogen and sulphur (N$^+$ and S$^+$), each measured relative to hydrogen recombination lines (\Ha\ and \Hb\ in our case), except for O32 or Ne3O2.
The strengths and limitations of each line diagnostic are summarised in Table~1 of \citet[see also \citealt{Kewley2019}]{MaiolinoMannucci2019}.
In brief, some diagnostics, such as R3, are double-valued with respect to metallicity and thus require auxiliary diagnostics, such as O32 and N2, that varies monotonically with O/H to resolve this degeneracy. 
However, because nitrogen is not an $\alpha$-element and, at the metallicities relevant to our sample, is produced mainly as a secondary element, nitrogen-line indices probe the oxygen abundance only indirectly through the N/O–O/H correlation. This makes them more sensitive to variations in chemical evolution history than oxygen-based indices. 
Therefore, since some diagnostics are more sensitive to the ionisation parameter, reddening corrections, and nitrogen enrichment than others, we can better marginalise over different systematic effects by combining the various diagnostics. 
We are aware, however, that the uncertainty estimates from such a Bayesian approach may not be statistically meaningful, since the line ratios are not independent of each other. 
As discussed later in this section and Appendix~\ref{app:method_II}, to estimate uncertainties in an alternative way, we compute the metallicity gradients independently using each diagnostic and calculate the standard deviation of these gradients.

We utilised the empirical relations from \citet{Sanders2025}\ that are calibrated 
based on a sample of $\sim$\,$140$ SFGs at $z\sim3$--$10$ drawn from the AURORA survey (PID: 1914, PIs: A. Shapley and R. Sanders, \citealt{Shapley2025}) and the literature, covering the widest range in metallicity where multiple auroral lines are available (e.g. \oiii\,$\lambda4363$).
It leverages the latter 
to infer metallicities via the direct electron-temperature ($T_e$) method,
in which elemental abundances are derived from the exponential dependence of line emissivity on $T_e$ measured through highly temperature-sensitive auroral lines.
The metallicity range where each diagnostic can be reliably applied is listed in Table~\ref{tab:line_ratios}.
Unfortunately, neither \oiii\,$\lambda$4363 nor H$\gamma$ is detected in every annulus of every galaxy, and even the integrated spectra lack robust \oiii\,$\lambda$4363 measurements for a number of sources (see \citet{Faisst2026}, Table~A4, and \cite{Faisst2025}, Table~1, for direct-$T_e$ global metallicity, which is only possible for five sources, with an average uncertainty of $\approx0.3$\,dex).

In an absolute sense, different calibrations can introduce systematics of up to $0.7$\,dex \citep{Kewley2008} in absolute metallicity measurements, due to variations between empirical and theoretical methods, sample selections, and line tracers.
However, these uncertainties are less problematic when measuring metallicity gradients, because gradients rely on relative (differential) abundances rather than absolute values, and the relative systematics are typically within $\sim$\,$0.15$\,dex. 
A more recent study by \citet[see their Fig.~16]{Li2025} found $<$\,$0.1\,$dex scatter around equality when using different strong-line calibrations \citep[e.g.][]{Maiolino2008,Curti2020,Nakajima2022,Sanders2024}, which is within their measurement uncertainty; there are no obvious systematic offsets between different calibrations.
We verified that our derived metallicity gradients with the calibrations of \citet{Bian2018}\footnote{S2 and O2 are not included.} and \citet{Nakajima2022}\footnote{We adopt the calibration for the `All' sample and omit Ne3O2 because of its nearly flat dependence on \lOH.}
are in general consistent with those obtained using the \citet{Sanders2025} calibration, 
differing on average by only $-0.01\,$\dexkpc and $-0.02$\,\dexkpc, respectively. 
In no case did the sign flip when an alternative calibration was applied.
Both \citet{Bian2018} and \citet{Nakajima2022}, however, are based on local-Universe analogues of high-redshift galaxies. Now that JWST increasingly provides direct auroral-line measurements in high-$z$ galaxies, calibrations anchored to these new observations are preferable to those relying on local analogues.
We also investigated the systematics associated with using different line tracers from the \citet{Sanders2025} calibration
to infer metallicity gradients later in this section.

We expected the contribution from the diffuse ionised gas (DIG), also known as the warm ionised medium, with $n_e$\,$\lesssim$\,$0.1\,$cm$^{-3}$, $T$\,$\sim$\,$8000$\,K, 
in our galaxies to be minimal, although it could bias metallicity measurements in nearby disks \citep[e.g.][]{Poetrodjojo2019,ValeAsari2019, Chevance2020,Belfiore2022}. 
Our systems have, on average, $\Sigma_{\mathrm{SFR}}$\,$\sim$\,10\,M$_\odot$\,yr$^{-1}$\,kpc$^{-2}$, about two orders of magnitude higher than in local disks where DIG dominates.  
As demonstrated in \citet{Sanders2017} and \citet{Shapley2019}, given the observed anticorrelation between $\Sigma_{\mathrm{SFR}}$ and $f_{\mathrm{DIG}}$ \citep[][and see also \citealt{Belfiore2022}]{Oey2007}, any DIG contribution to the spectra for our galaxies is likely to be small ($\lesssim20\%$).

The specific line ratios used for each system in each annuli depend on the detection of the involved lines at S/N $\gtrsim$\,2.5.
When lines such as \oiii\ and \Hb\ are covered by both gratings (G235M and G395M), we select the grating that also includes the other relevant lines. 
For example, for R3, we use G395M for \oiii\ and \Hb\ for its higher S/N, if possible, but for O32, we switch to G235M to evaluate the line ratio involving \oiii\ and \oii, as both lines fall within this grating. This is to avoid potential differences in the absolute calibration of data from different gratings.
The fluxes of the same line are consistent with each other in both gratings, but with larger uncertainty in G235M as expected.

\begin{table}
\caption{Line diagnostics and the calibrations used.}
\label{tab:line_ratios}
\centering
\begin{tabular}{l p{5cm} c}
\hline\hline  
Indicators & Emission line ratio & $Z$ Range\tablefootmark{a} \\
\hline  
N2\tablefootmark{$^\dagger$} &  \Ntwo & $[7.8,8.6]$ \\
R3{$^\dagger$} & \Rthree &  $[7.3,8.6]$\\
Ne3O2{$^\dagger$} & \neiii$\,\lambda$3869/\oii\,$\lambda\lambda3726,3729$ & $[7.4,8.6]$\\
O32 & \oiii $\,\lambda$5007/\oii\,$\lambda\lambda3726,3729$ & $[7.3,8.6]$\\
O2 & \oii\,$\lambda\lambda 3726,3729$/\Hb  & $[7.3,8.6]$\\
S2 & \sii\,$\lambda\lambda6716,6731$/\Ha & $[7.9,8.6]$ \\
  \hline
\end{tabular}
\tablefoot{
The calibrations are based on \citet{Sanders2025}.
\tablefoottext{a}{The metallicity range where each diagnostic can be reliably applied.}
\tablefoottext{$^\dagger$}{Relatively unaffected by dust reddening.}
}
\end{table}

\begin{table}
    \caption{Metallicity gradients inferred for the CRISTAL samples using multiple strong lines.}
    \label{tab:gradients}
    \centering
\begin{tabular}{>{\small}l>{\small}c>{\small}c>{\small}c}
\hline\hline  
ID  & Class.\tablefootmark{a} & Method I \tablefootmark{b} & Method II\tablefootmark{c} \\  
& &{\scriptsize (dex\,${\rm kpc^{-1}}$)} & {\scriptsize (dex\,${\rm kpc^{-1}}$)} \\ \hline  
01a & Non-Disk &${0.107_{-0.055}^{+0.054}}$ & ${0.107_{-0.028}^{+0.028}}$ \\ 
02 & Disk &${0.031_{-0.017}^{+0.016}}$ & ${0.029_{-0.015}^{+0.015}}$ \\ 
03 & Best Disk &${0.065_{-0.031}^{+0.030}}$ & ${0.080_{-0.039}^{+0.039}}$ \\ 
04a & Non-Disk &${0.031_{-0.020}^{+0.020}}$ & ${0.041_{-0.058}^{+0.058}}$ \\ 
05a & Disk &${0.055_{-0.057}^{+0.058}}$ & ${0.049_{-0.051}^{+0.051}}$ \\ 
06a & Non-Disk &${-0.054_{-0.081}^{+0.083}}$ & ${-0.026_{-0.054}^{+0.054}}$ \\ 
06b & Disk &${-0.035_{-0.292}^{+0.285}}$ & ${-0.004_{-0.032}^{+0.032}}$ \\ 
07a & Disk &${0.099_{-0.026}^{+0.025}}$ & ${0.125_{-0.027}^{+0.027}}$ \\ 
07b & Non-Disk &${0.166_{-0.090}^{+0.091}}$ & ${0.079_{-0.088}^{+0.088}}$ \\ 
09 & Disk &${0.101_{-0.114}^{+0.115}}$ & ${0.101_{-0.099}^{+0.099}}$ \\ 
10a-E & Disk &${0.017_{-0.027}^{+0.026}}$ & ${0.020_{-0.018}^{+0.018}}$ \\ 
11 & Best Disk &${0.025_{-0.040}^{+0.042}}$ & ${0.025_{-0.046}^{+0.046}}$ \\ 
13 & Non-Disk &${0.069_{-0.038}^{+0.037}}$ & ${0.029_{-0.033}^{+0.033}}$ \\ 
14 & Non-Disk &${0.036_{-0.064}^{+0.067}}$ & ${0.013_{-0.025}^{+0.025}}$ \\ 
15 & Best Disk &${0.057_{-0.044}^{+0.044}}$ & ${0.033_{-0.061}^{+0.061}}$ \\ 
19 & Best Disk &${0.038_{-0.036}^{+0.036}}$ & ${0.024_{-0.045}^{+0.045}}$ \\ 
20& Best Disk &${-0.029_{-0.056}^{+0.053}}$\tablefootmark{d} & $\ldots$ \\ 
22a& Disk &${-0.019_{-0.043}^{+0.042}}$\tablefootmark{d} & $\ldots$ \\ 
25 & Non-Disk &${0.051_{-0.100}^{+0.099}}$ & ${0.044_{-0.029}^{+0.029}}$ \\ 
VC-7875 & Disk &${0.040_{-0.023}^{+0.024}}$ & ${0.039_{-0.029}^{+0.029}}$ \\ 
\hline
\end{tabular}
\tablefoot{
\tablefoottext{a}{Same as Table~\ref{tab:main_table}}
\tablefoottext{b}{Best-fit metallicity gradients derived from the Bayesian fitting that combines the different diagnostics listed in Table~\ref{tab:line_ratios}.} 
\tablefoottext{c}{Median of the best-fit metallicity gradients derived from each individual diagnostic.
The individual values of the gradients are listed in Table~\ref{tab:gradients_eachdiag}.}
\tablefoottext{d}{Inferred from the metallicity maps in \citet{Parlanti2025} and \citet{Jones2025}.}
}
\end{table}

\begin{figure*}
  \centering
  
  \includegraphics[width=0.7\textwidth]{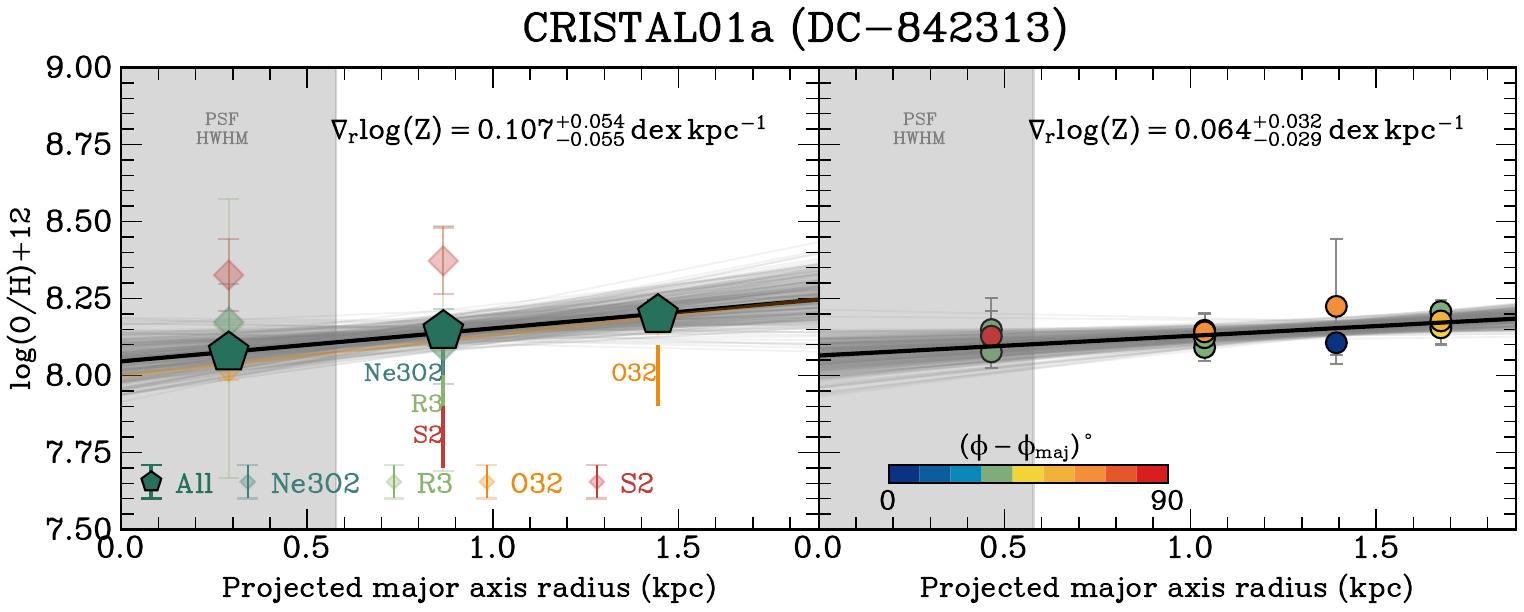}
  \includegraphics[width=0.7\textwidth]{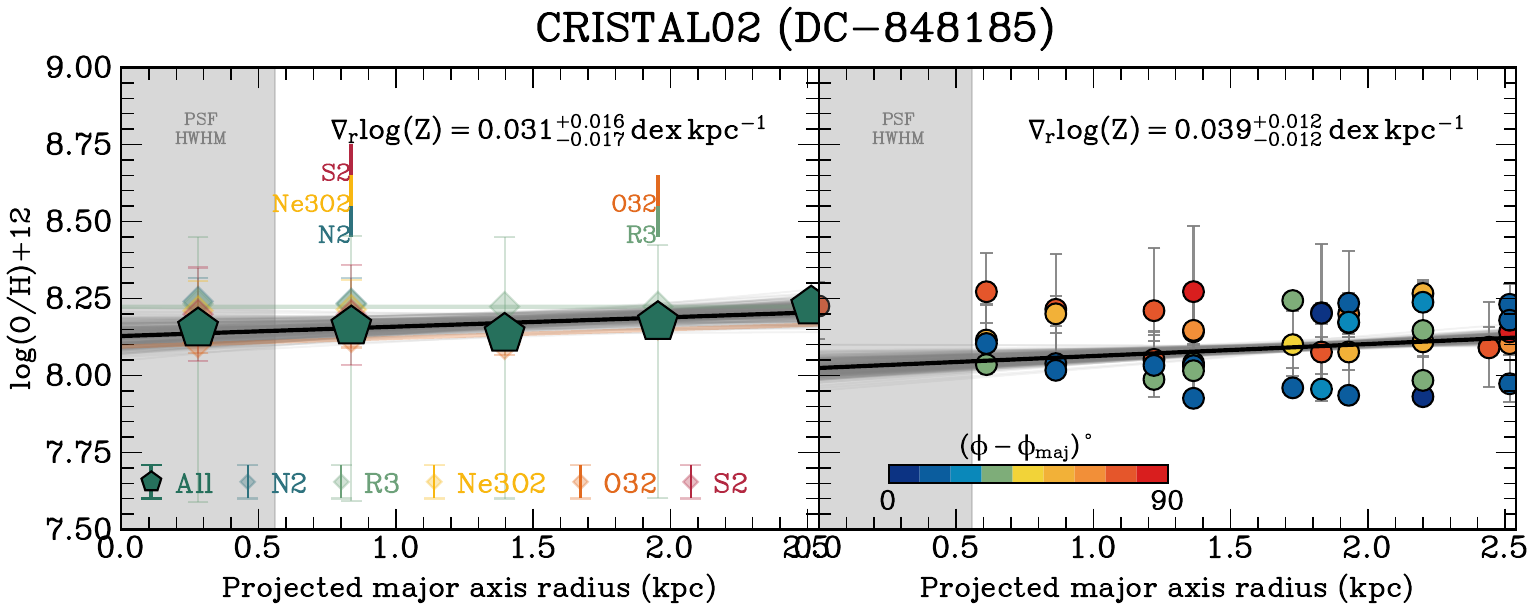}
  
  \includegraphics[width=0.7\textwidth]{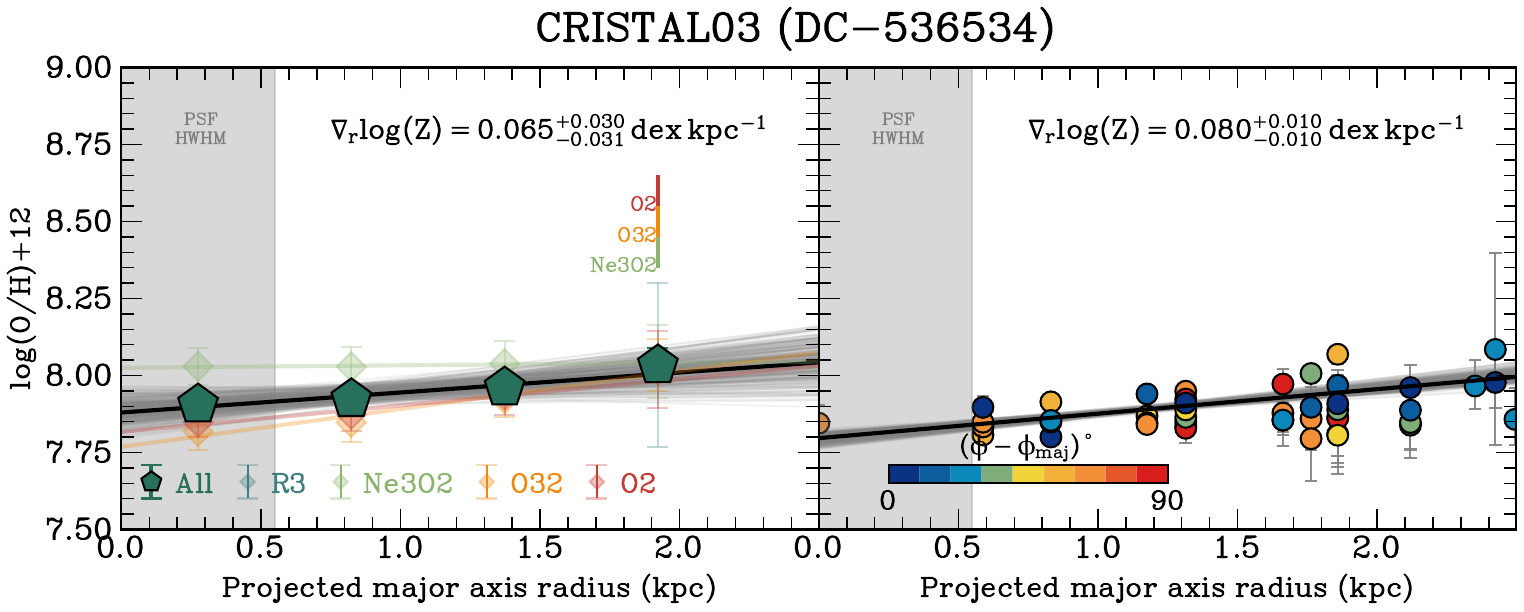}
  \includegraphics[width=0.7\textwidth]{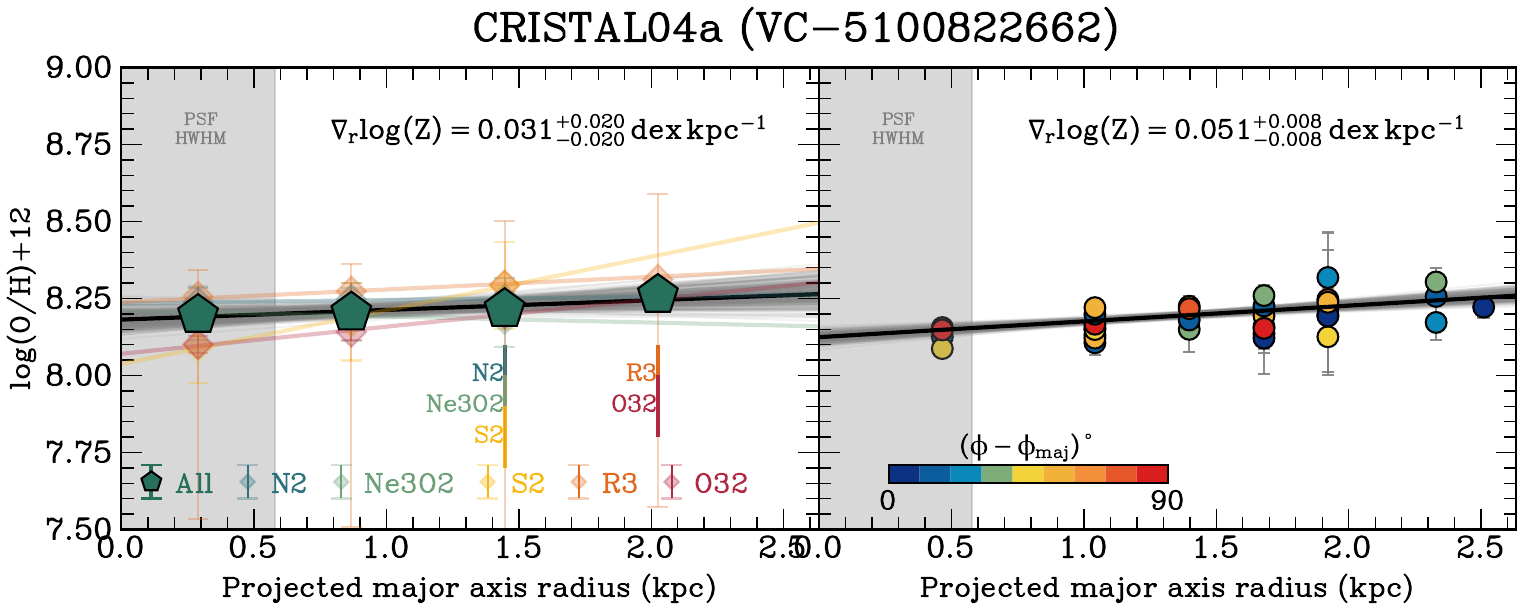}
  \caption{
Radial profiles of metallicities and their best-fit linear models (black lines) for CRISTAL-01a to 04a.
\textit{Left}: Azimuthally averaged profiles for each galaxy. Dark green pentagons with black outlines represent the metallicities derived by combining the listed line diagnostics using a Bayesian approach (Method I). 
The translucent grey lines are 300 random draws from the posterior distribution of the model parameters.
The outermost radius at which each diagnostic is used is annotated at the corresponding location on the profile.
The fainter coloured diamonds represent the metallicities obtained from individual diagnostics via Method~II, while the lines in the corresponding colours show the best-fit models for each diagnostic. \textit{Right}: Pixel-based metallicity profile. Each data point is coloured coded by its (absolute) acute azimuthal angle difference from the major axis.
Line diagnostics of which only two annuli are available are shown (coloured diamonds) but not fitted.
The profiles for the full samples are shown in Appendix~\ref{app:remaining_profiles}.
}
\label{fig:gradient_fit}
\end{figure*}

\begin{figure*}
\centering
    \includegraphics[width=\textwidth]{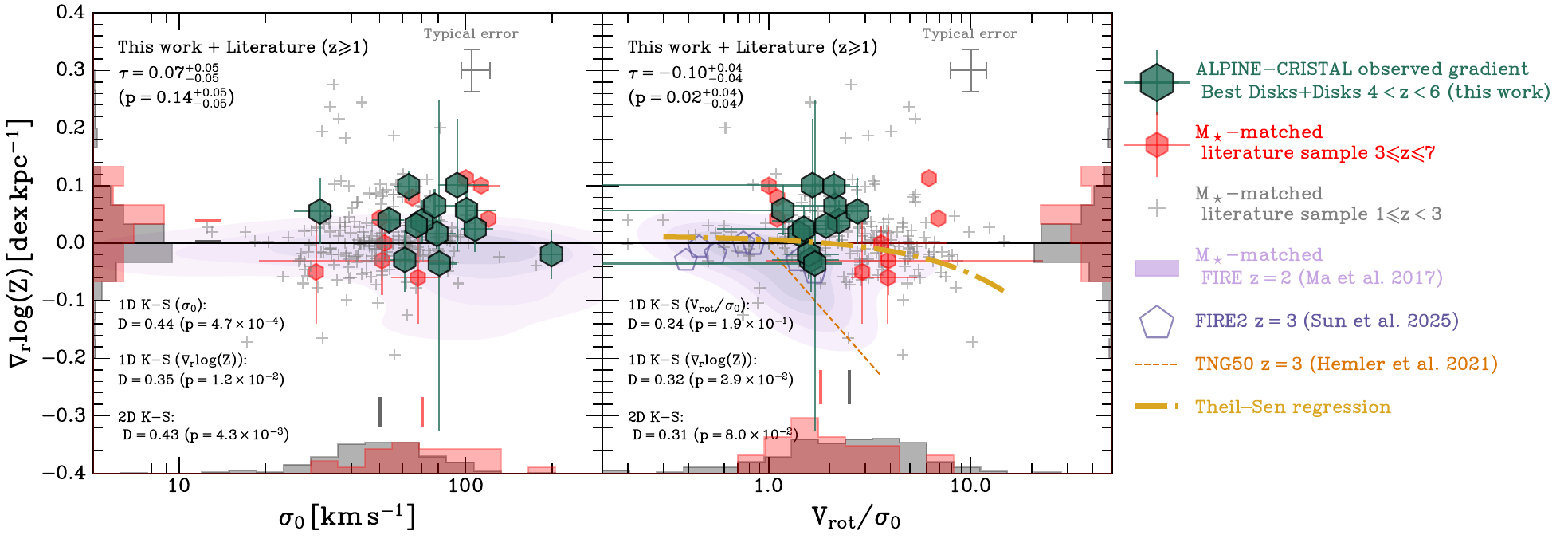}
      \caption{
      Distributions of metallicity gradients, \dlOH\ as a function of the velocity dispersion, $\sigma_0$ (\textit{left}), and the ratio of rotational velocity to velocity dispersion, $V_{\rm rot}/\sigma_0$ (\textit{right}), for the CRISTAL disk galaxies (green hexagons). 
    We also show \mstar-matched literature samples at $3$\,$\lesssim$\,$z$\,$\lesssim7$ as red hexagons, and $z$\,$<$\,$3$ as grey crosses. 
    The references for these literature samples are listed in Table~\ref{tab:literature_sample}. 
    The Kendall's $\tau$ correlation coefficients and the $p$-values are annotated in the top left in each panel.
    Histograms of the distributions of the $3$\,$\lesssim$\,$z$\,$\lesssim7$ and $z$\,$<$\,$3$ samples are shown along the $x$- and $y$-axes in red and grey, respectively.
    The combined CRISTAL and literature sample exhibits a shallow negative correlation between \dlOH\ and $V_{\rm rot}/\sigma_0$, as indicated by the Kendall's $\tau$ value. 
    The dot-dashed yellow line is the Theil-Sen median-slope regression \citep{Sen1968}, included solely to guide the eye; it is not used in any statistical tests.
    In contrast, the Kendall's correlation test reveals no significant correlation with $\sigma_0$, which could be due to the inherently shallow trend and the large scatter.
    However, comparing the overall distribution of the two populations at $z$\,$<$\,$3$ (grey histogram) and $z$\,$\geqslant$\,$3$ (red histogram), 
    a clearer distinction emerges between the two populations. 
    Specifically, the $z$\,$\geqslant$\,$3$ population tends to have higher $\sigma_0$ values and more positive metallicity gradients, whereas the $z$\,$<$\,$3$ population has lower $\sigma_0$ values and, on average, lower metallicity gradients.
     This distinction is supported by the significant 2D K-S test statistics, which gives a nearly 3-$\sigma$ significance level (annotated at the bottom left of each panel).
    The predicted trends from cosmological simulations are shown, including FIRE at $z$\,$=$\,$2$ from \citealt{Ma2017} (purple contour), FIRE2 at $z$\,$=$\,$3$ from \citealt{Sun2025} (\textit{right}, open purple pentagons), and TNG50 at $z$\,$=$\,$3$ from \citealt{Hemler2021} (\textit{right}, dashed orange line). 
    While the simulations capture the slight positive and negative relationships between \dlOH\ and $\sigma_0$ and $V_{\rm rot}/\sigma_0$, respectively, the normalisation and steepness of these relationships differ from the observations. 
      }
      \label{fig:DelZ_s_Vs}
\end{figure*}

\begin{figure}
\centering
    \includegraphics[width=0.4\textwidth]{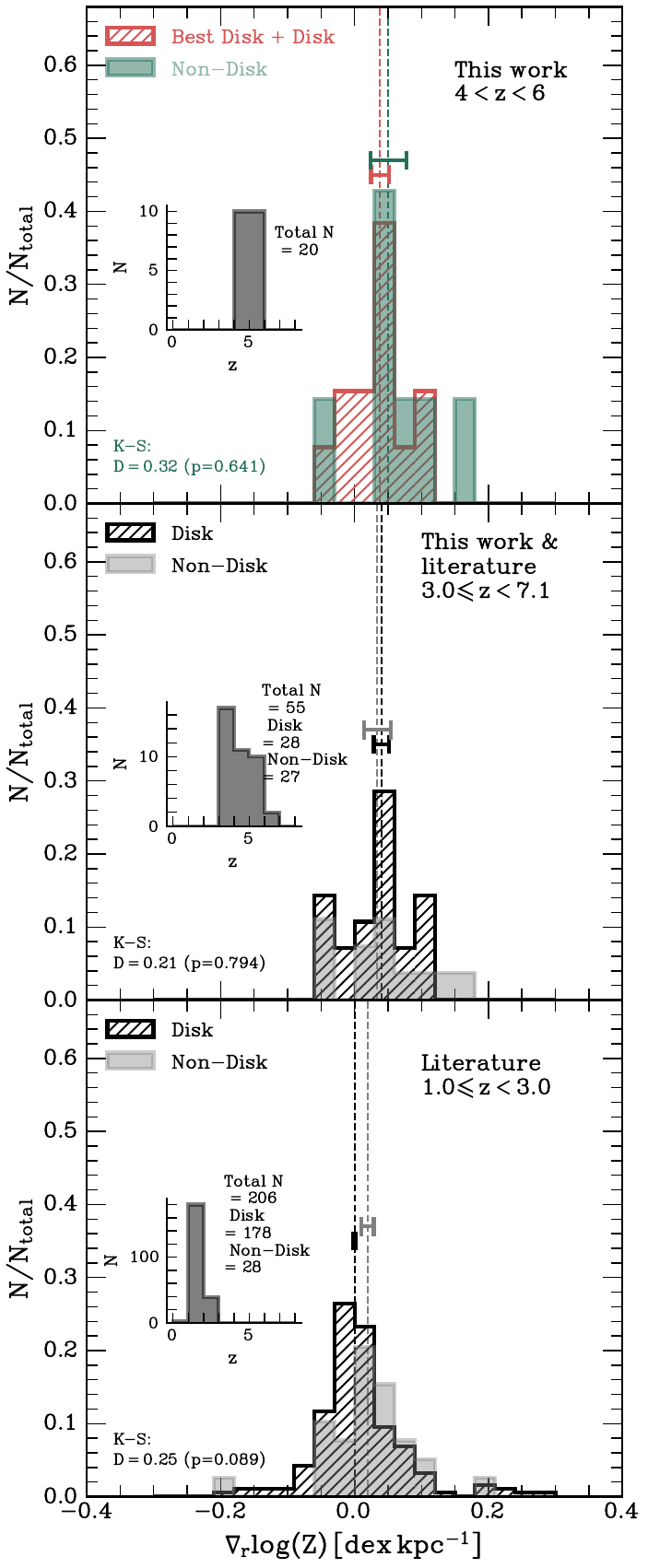}
      \caption{Metallicity gradients in disk and non-disk subsets of our sample (\textit{top}) and compiled stellar mass-matched sample from literature at $3<z\lesssim8$ (\textit{middle}) and $1<z<3$ (\textit{bottom}). The median values of the distributions are indicated by the vertical dashed line of the corresponding colours, and the errors of the median (obtained by bootstrapping) are represented by the horizontal error bars of the same colour. 
      The literature references are listed in Table~\ref{tab:literature_sample}.
      }
      \label{fig:DelZ_disknondisk}
\end{figure}

We determined the metallicities in each radial bin using two methods. The first approach employs Bayesian inference with the \texttt{emcee} package \citep{ForemanMackey2013} to sample the posterior probability distribution of \lOH, i.e. to find the metallicity that best satisfies all calibrations at once.
We name this method `Method I', and provide more description in Appendix~\ref{app:bayes}.
The inferred radial profile of these metallicities is then used to derive a gradient (denoted as `Method I’ in Table~\ref{tab:gradients}).
This method is similarly applied in the spaxel-to-spaxel fitting in Sect.~\ref{subsec:azimuthal_variation} to investigate azimuthal variations in metallicity.

For CRISTAL-20 and CRISTAL-22a, we utilised the metallicity maps presented in the works of \citet{Parlanti2025} and \citet{Jones2025}, respectively. 
In addition 
The metallicities were inferred from N2, R3, and O3N2 diagnostics \citet{Parlanti2025}, 
while \citet{Jones2025} additionally used RS32, O3S2, and O3N2, employing the calibrations 
from \citet{Curti2017} and \citet{Curti2020}, respectively.\footnote{O3N2\,$\equiv$\,([\ion{O}{iii}]\,$\lambda$5007/\Hb)/([\ion{N}{ii}]\,$\lambda$6583/\Ha); 
RS32\,$\equiv$\,([\ion{O}{iii}]\,$\lambda$5007/\Hb)+(\sii\,$\lambda\lambda$\,6716,6731/\Ha); 
O3S2\,$\equiv$\,([\ion{O}{iii}]\,$\lambda$5007/\Hb)/(\sii\,$\lambda\lambda$\,6716,6731/\Ha)}  
We share the same R3, N2, and S2, but not the RS32, O3S2, and O3N2 diagnostics. 
However, the extra indices are built from the same set of bright lines, 
(\oiii, \nii, \sii, H$\alpha$, H$\beta$) and are therefore not independent. 
Within the metallicity ranges spanned by the two objects ($\log(Z)$\,$\in$\,$[8.32,8.38]$ for CRISTAL-20 and $\log(Z)$\,$\in$\,$[8.44,8.47]$ for CRISTAL-22a), the \citet{Curti2017,Curti2020} and \citet{Sanders2025} calibrations are nearly parallel and share the same slope sign.
Hence, a negative metallicity gradient would remain negative if \citet{Sanders2025} were used, and vice versa.
However, the absolute metallicity values may differ by up to $0.3$\,dex higher \citep[see Fig.~15 of][]{Sanders2025}.

To quantify the systematic differences between the gradients inferred from different line diagnostics, and to provide an alternative estimate of the uncertainties associated with the metallicity gradients, we also derive metallicities from individual line diagnostics by directly inverting the empirical calibration curves for the measured line ratios.
We name this method `Method II'.
The gradient for each diagnostic is fitted individually only when there are more than two annuli available, and the median of these gradients is reported in Table~\ref{tab:gradients}.

Following other $z$\,$\gtrsim$\,$1$ studies \citep[e.g.][]{Stott2014,EWuyts2016,Carton2018}, we parametrised the radial metallicity profile using a linear function, 
\begin{equation}
\log_{10}Z(r)=\nabla_r\log_{10}(Z) \cdot r + Z_0, 
\label{eqn:Zeqn}
\end{equation}
where $\log_{10}Z=12+\log(\rm{O/H})$, and \dlOH\ is defined as the metallicity gradient and $Z_0$ is the central metallicity. 
For both Methods, we fit the radial profiles with \texttt{emcee}, assuming flat priors for both \dlOH\ and $Z_0$. The chains are initialised with the best-fitting solutions obtained from \texttt{mpfit} and then evolved with 64 walkers for 2000 steps, discarding the first 500 steps as burn-in. 

The left panels in 
Fig.~\ref{fig:gradient_fit} plot the resulting best-fit metallicities profiles from Methods I and II for CRISTAL-01a to 04a. We show the profiles of the full sample in Appendix~\ref{app:remaining_profiles}. 
The values are summarised in Table~\ref{tab:gradients}. 
The typical value of measurement uncertainties is on average $\sim$\,$0.06\,$\dexkpc.
We find that in general \dlOH\ and $Z_0$ obtained from Method II 
for individual diagnostics agree well with those from Method I, 
except in some cases where N2 (from Method II) prefers a flatter gradient. 
The gradients inferred from Method I and II, in terms of median for the latter, 
agree within 1$\sigma$ as shown in Fig.~\ref{fig:medianvsall}. 
We discuss further the difference between the two methods in Appendix~\ref{app:method_II}.
Henceforth, we adopt the values from Method I as our fiducial values. 

The mean metallicities of our radial profiles agree within 0.3\,dex with \citet{Faisst2025}, which uses \citet{Sanders2024} calibrations. 
When comparing our metallicity gradients with those inferred by \citet{Fujimoto2025}, who used the R3 and O32 diagnostics based on the same calibrations from \citet{Sanders2025}, we find good agreement if we restrict our analysis to the same two diagnostics.
 Minor differences between our results can be attributed to the choice of apertures in multiple-component systems, such as CRISTAL-01a. The choices are made because of the different science goals: our study focuses on the component whose kinematics can be reliably classified and studied, whereas \citet{Fujimoto2025} emphasise the full system.

Across the sample, the median \dlOH\,$=$\,\mediandlOH\,\dexkpc\ (standard deviation $=$\,\sdlOH).
All galaxies except CRISTAL-06a, 06b, 20 and 22a have positive metallicity gradients. However, there are no galaxies with \dlOH\,$>$\,$+0.05$\,\dexkpc\ at $3\sigma$ level, 
and only three galaxies at $1\sigma$. 
If we consider strictly flat metallicity gradient, 
i.e. \dlOH\,$=$\,$0$\,\dexkpc, 
there are 11 galaxies with \dlOH\,$>$\,$0$\,\dexkpc, but only at 1$\sigma$ level.
For CRISTAL-06a, 06b, 20 and 22a, none of them have gradients significantly steeper than 
$-0.05$\,\dexkpc\ (typical values of the Milky Way and local spirals) and they are consistent with $0$\,\dexkpc\ at 1$\sigma$ level. 
For CRISTAL-20 and -22a, we have verified the sign of the gradients would not reverse if \citet{Sanders2025} calibration was used, as discussed in Sect.~\ref{sec:strong_line_method}.

For the galaxies that exhibit mild positive gradients, the gradients could be produced by (at least) two non-exclusive mechanisms: (i) the accretion of metal-poor filaments that stream gas penetrating the galaxy halo and deliver pristine gas directly to the inner kiloparsec \citep{Dekel2009,Ceverino2016}, (ii) galactic outflows with high mass loading factors could preferentially remove metals from the central regions, and a fraction of this enriched gas could cool and be re-accreted at the outer regions of the galaxy, the so-called galactic fountain \citep{Gibson2013,Ma2017}, although direct evidence of metal-rich fountains at $z > 4$ is still lacking. These processes could lead to an increase in the outer metallicity, resulting in positive metallicity gradients .

The flatness of our observed metallicity gradients could be potentially an observational effect of angular resolution \citep{Yuan2011,Wuyts2016}. We explore whether beam smearing could contribute to the shallow gradients observed in most of our sample in Sect.~\ref{sec:effects_of_beam_smearing}.

Notably, the three galaxies with the steepest gradients are CRISTAL-01a, 07a and 07b, which are known to have an immediate neighbour within $\sim$\,$10$\,kpc.
CRISTAL-01a resides in a dense region with multiple companions and a massive neighbour, SMG\,J1000+0234 \citep{JimenezAndrade2023,Solimano2024,Solimano2025,Rubet2025}. 
The observed kinematic features 
suggest that this system may be in the later stages of a merger, and the misaligned velocity gradient hints at gas inflow triggered by the merger \citep{Lee2025b}. 
CRISTAL-07a and 07b is an interacting pair, with CRISTAL-07a retaining a lot of its disk rotation signatures.

The interacting nature of CRISTAL-01a, 07a, and 07b may contribute to their observed positive metallicity gradients as found at lower redshifts \citep[e.g.][]{Kewley2010,Rupke2010a,Queyrel2012,Carton2018,XWang2022}, and also reported in one case at $z$\,$\sim$\,$8$ by \citet{Venturi2024}. 
However, not all interacting pair systems in our sample exhibit positive gradients; for example, CRISTAL-06a and 06b have slight negative gradients instead. The uncertain stage of merger for these systems complicates a direct association between interaction and metallicity gradients.

\section{Gas-phase abundance gradients and their potential drivers}\label{sec:gradients_fitting}
\subsection{Metallicity gradients in disks}\label{subsec:Z_disks}
Studies of star-forming disks at cosmic noon have found that galaxies with 
higher gas velocity dispersions $\sigma_0$ tend to have flatter and in some cases even positive radial metallicity gradients \citep{Queyrel2012,Jones2013,Leethochawalit2016,Gillman2021,XWang2022}. 
Consistent with this trend, the metallicity gradients exhibit a mild anti-correlation with the ratio $V_{\rm rot}/\sigma_0$, where $V_{\rm rot}$ is the rotation velocity.\footnote{As discussed in Sect.~6.1 in \citet{Lee2025b}, the definition and methodology used to derived $V_{\rm rot}$ and $\sigma_0$ varies across literature.}
Both trends between metallicity gradients and $\sigma_0$ and $V_{\rm rot}/\sigma_0$ observed in the literature
are illustrated in Fig.~\ref{fig:DelZ_s_Vs}.

We investigated the relationships between metallicity gradients, \dlOH, and the gas-phase velocity dispersion, $\sigma_0$, 
as well as $V_{\rm rot}/\sigma_0$, 
from \cii\ for the 13 disk galaxies \citep{Fujimoto2020,Lee2025b}
and samples compiled from the literature at $z$\,$\gtrsim$\,$1$ in Fig.~\ref{fig:DelZ_s_Vs}.
The literature samples included are listed in Table~\ref{tab:literature_sample} (those with kinematics references, and see the discussion in Appendix~\ref{app:literature_sample}).
We select galaxies that fall within the mass range of our sample. 
The resulting distribution of stellar masses is shown in Fig.~\ref{fig:mstar_dist_lit}. 
 The stellar mass values are taken from the literature, which we discuss in detail in Sect.~\ref{subsec:mass_and_ssfr}.

The literature data were obtained using a variety of instruments, ranging from ground-based to space-based, and from grism to IFU spectrographs, 
with widely varying spatial and spectral resolutions.
Furthermore, the metallicities were inferred using different strong-line diagnostics and calibrations.
However, as discussed in Appendix~\ref{app:literature_sample}, the current sample size does not allow for a more unified sample selection, so we chose to retain the literature sample as is, with only the stellar mass selection. 
To maximise statistical coverage in our kinematics measurement, we do not unify the gas phase from which the measurement is taken. In addition, the different beam-smearing correction method and modelling methodologies  \citep{Uebler2019,Lee2025a,Wisnioski2025} could introduce systematic scatter in the kinematics data, which we could not correct here.
Nevertheless, since the kinematic data are predominantly derived from IFU observations, the known bias in using long-slit or grism data \citep{Uebler2019} is mostly mitigated.

We find a shallow yet statistically significant ($p$\,$<$\,$0.05$) negative correlation between \dlOH\ and $V_{\rm rot}/\sigma_0$ among the CRISTAL disks and literature values, with a Kendall's $\tau$ correlation coefficient of $-0.10^{+0.04}_{-0.04}$. The quoted uncertainties are obtained from the empirical distributions of Monte-Carlo bootstrapping for 1000 resamples. To further verify the trend is not driven by one object, we applied jack-knife resampling that gives a distribution of $\hat{\tau}$ values whose $[5,95]^{\rm th}$-percentile fall within $[-0.13,-0.07]$. These intervals lie well within the confidence intervals estimated from bootstrapping.

Observations of MS SFGs from $z\lesssim8$ to $z=1$ suggest that $V_{\rm rot}/\sigma_0$ increases with cosmic time \citep[Fig.~9 in][]{Lee2025b}, a trend that is also seen in simulations \citep[e.g.][]{Pillepich2019,Sun2025}.
Therefore, the correlation between \dlOH\ and $V_{\rm rot}/\sigma_0$ 
likely reflects the cosmic evolution of \dlOH\ with redshift, where \dlOH\ becomes more negative towards lower redshift, as discussed in \citet{Fujimoto2025} 
in the context of our sample and 
\citet{Curti2020a,Venturi2024} 
for a general discussion.

Since $V_{\rm rot}/\sigma_0$ is an indicator of disk maturity in terms of rotational support, 
the coherent picture of metallicity gradients decreasing with both $V_{\rm rot}/\sigma_0$ and cosmic time implies that disk maturity plays a dominant role in driving the negative metallicity gradient, among also other evolving variables with redshift.
This is consistent with simulations, such as those presented in \cite{Ma2017,Bellardini2021,Sun2025}, which show that disks become increasingly rotationally supported over time and consequently, suppress radial mixing of metals, leading to a gradual steepening of the radial metallicity gradients.

In contrast, no discernible correlation is observed between \dlOH\ and $\sigma_0$, as indicated by a Kendall's $\tau$ of $0.06$ and a $p$-value $=0.18$, which rules out any strong trends. 
The lack of trend between \dlOH\ and $\sigma_0$ could be attributed to the inherently shallow trend as predicted by FIRE simulation \citep[for $z=2$,][]{Ma2017}, and the large scatter preclude any trends.
Nevertheless, the overall distribution of the two populations at $z$\,$<$\,$3$ and $z$\,$\geqslant$\,$3$, a clearer distinction emerges between the $z$\,$<$\,$3$ and $z$\,$\geqslant$\,$3$ populations. 
Observationally, $\sigma_0$ exhibits a decreasing trend with decreasing redshift \citep[Fig.~9 in][]{Lee2025b}. 
Consequently, the distinction between the $z$\,$<$\,$3$ and $z$\,$\geqslant$\,$3$ populations could be, in part, reflecting the trend between $\sigma_0$ and the metallicity gradient.
This distinction is supported by the two-dimensional Kolmogorov-Smirnov (K-S) D-statistics \citep{Peacock1983,Fasano1987,Press2007} test statistics, which gives a nearly 3-$\sigma$ significance level.
For reference, the D-statistics from the 1D K-S test of the galaxies in our sample for each variable are also provided in Fig.~\ref{fig:DelZ_s_Vs}.

The $\sigma_0$ and $V_{\rm rot}/\sigma_0$ values could be used to gauge the local stability of a gaseous disk through the Toomre $Q$ parameter, 
in which $Q\propto (\sigma_0/V_{\rm rot})f_{\rm gas}^{-1}$ \citep[e.g.][]{Genzel2011}.\footnote{An approximation neglecting the contribution from the stellar and other gaseous components, a multi‐component or effective $Q$ would give a slightly different numerical value but the same qualitative behaviour \citep{WangSilk1994,RomeoFalstad2013}.} The lower the $Q$, the more unstable the disk against local collapse.
CRISTAL galaxies have high gas fractions, $f_{\rm gas} \approx 0.5$ on average, and a median $Q \sim 0.6$ \citep{Lee2025b}, close to the critical value for a marginally stable thick disk \citep[e.g.][]{Behrendt2015}.  
In such low-Q conditions, the disk is expected to fragment into giant, star-forming clumps that could contain $\sim$\,$7\%$ of the total stellar mass \citep[][and references therein]{nmfs2020}.  
The gravitational torques exerted by these clumps, together with global non-axisymmetric modes, drive radial gas inflow that can efficiently flatten the metallicity gradient \citep[e.g.][]{Bournaud2007,Bournaud2011,Goldbaum2016}.
Simulations suggest, however, that only the most massive and densest long-lived clumps are likely to survive the inward migration across the disk \citep{Dekel2022,Ceverino2023}. Observations have yet to provide clear evidence of such clump migration \citep{Claeyssens2025,Kalita2025b}.

Morphologically, many CRISTAL galaxies exhibit pronounced clumpy structure in rest-frame optical JWST/NIRCam images \citep[][see also \hc]{Lee2025b}, whereas their stellar-mass distributions, 
inferred from resolved spectral energy distribution (SED) modelling \citep{JLi2024,Lines2025} and ALMA \cii\ data \citep{Ikeda2025,Lee2025b}, appear relatively smooth.

Most of the disks in our sample lack a massive bulge, as inferred from kinematic analysis \citep{Lee2025b}. The flat metallicity gradients observed in these disks lend additional support for the absence of a strong and mature bulge component at \zcristal.

The empirical trends between metallicity gradients and $\sigma_0$ and $V_{\rm rot}/\sigma$ are broadly reproduced by cosmological simulations from FIRE and TNG50 \citep{Ma2017,Hemler2021} for $z\gtrsim2$ (see also the analytical model by \citealt{Sharda2021}). They suggest that a galaxy requires a highly rotation-dominated cold disk to develop a steep negative metallicity gradient, whereas galaxies lacking ordered rotation tend to exhibit flat gradients. 
More recently in \citet{Sun2025}, they also found galaxies with weak rotational support ($V_{\rm rot}/\sigma_0$\,$\lesssim$\,$1$) are more likely to develop positive metallicity gradients. 
In these simulations, the observed correlation with $V_{\rm rot}/\sigma_0$ is driven by the effective redistribution of metal-rich gas in dispersion-dominated galaxies through processes such as rapid gas infall, which flattens the metallicity gradients.

\subsection{Metallicity gradients in disks versus non-disks}\label{subsec:disk_nondisk}
Figure~\ref{fig:DelZ_disknondisk} shows the distribution of \dlOH\ of disks and non-disks of our sample.
We find no significant distinction between disks and non-disks as classified in \citet{Lee2025b}, with median \dlOH\,$=$\,\dlOHdisk\,\dexkpc\ and \dlOHnondisk\,\dexkpc, respectively, with a negligible difference of $<$\,$1\sigma$ level, and a K-S test does not indicate the two populations are distinct with score of \dlOHdisknondiskKS\ ($p=$\,\dlOHdisknondiskKSp).

We also compare with the disks and non-disks in the literature listed in Table~\ref{tab:literature_sample}.
Galaxies are classified as disks or non-disks following the criteria adopted in the original studies, using either (i) their kinematic designations (`rotating' versus `non-rotating') 
or (ii) their morphological classifications from imaging data, including \textit{Hubble} Space Telescope (HST) observations.

For the literature sample at $3$\,$\leqslant$\,$z$\,$\lesssim$\,$7$, we find no significant distinction between the disk and non-disk populations, with an insignificant difference in their median values (disk: \dlOHdiskCRILIT\,\dexkpc, non-disk: \dlOHnondiskCRILIT\,\dexkpc). 
When combining our sample with the existing literature at $3$\,$\leqslant$\,$z$\,$\lesssim$\,$7$ (Fig.~\ref{fig:DelZ_disknondisk}), increasing the number of objects from $35$ to $55$, the difference between the disk and non-disk populations remains negligible in terms of median difference, with high $p$-value $=0.80$ from K-S test.

On the other hand, a more noticeable difference in metallicity gradients between disks and non-disks is observed at 
$1$\,$\leqslant$\,$z$\,$<$\,$3$ in our compiled sample of \Ngallowz\ galaxies shown in Fig.~\ref{fig:DelZ_disknondisk}, 
with median values of \dlOHdiskCRILITlowz\,\dexkpc\ and \dlOHnondiskCRILITlowz\,\dexkpc, 
respectively, 
although this difference in median values is only significant at $2\sigma$ level.
This significance is comparable to that reported in the 21 disk and non-disk samples in SINS/zC-SINF at $z$\,$\sim$\,$2$ by \citet{nmfs2018}.
Similarly, \citet{Gillman2021} found no difference in the metallicity gradients between morphologically irregular systems
and disk galaxies in a sample of $\approx650$ SFGs at $z$\,$\sim$\,$1$ based on HST data available at the time.

The lack of a strong distinction between disks and non-disks in our sample may be attributed to the ambiguous nature of non-disks, as discussed in \citet{Lee2025b}. 
Non-disks could be potential mergers or disks with strong non-circular motions that are not resolved with the current data quality. The merger stage for the former scenario is largely unknown, while the latter scenario would make non-disks less distinct from disks.

Furthermore, the overall high molecular gas fraction  ($f_{\rm gas}=M_{\rm gas}/(M_{\rm gas}+M_\star) \gtrsim 50\%$) of the sample with little systematic difference between disks and non-disks \citep{Lee2025b} imply rapid and efficient inward gas streaming from the pristine circum-galactic medium (CGM) and inter-galactic medium (IGM), as well as minor mergers, which feed the galaxy to maintain a large gas fraction \citep[e.g.][]{Tacconi2020}. 
As discussed in Sect.~\ref{subsec:Z_disks}, the gas-rich nature of these systems could induce efficient formation of clumps, 
and their migration and further gas inflow driven by tidal torques exerted on the gas 
could flatten metallicity gradients. 
Nevertheless, observational evidence remains unclear on whether the clumps can actually survive this migration \citep{Claeyssens2025,Kalita2025b}.

Further supporting the above picture is the growing evidence of rapid radial gas inflow ($v_{\rm inflow}\approx50\,$\kms) in star-forming disks at $z$\,$\sim$\,$2$ \citep[][]{Genzel2023,Huang2025,Pastras2025,Jolly2026}. This would imply material migrates over a flow (dynamical) time of $<$\,$100\,$Myr, much shorter than that of local disks.
Because this inflow timescale is shorter than, or at best comparable to, the gas-consumption (star-formation) timescale \citep[$\lesssim$\,Gyr, ][]{Tacconi2020}, relatively small in situ metal enrichment can occur during a single episode of inward transport. 
Consequently, these galaxies are expected to develop relatively flat metallicity gradients, and with little azimuthal variations \citep{Petit2015}, which we investigate further in Sect.~\ref{subsec:azimuthal_variation}.

Comparable gradient flattening could also take place in interacting and merging systems at all epochs, where tidal torques likewise drive efficient radial gas mixing  \citep[e.g.][]{Rupke2010a,Croxall2015,XWang2019,Venturi2024}. 
Hence, both high-redshift gas-rich disks and mergers may share a common, rapid, large-scale gas redistribution that suppresses steep abundance gradients.

The high gas fraction at $z$\,$\sim$\,$5$ also sustain elevated star-formation rates compared to the local Universe. The ensuing strong stellar feedback launches outflows that could redistribute metals into the outer disk \citep{Gibson2013,Uebler2014,Tissera2022}. 
Subsequent re-accretion of this material, together with turbulent mixing, 
is expected to further moderate radial variations in chemical composition, 
which could ultimately diminish the difference between disk and non-disk galaxies at \zcristal.

\subsection{Azimuthal variations}\label{subsec:azimuthal_variation}
As found in many \znoon\ galaxies, metallicity distribution is often very irregular, with considerable local variations even on kpc scales \citep[e.g.][]{nmfs2006,nmfs2018,Gillman2022}, as well as in simulations \citep{Bellardini2021}.
These irregularities may reflect the non-symmetric processes such as chaotic accretion, feedback, gravitational instabilities and transport processes during these early galaxy evolution phases. 
The presence of more metal-poor clumps or unresolved small companions will also introduce local variations in metallicities.
Hence, averaging large azimuthal variations into the same radial bins could wash out such features and result in an apparently flat radial gradient.

To quantify the azimuthal variations in our sample, 
we apply the same procedure used for the annular-binned profile in Sect.~\ref{sec:strong_line_method}, 
but on a spaxel-to-spaxel basis. 
We select only those pixels where the line ratios exceed the $2.5\sigma$ limit 
(see Sect.~8.3 of \citealt{nmfs2018}), 
which implies the line flux of the fainter line (e.g. \Hb\ in R3) must also exceed $2.5\sigma$.
Note that the $2.5\sigma$ threshold is applied on the line ratio, which is a more stringent requirement than the same threshold on the individual line fluxes; the former is effectively equivalent to raising the per-line S/N threshold much above 2.5$\sigma$.

Here we focus on oxygen-based diagnostics (e.g. O32) if available, which typically have higher S/N and larger radial coverage. For CRISTAL-06a and 06b, only N2 is available. 
The Fig.~\ref{fig:gradient_fit} shows the per-pixel metallicities profile for our sample, the data points are colour-coded by their acute azimuthal angle difference relative to the position angle of the morphological major axis ($\Delta\phi = |\phi - \phi_{\rm maj}|$). 

\begin{figure}
\centering
    \includegraphics[width=0.5\textwidth]{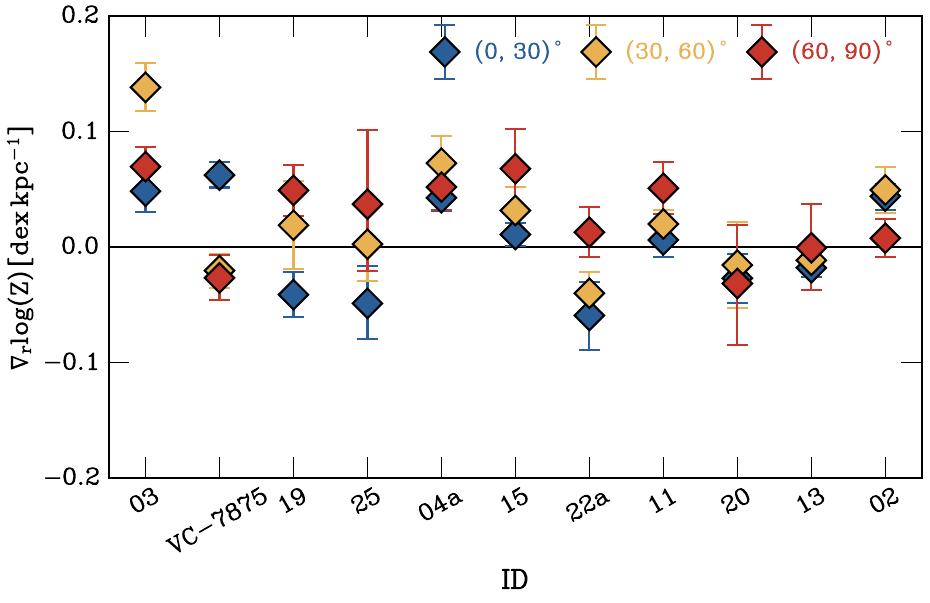}
      \caption{Metallicity gradients inferred in three azimuthal bins: $[0^\circ,30^\circ)$ (blue), $[30^\circ,60^\circ)$ (yellow), and $[60^\circ,90^\circ)$ (red). 
      The galaxy IDs are sorted from left to right by their azimuthal variations in terms of standard deviation.
      }
      \label{fig:azibin}
\end{figure}

We then fit the metallicity gradients for pixels within an azimuthal bin of $\Delta\phi=30^\circ$, dividing the galaxies into three bins spanning $[0,90]^\circ$. Out of the 20 galaxies in our sample, 11 have sufficient pixels within an azimuthal bin for this measurement.
There are typically 17 spaxels fall within each azimuthal bin. 
Figure~\ref{fig:azibin} presents the distribution of metallicity gradients for the 11 galaxies in each bin, and Table~\ref{tab:azibin} summarises the values.
The scatter among the azimuthal-binned values around their mean values is relatively low, averaging $0.02$\,\dexkpc, with a maximum scatter of $0.04$\,\dexkpc\ observed for CRISTAL-03, 19, 25 and VC-7875, which is still smaller than the typical measurement uncertainties of the annular-binned gradients of $\sim$\,$0.06\,$\dexkpc\ from Method I in Sect.~\ref{sec:strong_line_method}

Comparing the mean difference between the azimuthally binned above and annular-binned metallicity gradients inferred from Method I, we find that none of the galaxies show a significant deviation from the annular-binned metallicity gradient, with no galaxies showing a mean difference exceeding $3\sigma$. 
Only four galaxies (CRISTAL-03, 04a, 13, and VC-7875) show a difference at $1\sigma$ level, indicating there could be mild azimuthal variations these cases, given the uncertainties of this pixel-based method.

\begin{figure}
\centering
    \includegraphics[width=0.4\textwidth]{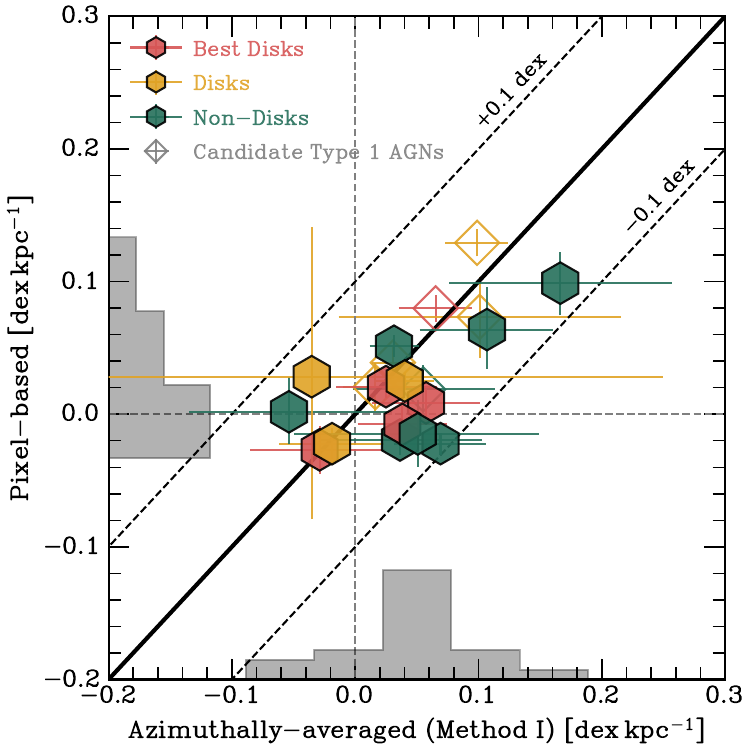}
      \caption{Comparison between the azimuthally averaged (from Method I) and pixel-based metallicity gradients.
      The disks and non-disks classified in \citet{Lee2025b} are shown in pink or yellow and green, respectively.
      The candidate Type-1 AGNs identified in \citet{Ren2025} are marked with the diamond symbol.
      }
      \label{fig:pixvsazi}
\end{figure}

\begin{figure*}
\centering
    \includegraphics[width=0.7\textwidth]{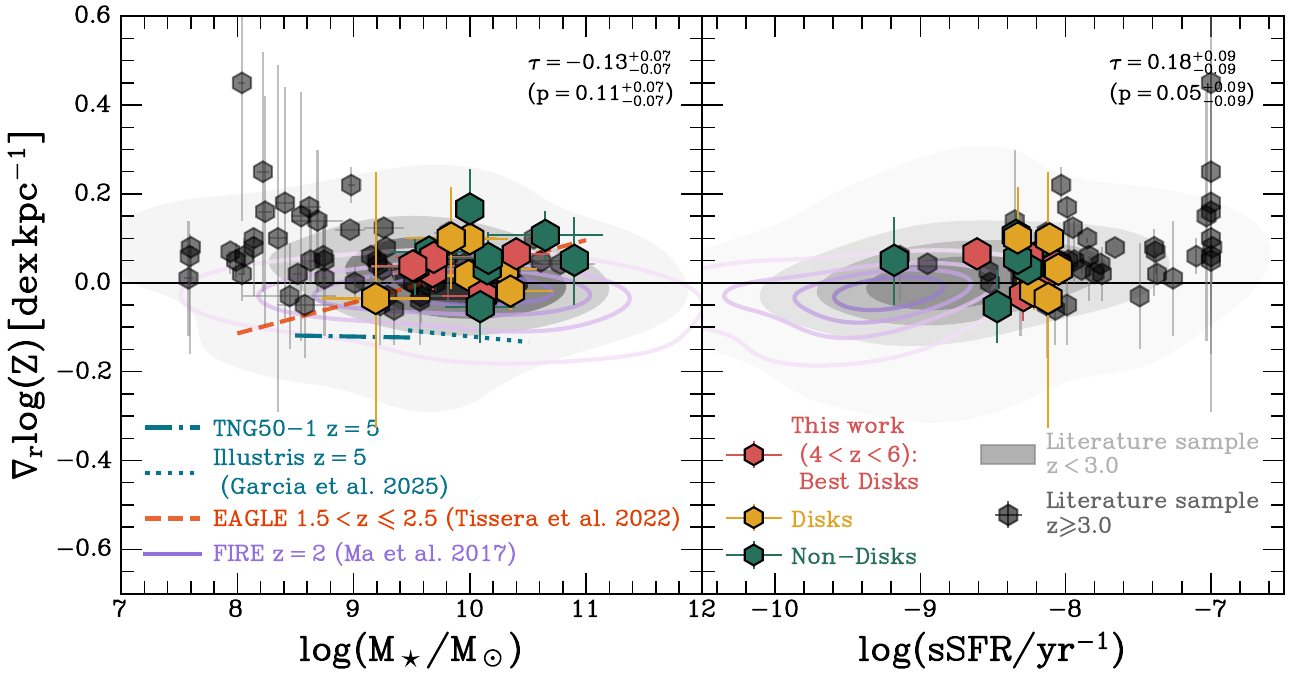}
      \caption{
        Observed metallicity gradients as a function of \mstar\ (\textit{left}) and sSFR (\textit{right}). The literature samples at $z$\,$<$\,$3.5$ and $z$\,$\geqslant$\,$3.5$ are represented by black contours and black hexagons, respectively. The references for the literature sample are listed in Table~\ref{tab:literature_sample}. Predictions from the cosmological simulations TNG50-1 and Illustris \citep{Garcia2025} are plotted as dot-dashed and dotted blue lines, respectively. Additionally, we show the trends from other simulations at $z\approx2$, including EAGLE \citep{Tissera2022} as an orange dashed line (\textit{left} only) and FIRE \citep{Ma2017} as purple contours. The values from FIRE2 \citep{Sun2025} are within the FIRE contours and are therefore not plotted.
        The Kendall's $\tau$ correlation coefficient and the $p$-value for the combined distribution of our sample and the literature sample are annotated in the top-right corner of each panel.
      }
      \label{fig:DelZ_lMstar_sSFR}
\end{figure*}

To assess the azimuthal variation in the whole sample, especially in cases where the azimuthal binning method above could not be applied, we fit the radial profile of these per-pixel metallicities in the full azimuthal range, for example following \citet{nmfs2018}, using the same procedure as Method I in Sect.~\ref{sec:strong_line_method}.
The derived gradients are presented in Fig.~\ref{fig:gradient_fit} on the right of each panel. 
Figure~\ref{fig:pixvsazi} compares the gradients from Method I and the pixel-based metallicity profiles derived here.

In all cases, the two profiles agree within $1\sigma$, indicating that there are generally no strong azimuthal variations in the radial profiles.
The pixel-based metallicity gradients are only slightly flatter than those derived from azimuthal averaging, with a mean \dlOH\ of \meandlOHpix\,\dexkpc, compared to \meandlOH\,\dexkpc, with a marginal difference of $\approx1\sigma$.
Such a small discrepancy could be attributed to several factors, 
including the different radial coverage of the azimuthally averaged and 
pixel-based metallicity gradients, and the inherently light-weighted nature of the azimuthally averaged method.

Overall, we do not find strong azimuthal variations that are statistically significant above the measurement uncertainties. 
However, four galaxies, namely CRISTAL-03, 19, 25, and VC-7875, exhibit slightly stronger azimuthal variations compared to the rest of the sample.
CRISTAL-03 was identified by \citet{Ren2025} as candidate Type-1 AGN because their spectra exhibit broad Balmer components. 
However, in their BPT diagrams, as well as in diagnostics based on \heii\,$\lambda4686$ \citep{Shirazi2012} and the \oiii\,$\lambda4363$/H$\gamma$ ratio --- the latter being more suitable at high-$z$ \citep{Mazzolari2024}, CRISTAL-03 consistently lies around the star-forming/composite locus, showing no significant excess of hard ionising photons.
VC-7875 is a disturbed disk galaxy, as identified by \citet{Fujimoto2020}, and further supported by JWST NIRCam imaging data that it is part of an interacting system.
For CRISTAL-25, \citet{JLi2024} suggested the potential presence of a Type-1 AGN based on NIRCam broadband photometry, which hinted at an excess of flux from the broad component of \mgii\,$\lambda2798$ and \Hb+\oiii\,$\lambda5007$. However, our G235M data are unable to confirm this due to the low S/N of \Hb+\oiii\ (even \oiii\,$\lambda4959$ is barely detected), and \mgii\ being shifted out of the spectral coverage.
For CRISTAL-19, a disk system, the origin of the mild azimuthal variation is unclear.

The overall lack of strong azimuthal variations is also found in some studies in the local Universe \citep{Kreckel2020}, which, despite having much better statistics based on over hundreds of \Hii\ regions, report RMS scatter of $\lesssim0.05\,$dex after removing the large-scale gradient, suggesting a high level of chemical homogeneity over large spatial scales. 
\citet{Williams2022} also finds no clear systematic enrichment in the spiral arms, which is somewhat unexpected, given that the primary source of oxygen is Type II supernovae, which are typically more densely populated along the spiral arms.

The lack of azimuthal variations in disks could be 
explained if the turbulent mixing timescale 
is much shorter than the orbital time 
required for differential rotation to wind non-axisymmetric abundance patterns into tight spirals.
The resulting shear-mixing could therefore erase most of the contrast, resulting in the weak azimuthal scatter observed at fixed galacto-centric radius \citep{Yang2012}. 
Given that the disk galaxies in our sample have higher velocity dispersions, averaging $\sim70\,$\kms, than local disks, the turbulence-driven mixing is likely even more efficient, which could explain the absence of strong azimuthal variations.

Nevertheless, we are cautious that observational limitations may also explain why the majority of the galaxies do not exhibit a large azimuthal variation in metallicities.
A single pixel in the NIRSpec IFU of size $0\farcs1$ corresponds to a physical scale of $0.5$--$0.7$\,kpc at redshifts $z$\,$=$\,$4$ to $6$. Because our observations employ only the standard 2-point dither pattern
\citep[with the exception of CRISTAL-01, 20 and 22a see][]{Fujimoto2025,Jones2025,Parlanti2025}, the data provide only modest sub-pixel sampling.
It is further exacerbated by the limited angular resolution of a similar order, 
which altogether limits our view to the non-axisymmetric features that may bring about 
local metallicity variations, such as clumps \citep[e.g.][]{XWang2017,EstradaCarpenter2025,Sok2025,Parlanti2025b} 
and spiral arms, which have been observed in some other local studies \citep{Vogt2017, Bresolin2025} and simulations \citep{Spitoni2019,Orr2023}. 
Hence, it is difficult to confirm or rule out the presence of these features as drivers of metallicity variations \citep{Genzel2011,Newman2012,Khoperskov2023} within the physical scales probed by our current data.

\subsection{Mass and sSFR}\label{subsec:mass_and_ssfr}
A positive correlation between sSFR and radial metallicity gradients was identified by studies of SFGs at cosmic noon \citep[e.g.][]{Stott2014,EWuyts2016,Curti2020}.
A negative correlation with \mstar\ \citep[e.g.][but see also \citealt{Li2025} for \mstar$\lesssim10^9M_{\odot}$ at $z>6$]{EWuyts2016,XWang2020,Cheng2024} has also been found.
In general, these studies found that high-sSFR and low-mass galaxies tend to have flatter to even positive gradients, 
as shown in Fig.~\ref{fig:DelZ_lMstar_sSFR}. 
This may be due to the higher mass loading of starburst-driven galactic winds at early times in those low-mass systems, which eject metal-enriched gas from galaxy centres and may deposit it at larger radii.
In simulations, this is attributed to efficient feedback in such systems, which keeps them in the `bursty' star formation mode \citep{Ma2017}. 

The stellar masses and SFRs were derived from SED fitting using various tools in the literature. 
For the single literature source that adopted the Salpeter IMF, we scale its stellar mass and SFR down by a factor of $1.7$ to place them on the Chabrier IMF \citep{Speagle2014}. 
For the 17 galaxies already adopting a Kroupa IMF, no adjustment is applied since the Kroupa-Chabrier offset is negligible relative to our error budget.

We cannot confirm any significant trends between metallicity gradients and $M_\star$ or sSFR in our sample, with low Kendall's $\tau$ correlation coefficients ($\tau$\,$\lesssim$\,$0.15$, $p$\,$\gg$\,$0.05$). 
We did not observe any significant correlation with the surface density of star formation, $\Sigma_{\rm SFR}$, either.
Even when combining our sample with literature samples compiled at $z$\,$>$\,$3.0$ (Table~\ref{tab:literature_sample}), which extend to lower stellar masses ($\log(M_\star/M_\odot)<9.0$) and higher sSFR values ($\log({\rm sSFR}/{\rm yr^{-1}})\gtrsim-8$), the negative and positive trends with $M_\star$ and sSFR, respectively, are only marginally stronger. 
However, the statistical significance of these trends remains weak, as indicated by the low Kendall's $\tau$ values and high $p$-values annotated in Fig.~\ref{fig:DelZ_lMstar_sSFR}.

Comparing our results with cosmological simulations, the lack of a trend between $\log(M_\star)$ and metallicity gradient is not entirely unexpected. At $z$\,$\leqslant$\,$2.5$, both the FIRE \citep{Ma2017} and FIRE2 \citep{Sun2025} simulations found very shallow to no correlation with $\log(M_\star)$. Given the uncertainties present in our data and the literature, constraining a weak trend such as this remains challenging.
Although our results show a better match with the EAGLE simulation within the mass range of our sample, the relationship with $M_\star$ is not well reproduced for the literature-compiled sample at $z$\,$\geqslant$\,$3$ across a broader mass range, particularly at the low-mass end.
However, studies by \citet[EAGLE]{Tissera2022} and \citet[TNG]{Hemler2021} do suggest that lower-mass galaxies exhibit greater fluctuations, and our current limited sample may be too small to capture the full scatter.
A more recent study by \citet{Garcia2025} found that the TNG50-1 and Illustris simulations at $z$\,$=$\,$5$ capture the flat dependence on $M_\star$, but predict an overall more negative gradients than our observed results for our mass range.

Since the $M_\star$ and sSFR that we are comparing here are primarily SED-based quantities, and could give rise to systematic scatter 
due to different assumptions for example on the star-formation history, and SED-modelling tools \citep{Pacifici2023}, we also compare with SED-independent quantities such as dynamical mass ($M_{\rm dyn}$) and gas mass ($M_{\rm gas}$).
Within our sample, we do not find strong correlations between the metallicity gradients
and the dynamical mass and the gas mass or fraction (either $M_{\rm gas}/M_{\rm baryon}$ or $M_{\rm gas}/M_\star$) measured in \citealt{Lee2025b}), with Kendall's $\tau$\,$\lesssim$\,$0.3$ and a large $p$-value. 
Note that we only have dynamical mass for disks. In mergers, the $M_{\rm dyn}$ inferred from the integrated line width lacks physical meaning.
We could not expand the $M_{\rm gas}$ comparison much beyond our own sample because most galaxies lack direct $M_{\rm gas}$ measurements, nor did we.
If we use $V_{\rm rot}$ as a proxy for the $M_{\rm dyn}$ of literature disks, we find Kendall's $\tau$\,$=$\,$-0.09^{+0.05}_{-0.04}$ and $p$-value of $0.05^{+0.05}_{-0.04}$, no compelling trend.

\section{Effects of beam smearing}\label{sec:effects_of_beam_smearing}
The observed shallow gradients in our sample could be partly caused by observational effects \citep{Yuan2011,Wuyts2016,Carton2018,Acharyya2020}. 
Convolution of the intrinsic line flux distribution with the PSF acts as a spatial blurring kernel that reduces local contrast and any kind of gradients would be suppressed and appear shallower, an effect known as beam smearing. 
Intuitively, the steeper the intrinsic gradient and the broader the PSF relative to the size of the galaxy, the stronger the flattening. To the extreme, for a flat gradient, beam smearing has no effect.

To quantify the beam-smearing effect and understand the impact on our measurements, we generate 1500 mock galaxies with a given intrinsic metallicity gradient ranging between \dlOH\,$\in$\,$[-0.2,+0.2]$\,\dexkpc. 
The mock data cubes are built out of exponential disk models using \texttt{DysmalPy}\footnote{\url{https://www.mpe.mpg.de/resources/IR/DYSMALPY}} 
\citep{Davies2004a,Davies2004b,Cresci2009,Davies2011,Wuyts2016,Lang2017,Price2021,Lee2025a} with parameters listed in Table~\ref{tab:mock_params}. 
Since the purpose here is to test the effect of beam-smearing on metallicity gradient, we set the mass model to $\log(M_{\rm bary}/M_\odot)=10.5$, where $M_{\rm bary}$ is the total baryonic mass, and $R_{\rm e,disk}=2\,$kpc which are representative of the galaxies in our sample \citep{HerreraCamus2025}. We do not vary these physical properties for the mock models, but vary only the ratio of the effective radius to that of the beam size, as well as the S/N ratio. 
The inclination is fixed at $20\,^\circ$ because, 
as demonstrated by \citet{Wuyts2016}, in practice, 
the ellipticity of the apertures is typically tailored to match the observed isophotes, and hence variations with respect to inclination are less relevant.

\begin{table}
\caption{Parameters of mock data cubes at $z=5$.}
\label{tab:mock_params}
\centering
\begin{tabular}{l c c}
\hline\hline  
Parameter & (Range of) Value \\
\hline  
$\log(M_{\rm bary}/M_\odot)$ & 10.5 \\
$R_{\rm e, disk}$ (kpc) & 2 \\
$\sigma_0$\,(\kms) & 80 \\
$\sigma_{\rm LSF}\,$(\kms)\tablefootmark{a} & 125 \\
beam$_{\rm FWHM}$ & 0\farcs1--1\farcs0\\
$2\times R_{\rm e, disk}$/beam$_{\rm FWHM}$ & 0.6--6.4\\
S/N\tablefootmark{b} & 20--40 \\
Pixel scale & 0\farcs05 \\
\dlOH\,(\dexkpc) & [$-0.20,+0.20$] \\
Inclination & $20^\circ$ \\
  \hline
\end{tabular}
\tablefoot{
\tablefoottext{a}{In $\sigma$.}
\tablefoottext{b}{Average per-pixel S/N within an effective radius of the channel of the brightest line.}
}
\end{table}

\begin{figure*}
\centering
    \includegraphics[width=0.9\textwidth]{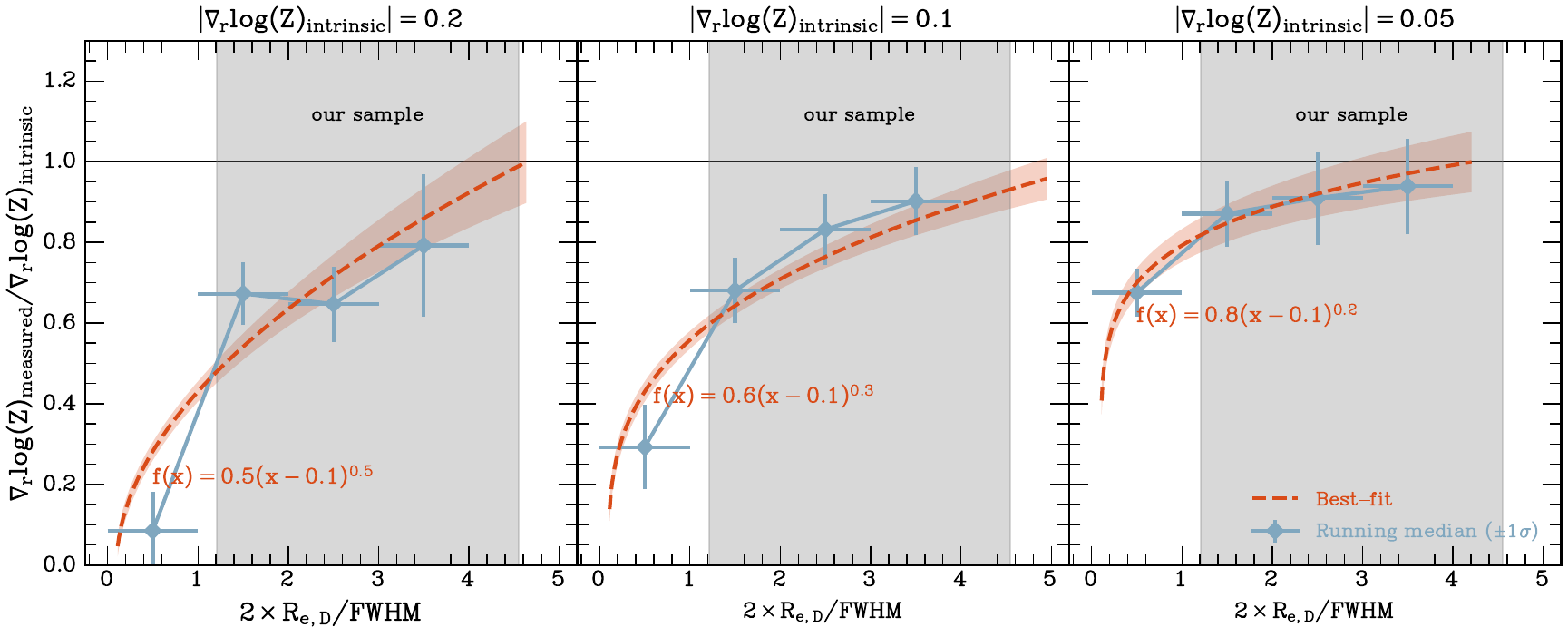}
      \caption{Ratio of the measured ($\nabla_r{\log}{(Z)}_{\rm measured}$) versus intrinsic ($\nabla_r{\log}{(Z)}_{\rm intrinsic}$) metallicity gradient for a set of mock exponential disk models imprinted with a linear intrinsic gradient in the range $\pm[0.05,0.10,0.20]$\,\dexkpc (from right to left) as a function of $2\times R_{\rm e, disk}$/beam$_{\rm FWHM}$.
      The dashed orange line denotes the best-fit polynomials through the distribution, with the functional form annotated.
      The light blue curve shows the running median through the same distribution.
      The vertical grey band indicates the range of $2\times R_{\rm e, disk}$/beam$_{\rm FWHM}$ spanned by our data.
      We infer the correction factor by inverting the orange curves for different intrinsic gradients.
      The median $2\times R_{\rm e, disk}$/beam$_{\rm FWHM}$ of our sample is $2.5$.
      }
      \label{fig:bmsc}
\end{figure*}

\begin{figure}
\centering
    \includegraphics[width=0.4\textwidth]{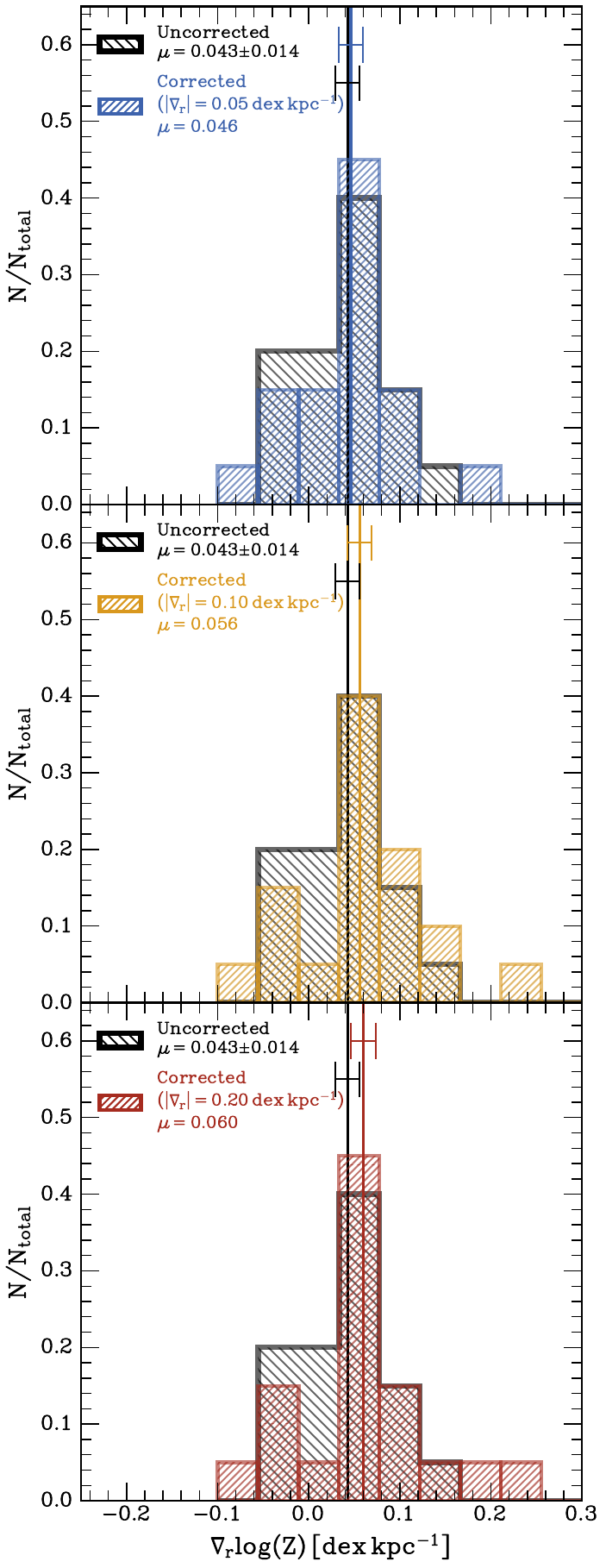}
      \caption{Distributions of the observed metallicity gradients as well as the corrected gradients obtained by the applying correction factors derived from correction curves that assume different intrinsic metallicity gradients presented in Fig.~\ref{fig:bmsc}.
      The error of the mean of the gradients is indicated by the horizontal error bar. Note that the error of the mean does not change after beam-smearing correction is applied.
      }
      \label{fig:bmsc_grad}
\end{figure}

We added \oii, \neiii, \Hb, and \oiii\ line emissions assuming a linear radial gradient (Eq.~\ref{eqn:Zeqn}) such that the line intensity of \Hb, \oii, and \neiii\ relative to \oiii\ are controlled via the R3, O32, and Ne3O2 relations listed in Table~\ref{tab:line_ratios}, assuming the spectra have already been de-reddened.
The light profile follows that of an exponential disk (S\'ersic index $=1$).
The central metallicity $Z_0$ is chosen such that over the extent of the profile, $\log_{10}Z(r)$ falls within the valid range of R3, O32 and Ne3O2 as listed in Table~\ref{tab:line_ratios}.

The cubes are convolved with a PSF of varying sizes, such that the galaxy extent in diameter ($2\times R_{\rm e,disk}$) is covered by $0.6$ to $6.4$ beams in FWHM (beam$_{\rm FWHM}$). 
The intrinsic effective radii of our galaxies range from $0\farcs11$ to $0\farcs40$, measured from 
\texttt{imfit} modelling of the F444W photometry \citep{Lee2025b}.\footnote{ 
For CRISTAL-10a-E, in which F444W is unavailable, we measure the effective radius from the \Ha\ line map.}
We assume that the line emission follows the stellar light traced by the F444W continuum.
Therefore, for our data, $2\times R_{\rm e, disk}$/beam$_{\rm FWHM}\sim2.5$ (median), with a range of $\sim1.2$--$4.5$. 

The cubes are also convolved with a LSF kernel of size (in $\sigma$) $125\,$\kms\ ($1.28\times10^{-3}\,$\micron), 
corresponding approximately to the spectral resolution of the $R$\,$\sim$\,$1000$ NIRSpec IFU at the observed wavelength of \oiii\ ($\sim$\,$3\,$\micron). 
The channel width is fixed at 106\,\kms.
The pixel scale is set to 0\farcs05$\,{\rm pix^{-1}}$ (compared to 0\farcs1$\,{\rm pix^{-1}}$ in the NIRSpec data) to sample the smaller PSF in some of the mock cubes. This finer pixel scale is chosen to avoid the undesirable situation where the pixel size is larger than the smallest beam FWHM.
Since the axis ratios of the actual PSF $\gtrsim$\,$0.9$, the PSF shape of the mock cubes is set to be circular.
To simulate a range of S/N covered by our data and slightly beyond to recover trends, we add Gaussian noise to the data cubes, 
resulting in average per-pixel S/N values of 20 to 100 within the effective radius of the brightest channel of the \oiii\,$\lambda$5007 emission line.
Such a definition of S/N means that the \neiii\,$\lambda$3869 line may be undetected in some of the mock cubes, particularly those with high metallicity. In such cases, we exclude the use of the Ne3O2 diagnostic to recover the metallicity, as in our treatment of the actual data. 

We then repeat the spectral line fitting and derive the azimuthally averaged metallicity gradients of the mock data cubes following the same procedure in Sect.~\ref{sec:emission_line_fitting} and Sect.~\ref{sec:strong_line_method}, respectively. 
The resulting distributions of the recovered versus the intrinsic metallicity gradients as a function of $2\times R_{\rm e,disk}/$beam$_{\rm FWHM}$
is shown in Fig.~\ref{fig:bmsc}.

As expected, we find a positive correlation between the number of beams covering a galaxy and the accuracy of the recovered intrinsic metallicity gradient. 
Additionally, there is a secondary dependence on the intrinsic metallicity gradient, as evidenced by the shallower correlation when the intrinsic gradient is $0.05\,$\dexkpc\ in absolute value, compared to $0.2\,$\dexkpc. This suggests that a shallower intrinsic gradient is less affected by beam smearing, consistent with the findings of \citet{EWuyts2016}. 
We do not observe a substantial difference in the correlation between the sign (negative vs. positive) of the gradients for a fixed magnitude. Furthermore, we find no strong dependence on S/N, which only affects the uncertainties and scatter, which is also reported by \citet{EWuyts2016} and \citet{Gillman2021}.

To quantify the correlation between the measured and intrinsic metallicity gradients, we fit a polynomial to the relationship between $\nabla_r{\log}{(Z)}_{\rm measured}/\nabla_r{\log}{(Z)}_{\rm intrinsic}$ and $2\times R_{\rm e,disk}$/beam$_{\rm FWHM}$. The resulting fit is a reasonably good match to the running medians of the distribution, as shown in Fig.~\ref{fig:bmsc}.

We can then invert the best-fit polynomials to derive a `correction factor' based on the $2\times R_{\rm e, disk}$/beam$_{\rm FWHM}$ ratio of an individual galaxy. 
Since the correction factor also depends on the unknown intrinsic metallicity gradient, we report 
the corrected gradients
in Table~\ref{tab:bmc_gradients} 
for the range of $|\nabla_r{\log}{(Z)}_{\rm intrinsic}|=[0.05,0.2]$ that we have examined. 
For our data, the correction factors as we defined, $\nabla_r{\log}{(Z)}_{\rm measured}/\nabla_r{\log}{(Z)}_{\rm intrinsic}$,
for $|\nabla_r{\log}{(Z)}_{\rm intrinsic}|=[0.05,0.10,0.20]$ range across 
$[0.8, 1.0]$,
$[0.6, 1.0]$ and 
$[0.5, 1.0]$, 
respectively.
It is an interesting ancillary finding that, regardless of the intrinsic steepness of the gradients, the correction factors all asymptotically approach unity when the ratio of $2\times R_{\rm e, disk}$/beam$_{\rm FWHM}$ reaches $\sim5$. In other words, one should aim for an angular resolution of this order to determine the intrinsic gradient that is not compromised too much by beam smearing.

After applying these correction factors, we find that the corrected metallicities remain largely consistent with a flat gradient, as found previously in Sect.~\ref{sec:strong_line_method}, with a maximum correction of up to $20\%$ (mean $\sim$\,$10\%$) correction needed.
For the 16/20 galaxies that exhibit positive metallicity gradients, applying the correction factors derived from an intrinsic gradient of 
$|\nabla_r{\log}{(Z)}_{\rm intrinsic}|$\,$=$\,$0.05$\,\dexkpc, we find that none of the galaxies have a metallicity gradient of $\nabla_r{\log}{(Z)}$\,$>$\,$+0.05\,$\dexkpc\ at $3\sigma$ level, and only three galaxies (CRISTAL-01a, 07a, and 07b) have gradients at $1\sigma$, consistent with the uncorrected results presented in Sect.~\ref{sec:strong_line_method}.
Even when applying the most extreme correction, assuming an intrinsic gradient of $|\nabla_r{\log}{(Z)}_{\rm intrinsic}|$\,$=$\,$0.2$, none of the galaxies have a metallicity gradient of $\nabla_r{\log}{(Z)}$\,$>$\,$+0.2\,$\dexkpc.
For the 4/20 galaxies that exhibit negative metallicity gradients, their gradients remain consistent with a strictly flat metallicity gradient ($\nabla_r{\log}{(Z)}$\,$=$\,$0\,$\dexkpc) within their $1\sigma$ uncertainty, regardless of the assumed intrinsic gradient. 
Figure~\ref{fig:bmsc_grad} presents the distributions of the corrected metallicity gradients, along with their mean values and the error of the mean, assuming different intrinsic gradients.

The investigation presented here demonstrates that, regardless of sign, the intrinsic metallicity gradients are unlikely to be extreme. This finding suggests that the intrinsic metallicity gradients of these galaxies are likely to be flat, rather than being solely an artefact of observational effects such as beam smearing.

Nevertheless, we note that our first-order beam-smearing correction has several caveats. 
First, we assume a smooth linear gradient when imprinting the mock data cubes, but observations of nearby galaxies have shown that their metallicity gradients may not be strictly linear. 
Instead, many local disks exhibit a double-linear slope within $\sim$\,$2R_{\rm e}$ (from \citealt{MartinRoy1995} to more recent IFU studies \citealt{SM18,Easeman2022,Cardoso2025}), 
but evidence at higher redshifts is still lacking.
Furthermore, high-$z$ disks are often characterised by actively star-forming clumps, 
which could introduce local metallicity irregularities \citep[e.g.][]{XWang2017,EstradaCarpenter2025,Sok2025}, 
and light-weighting effects distinct from those for our assumed exponential disk profiles, 
in addition to other mechanisms that would give rise to local variations as discussed in Sect.~\ref{subsec:azimuthal_variation}, that a smooth gradient as a function of galactic radius only cannot capture.

\section{Summary}\label{sec:summary}
In this work, we have analysed the gas-phase oxygen abundance (metallicity) gradients of \Ngal\ MS SFGs at \zcristal\ using JWST/NIRSpec IFU at $\sim$\,$1$\,kpc resolution. 
We measured the metallicity at each radial position using the strong-line method with $T_e$-based metallicity calibration \citep{Sanders2025} and leveraging the multiple optical emission lines available from \oii\,$\lambda\lambda3726,3729$ to \sii\,$\lambda\lambda6716,6731$. 
We find the metallicity gradients across the sample are consistent with a flat gradient, and only three galaxies show a gradient $>$\,$0.05\,$\dexkpc\ at only the $1\sigma$ level (Sect.~\ref{sec:strong_line_method}).
We did not find a significant systematic offset in the metallicity gradients inferred from different line diagnostics, and the Bayesian approach could tighten the overall constraint on the metallicity (Sect.~\ref{sec:strong_line_method}).
We subsequently examined the correlations between the metallicity gradient, kinematic properties, and other physical properties. We found the following:

\begin{enumerate}
\item 
By combining our disk sample with literature data, 
our study effectively increases the number of available samples at $3$\,$\leqslant$\,$z$\,$\lesssim$\,$8$ by $50\%$ within the comparable mass range.
The combined dataset reveals a shallow yet statistically significant negative relationship between metallicity gradients and $V_{\rm rot}/\sigma_0$ but not with $\sigma_0$.
This trend is consistent with the reported cosmic evolution of metallicity gradients, which become progressively more negative from early to recent epochs \citep[][and references therein]{Fujimoto2025}. 
Fundamentally, our results suggest that the dynamical maturity of a galaxy disk, as indicated by $V_{\rm rot}/\sigma$, plays a crucial role in shaping the radial metallicity gradients (Sect.~\ref{subsec:Z_disks}).

\item There is no strong distinction between disks and non-disks. This could be due to the frequent accretion events and efficient gas transport at $z$\,$\sim$\,$5$ as implied from their high gas fraction (Sect.~\ref{subsec:disk_nondisk}).

\item 
The majority of our galaxies do not exhibit significant azimuthal variations, which may be attributed to the efficient azimuthal mixing in turbulent disks. However, we also acknowledge the potential observational limitations in resolving the physical scales of structures, such as clumps and spirals, that could potentially drive local metallicity variations (Sect.~\ref{subsec:azimuthal_variation}).

\item There is no statistically significant correlation with \mstar\ and only a marginal positive trend with sSFR, even when including with literature samples at $z$\,$>$\,$3$ of a wider range in \mstar\ and sSFR (Sect.~\ref{subsec:mass_and_ssfr}).

\item Applying a beam-smearing correction to our sample using correction factors derived from a large suite of mock galaxies that mimic the observational sample indicates that beam smearing can account for a maximum of $20\%$ of flattening, but it typically accounts for only $10\%$. Given the uncertainties, we ruled out very steep metallicity gradients in the sample regardless of sign (Sect.~\ref{sec:effects_of_beam_smearing}).
\end{enumerate}

Overall, the lack of a strong correlation between the metallicity gradients and the kinematic nature and global physical properties, especially the former, may be due to the limited dynamic range of the sample and the substantial scatter in the empirical data.
Consequently, this permits the identification of only the steepest trends, while shallower trends, such as those presented here, require a much larger set of statistics to constrain them more robustly.
Our study, together with the two other works by \citet{Fujimoto2025} and \citet{Faisst2026}, nevertheless provides a homogeneously selected sample of MS SFGs available at \zcristal\ that allows for comparison with gas kinematics properties.

Ideally, future studies should aim to obtain an even higher spatial resolution to examine the local metallicity variations on smaller physical scales than resolvable by JWST $\lesssim\,$kpc.
Gravitational lensing can alleviate some observational limitations, yet strong lensing events are intrinsically rare, especially for sources at $z$\,$>$\,$4$, so assembling a statistically significant sample that will be studied at the population-level remains challenging.
The next-generation IFU HARMONI, which is to be equipped on the Extremely Large Telescope, will offer a significant improvement to the capabilities of studies at \zcristal. With a linear spatial resolution approximately six times better and at least five-fold higher spectral resolution, HARMONI will enable the study of local metallicity variations at physical scales of $<$\,$100$\,pc, although this will be restricted to rest-frame UV lines blueward of \neiii~$\lambda3869$ and \oii$~\lambda\lambda3726,3729$ at 
$z$\,$\sim$\,$5$ due to atmospheric transmission, which will require additional empirical calibrations for metallicity inference.

\begin{acknowledgements}
We thank the anonymous referee for the careful reading and constructive comments, which have improved the clarity of the manuscript.
We acknowledge Michele Perna and Francesco D'Eugenio for their work in developing reduction scripts for NIRSpec IFU data.
N.M.F.S., J.C. and G.T. acknowledge financial support from the European Research Council (ERC) Advanced Grant under the European Union's (EU's) Horizon Europe research and innovation programme (grant agreement AdG GALPHYS, No. 101055023).
H.Ü. acknowledges funding by the EU (ERC APEX, 101164796).
Views and opinions expressed are, however, those of the author(s) only and do not necessarily reflect those of the EU or the ERC. 
Neither the EU nor the granting authority can be held responsible for them.
R.H-C. thanks the Max Planck Society for support under the Partner Group project `The Baryon Cycle in Galaxies' between the Max Planck for Extraterrestrial Physics and the Universidad de Concepción.
T.N. acknowledges support from the Deutsche Forschungsgemeinschaft (DFG, German Research Foundation) under Germany’s Excellence Strategy – EXC - 2094 - 390783311 from the DFG Cluster of Excellence `ORIGINS'.
M.A. is supported by FONDECYT grant number 1252054, and gratefully acknowledges support from ANID Basal Project FB210003.
M.A. and J.M. gratefully acknowledges support from ANID MILENIO NCN2024\_112.
M.B. acknowledges support from the ANID BASAL project FB210003. 
This work was supported by the French government through the France 2030 investment plan managed by the National Research Agency (ANR), 
as part of the Initiative of Excellence of Université Côte d'Azur under reference number ANR-15-IDEX-01.
E.d.C. acknowledges support from the Australian Research Council (project DP240100589).
R.L.D. is supported by the Australian Research Council through the Discovery Early Career Researcher Award (DECRA) Fellowship DE240100136 funded by the Australian Government.
D.R. gratefully acknowledges support from the Collaborative Research Center 1601 (SFB 1601 subprojects C1, C2, C3, and C6) funded by the Deutsche Forschungsgemeinschaft (DFG) – 500700252.
M.R. acknowledges support from project PID2023-150178NB-I00 financed by MCIU/AEI/10.13039/501100011033, and by FEDER, UE.
A.N acknowledges support from the Narodowe Centrum Nauki (NCN), Poland, through the SONATA BIS grant UMO-2020/38/E/ST9/00077.
D.B.S. gratefully acknowledges support from NSF Grant 2407752.
W.W. also acknowledges support associated with program JWST-GO-03950. These observations are associated with programs JWST-GO-03045 and JWST-GO-04265. Support for program JWST-GO-03045 was provided by NASA through a grant from the Space Telescope Science Institute, which is operated by the Association of Universities for Research in Astronomy, Inc., under NASA contract NAS 5-03127. 
This work is based in part on observations made with the NASA/ESA/CSA James Webb Space Telescope. 
The data were obtained from the Mikulski Archive for Space Telescopes (MAST) at the Space Telescope Science Institute, which is operated by the Association of Universities for Research in Astronomy, Inc., under NASA contract NAS 5-03127.
The specific observations analysed can be accessed via
\href{http://dx.doi.org/10.17909/cqds-qc81}{10.17909/cqds-qc81}. 
This work made use of the following Python packages:
\texttt{Astropy} \citep{Astropy2022}, 
\texttt{corner} \citep{Foreman-Mackey2016},
\texttt{DysmalPy} \citep{Davies2004a,Davies2004b,Cresci2009,Davies2011,Wuyts2016,Lang2017,Price2021,Lee2025a},
\texttt{emcee} \citep{ForemanMackey2013}, 
\texttt{Matplotlib} \citep{Hunter2007}, 
\texttt{Numpy} \citep{harris2020}, 
\texttt{pymccorrelation} \citep{Curran2014,Privon2020},
and \texttt{Scipy} \citep{Virtanen2020}.
\end{acknowledgements}
\bibliographystyle{aau}
\bibliography{ref}

\begin{appendix}
\section{Inferring metallicity via a Bayesian method}\label{app:bayes}
\begin{figure}
\centering
    \includegraphics[width=0.45\textwidth]{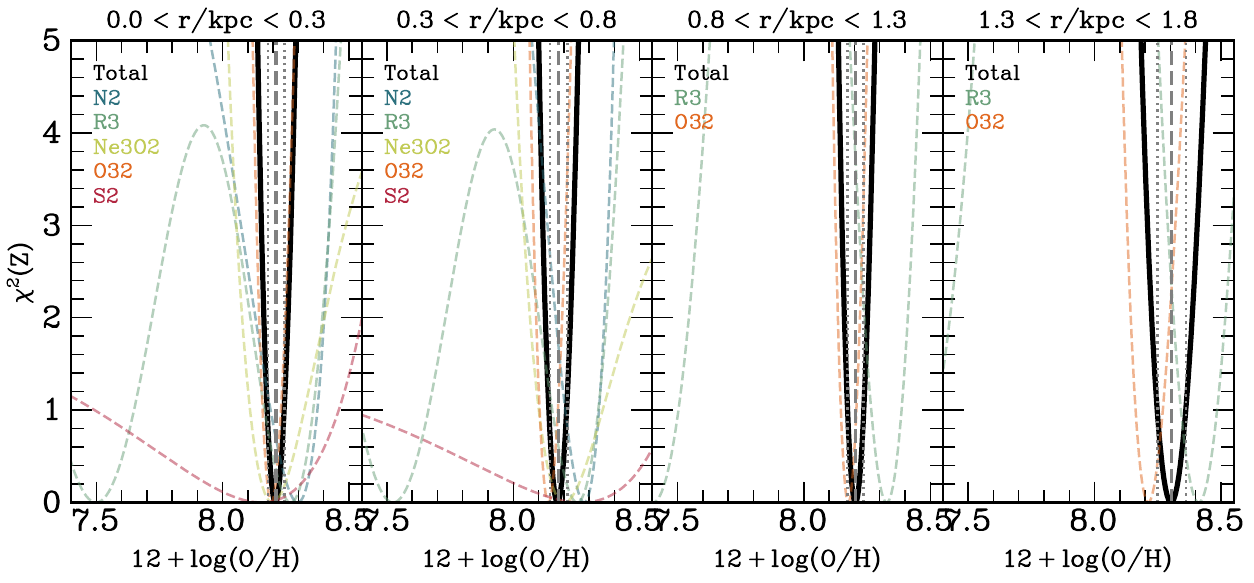}
      \caption{
      Example of the $\chi^2$ landscape as a function of metallicity for CRISTAL-19 for each radial bin. The $\chi^2$ values for each diagnostic are shown as coloured dashed lines, while the total $\chi^2$ from the combined measurements is represented by a black thick curve. 
      The 1$\sigma$ confidence interval is indicated by two vertical dotted lines, and the best-fitting value is marked with a thick vertical grey dashed line.
      }
      \label{fig:emcee_chi2}
\end{figure}

In Sect.~\ref{sec:strong_line_method} we describe how we derive metallicity using the various diagnostics in Table~\ref{tab:line_ratios} using a Bayesian approach (our Method I).
The $\chi^2$ value is computed from the residuals between the observed line ratios and the values predicted by the calibration relations at a given metallicity, 
and we use the \texttt{emcee} ensemble sampler to map the posterior distribution.
The metallicity that minimises $\chi^2$ (equivalently, maximises the likelihood for flat priors) is taken as our best-fitting value. 
Figure~\ref{fig:emcee_chi2} shows an example (CRISTAL-19) of the resulting $\chi^2$-\lOH\ curves, plotted for each diagnostic individually and in combination.
The figure demonstrates the advantage of combining multiple diagnostics, which provides greater constraining power than any single diagnostic on its own.

\begin{figure}
\centering
    \includegraphics[width=0.45\textwidth]{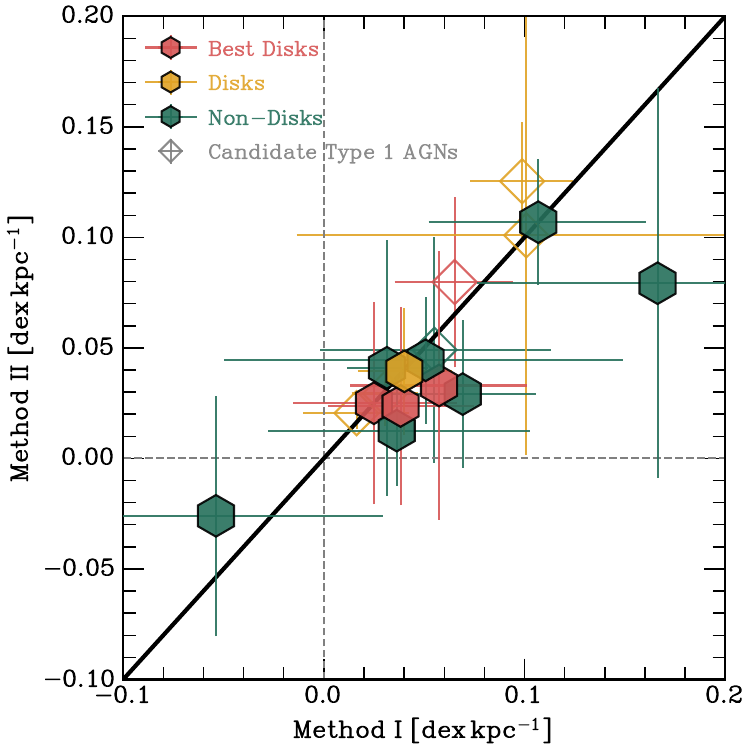}
      \caption{Comparison between the metallicity gradients inferred from the Bayesian method (`All') and the median from Method II (`Median').
      The disks and non-disks classified in \citet{Lee2025b} are shown in pink/yellow and green, respectively.
      The candidate Type-1 AGNs identified in \citet{Ren2025} are marked with the empty diamond symbol.
      CRISTAL-06b is not shown because only one diagnostic N2 is used for Method I.
      }
      \label{fig:medianvsall}
\end{figure}

\section{Metallicity gradients inferred from each diagnostic}\label{app:method_II}
As discussed in Sect.~\ref{sec:strong_line_method}, the line diagnostics we used are not all independent, which would underestimate the uncertainties of the best-fit gradients using the Bayesian approach (Method I).
To better estimate the uncertainties and to examine the systematics associated with different line tracers, we alternatively infer the metallicities by simply inverting the calibration relations to find the solution given a line ratio, which we refer to as Method II in Sect.~\ref{sec:strong_line_method}. 

Table \ref{tab:gradients_eachdiag} lists the metallicity gradients derived with Method II for each emission‐line diagnostic. 
Figure \ref{fig:m1m2indiv} compares these values with the gradients obtained using Method I. Overall, the two methods agree within $1\sigma$ uncertainties.
The figure also demonstrates that gradients inferred from a single diagnostic carry relatively large uncertainties, 
whereas the Bayesian method adopted in Method I provides tighter constraints.

\section{Metallicity gradient profiles for the full samples}\label{app:remaining_profiles}
Figure~\ref{fig:gradient_fit_2} continues from Fig.~\ref{fig:gradient_fit} to show the metallicity gradient profiles of their best-fit models with method described in Sect.~\ref{sec:strong_line_method}.

\begin{figure*}
  \centering
  \gradimage{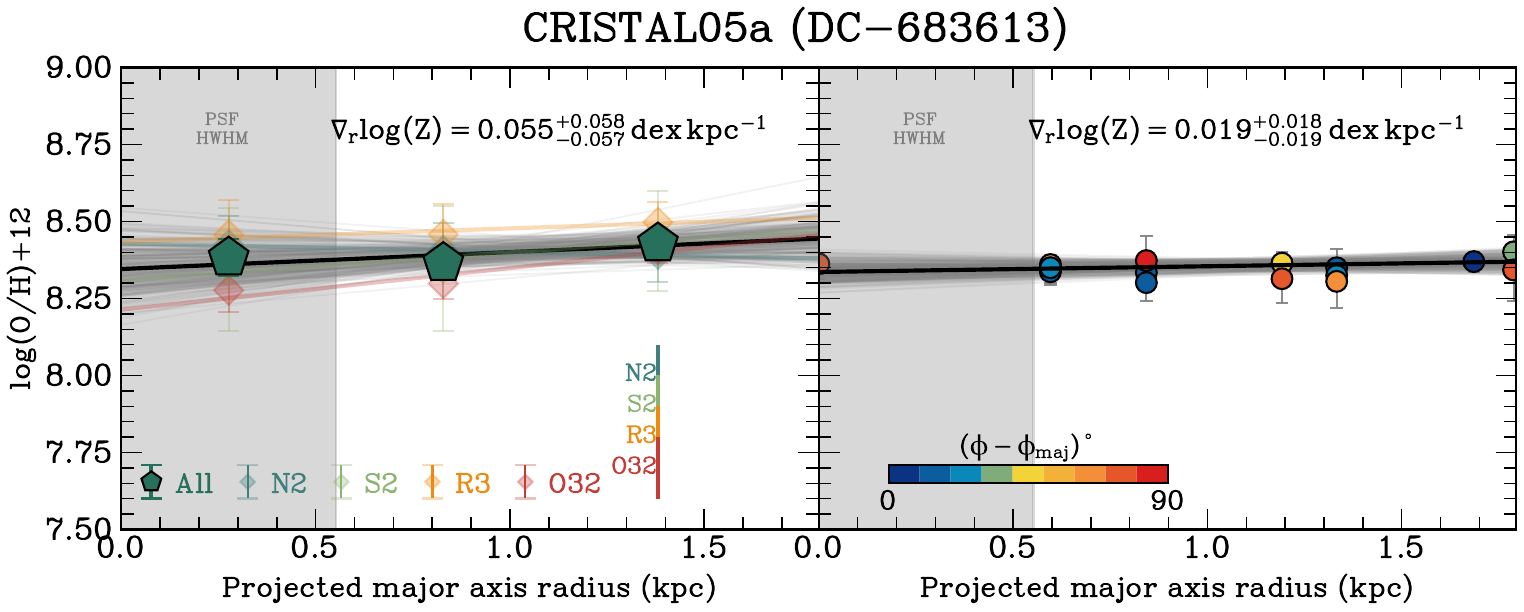}\hfill
  \gradimage{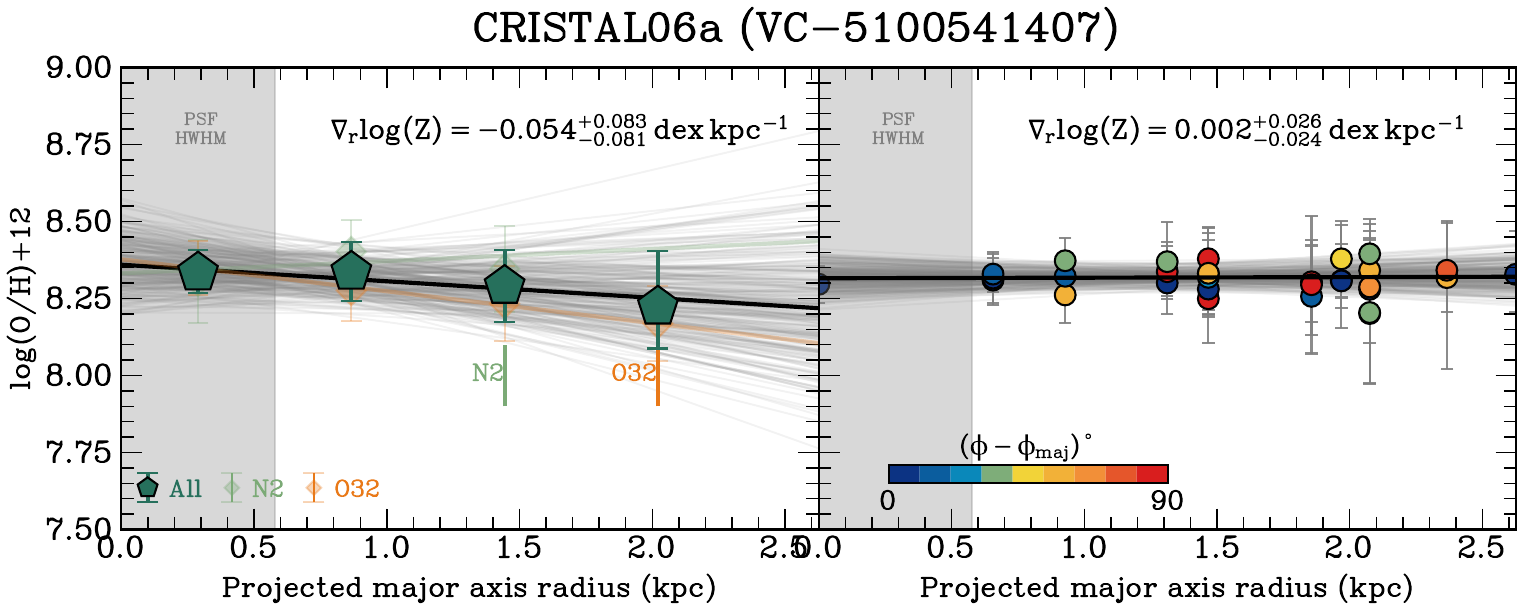}
  \gradimage{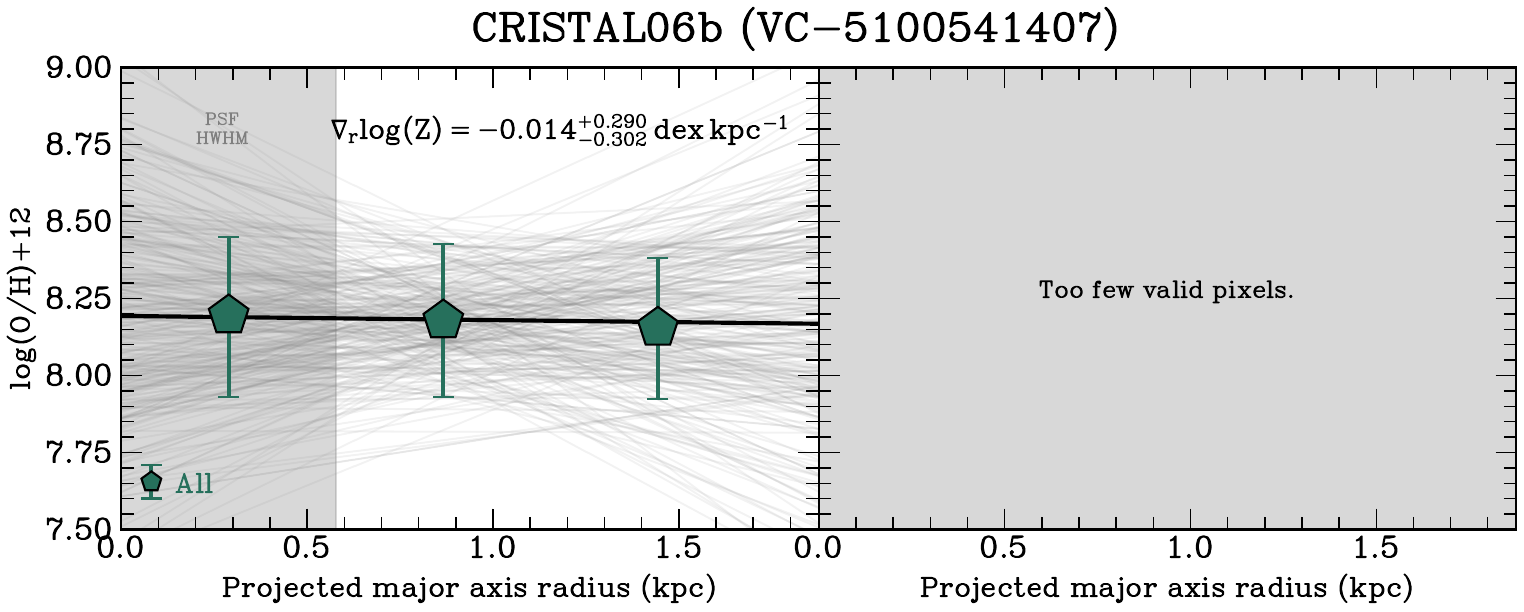}\hfill
  \gradimage{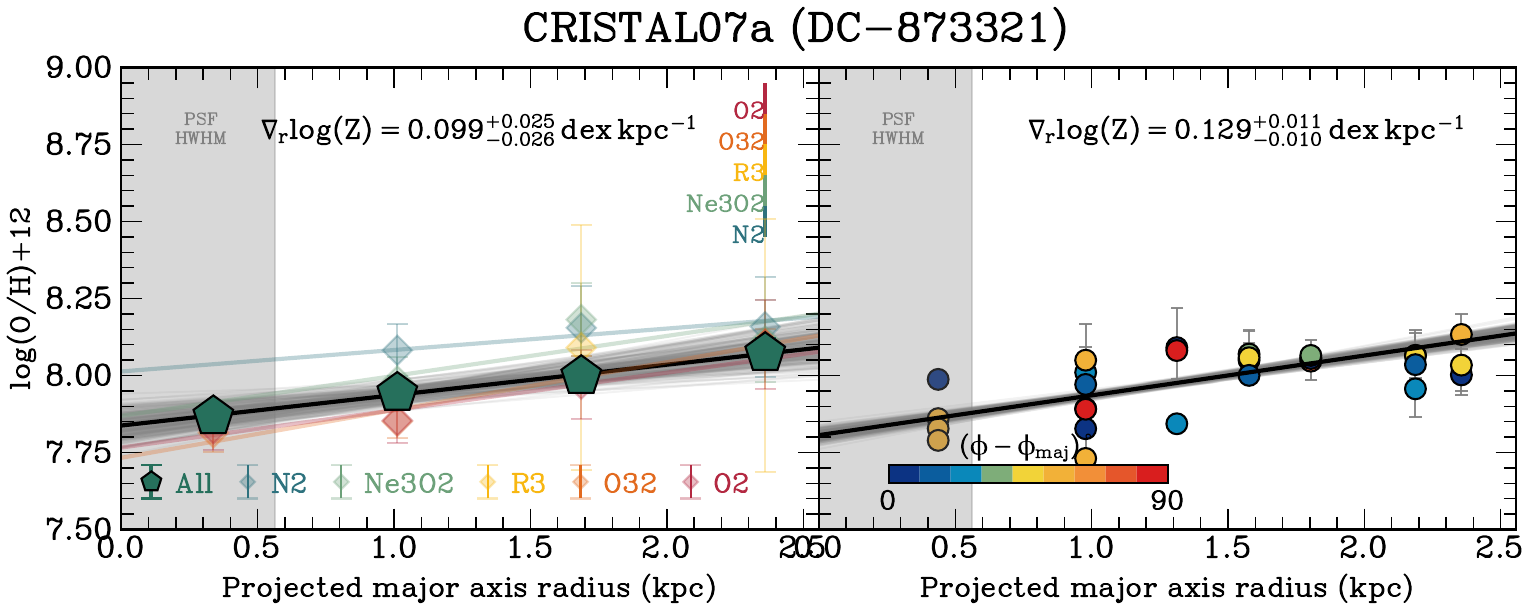}
  \gradimage{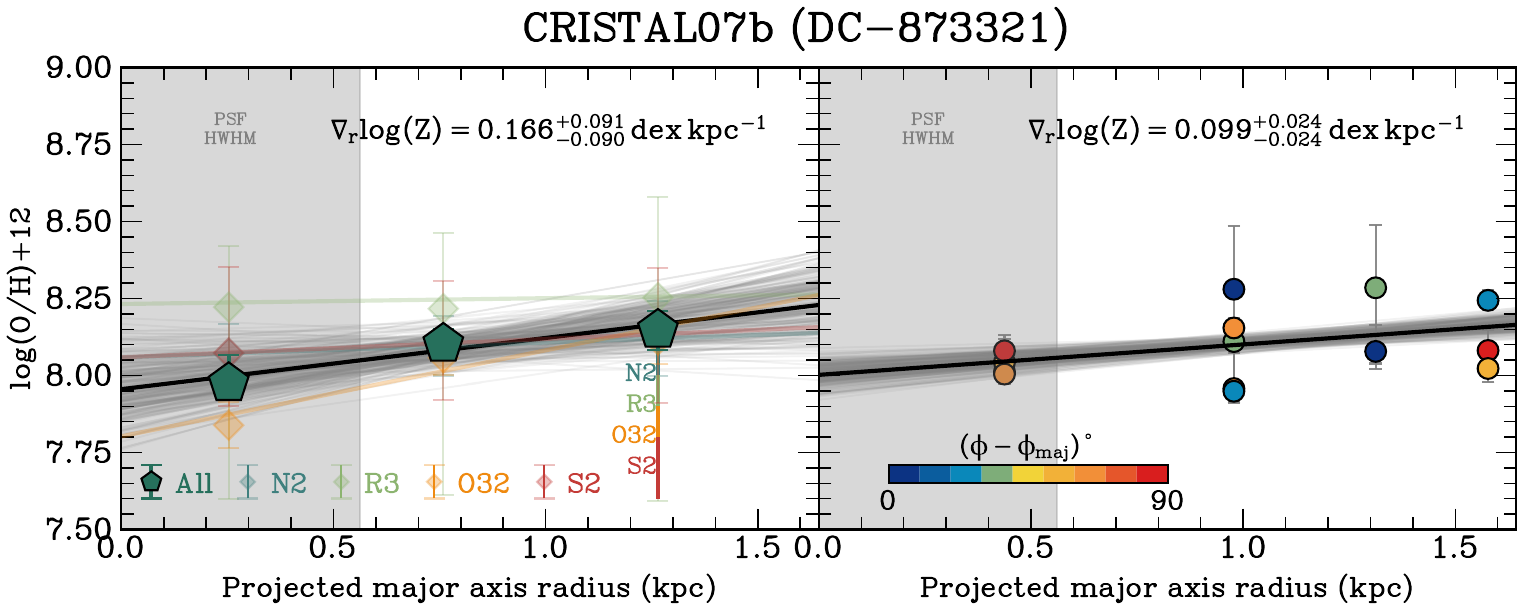}\hfill
  \gradimage{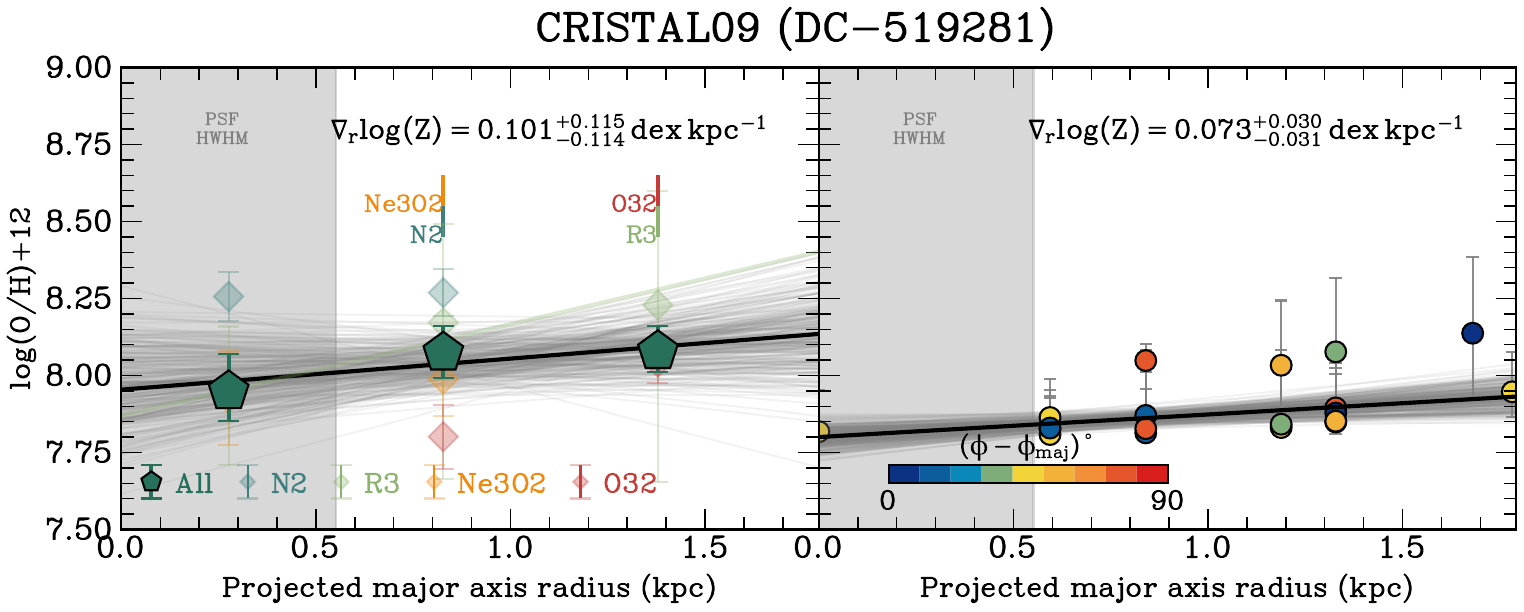}
  \gradimage{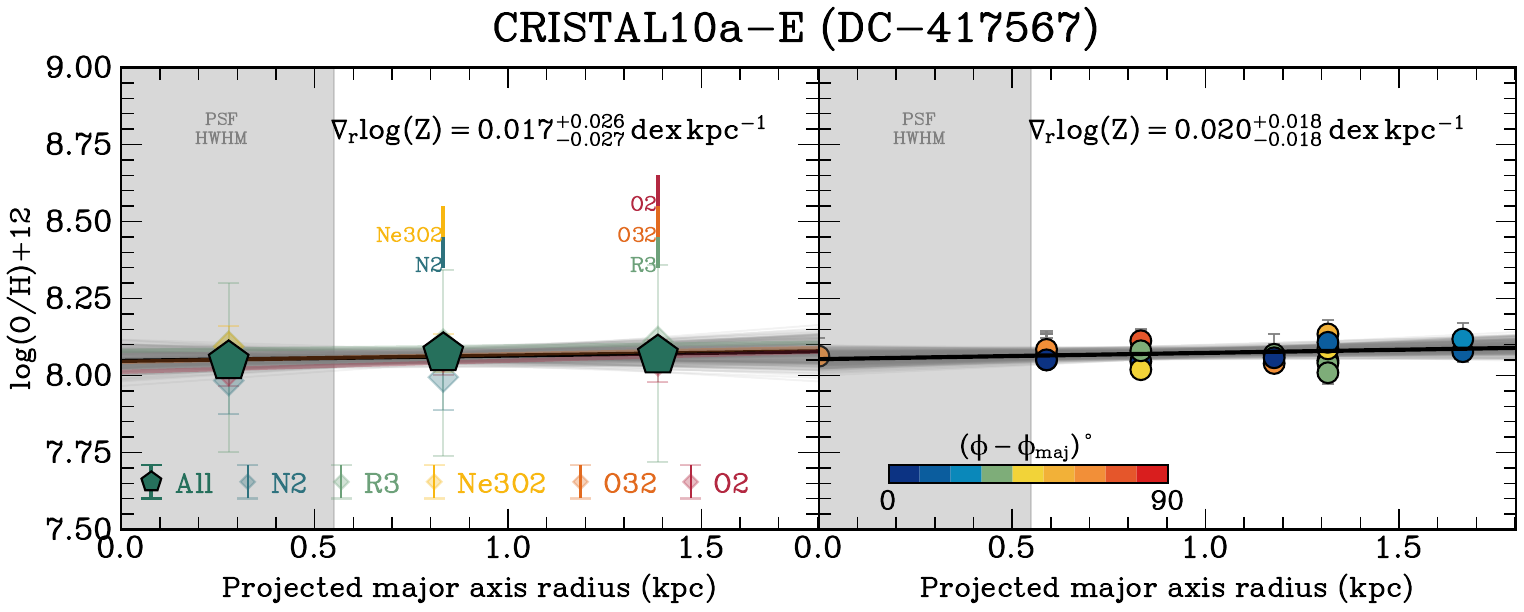}\hfill
  \gradimage{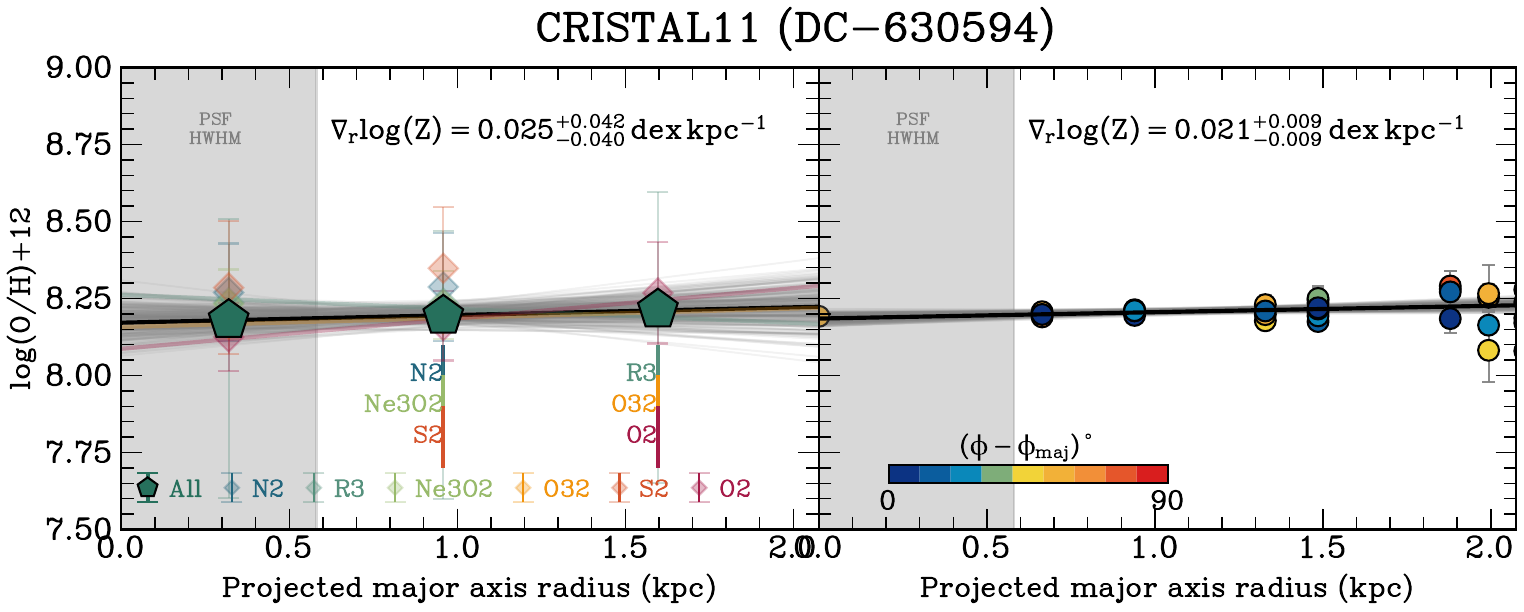}
  \gradimage{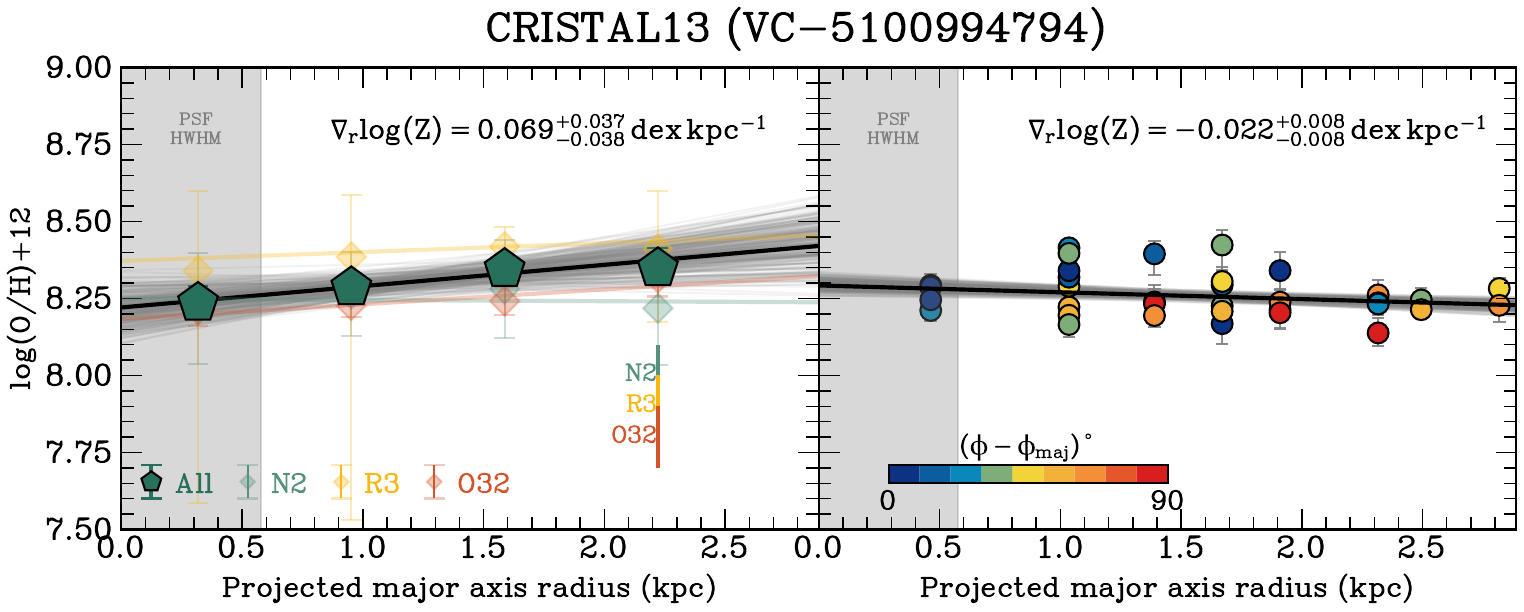}\hfill
  \gradimage{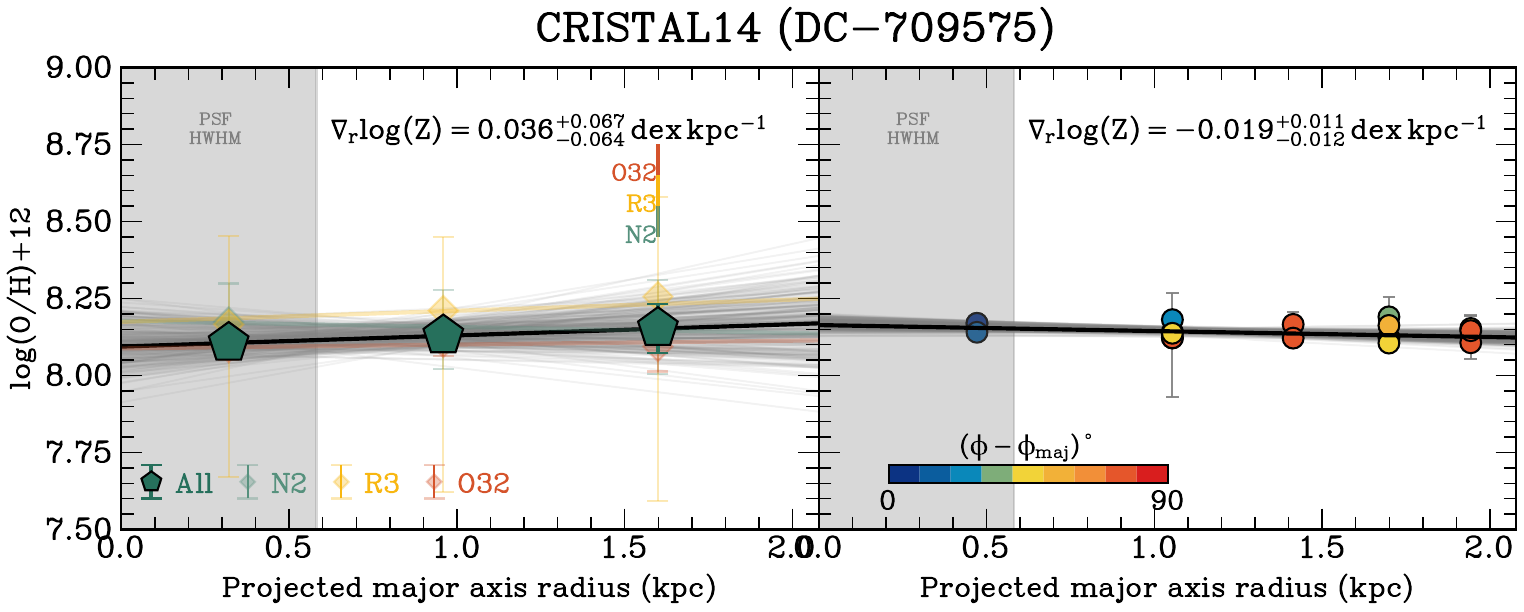}
  \caption{Continued from Fig.~\ref{fig:gradient_fit}.}
  \label{fig:gradient_fit_2}
\end{figure*}
\begin{figure*}[t]\ContinuedFloat
  \centering  
  \gradimage{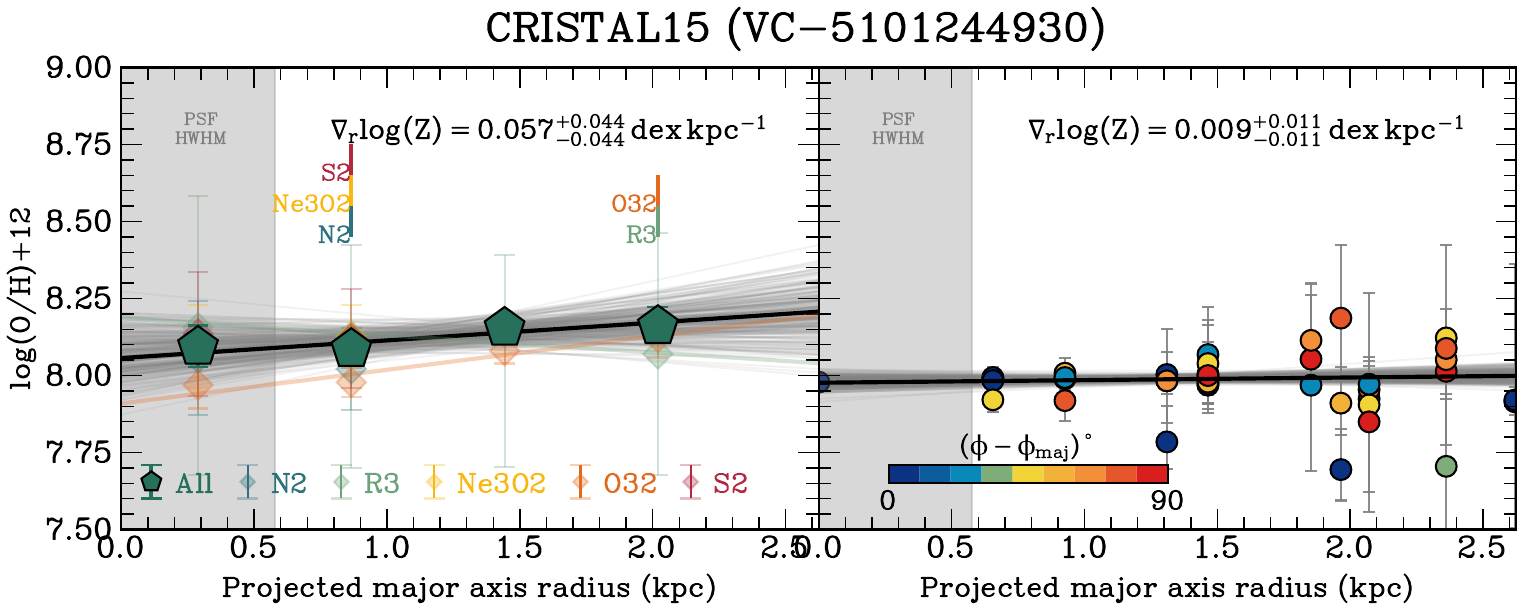}\hfill
  \gradimage{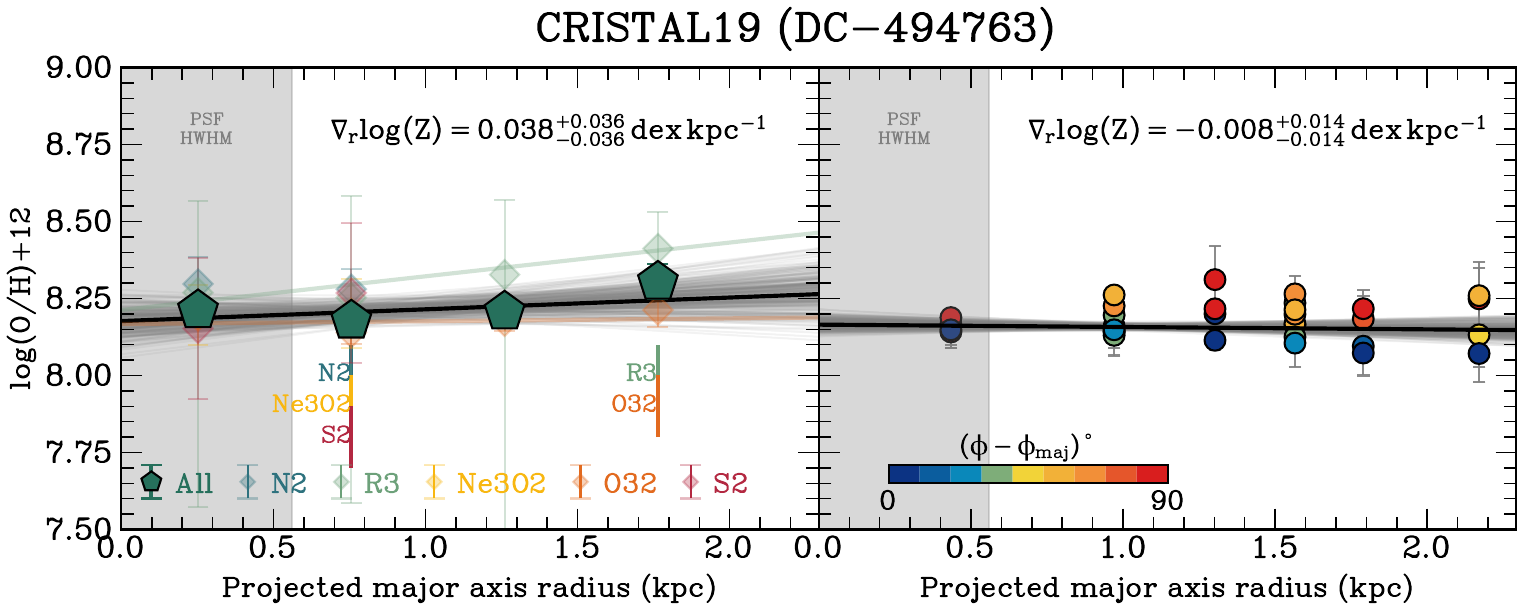}
  \includegraphics[width=0.496\textwidth]{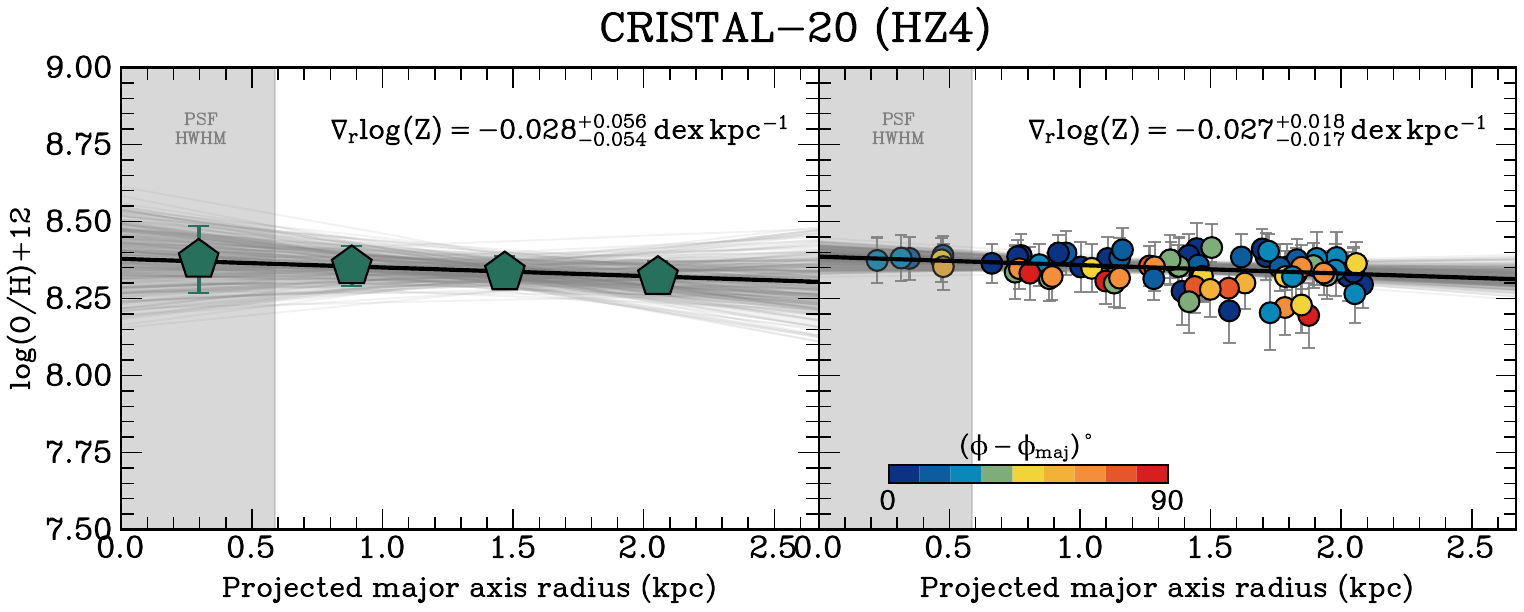}
  \includegraphics[width=0.496\textwidth]{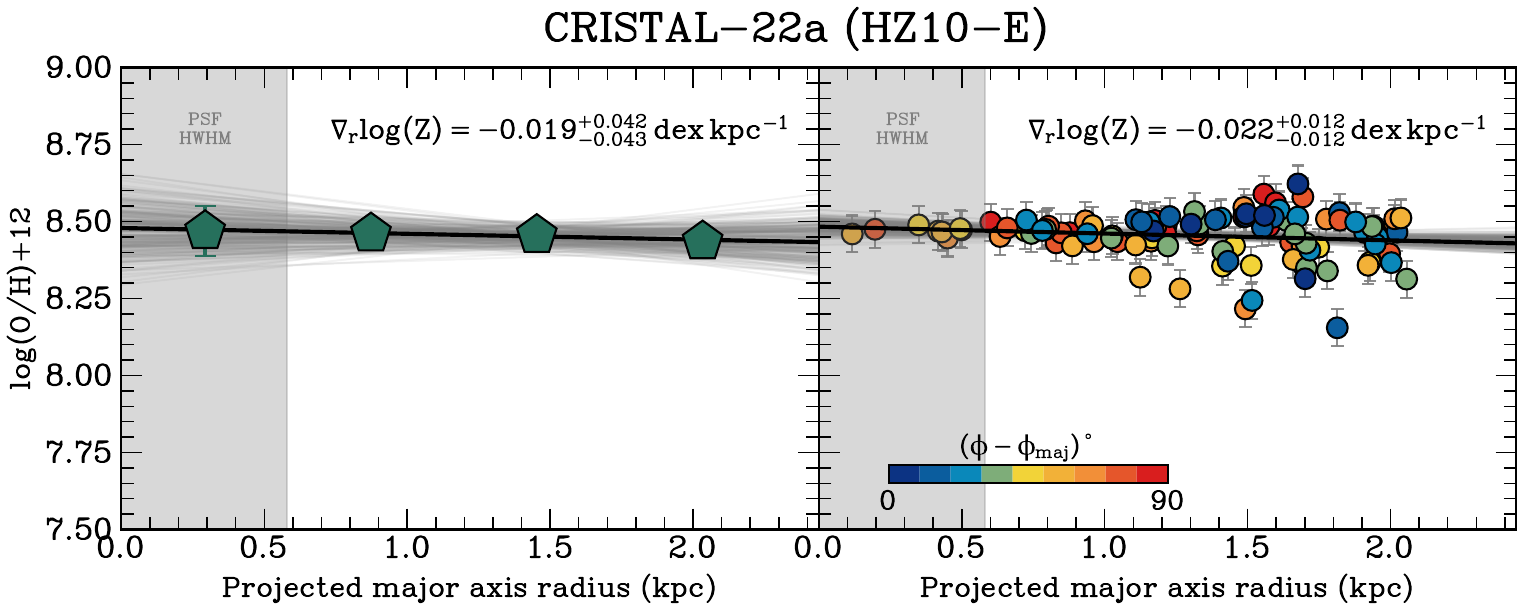}
  \gradimage{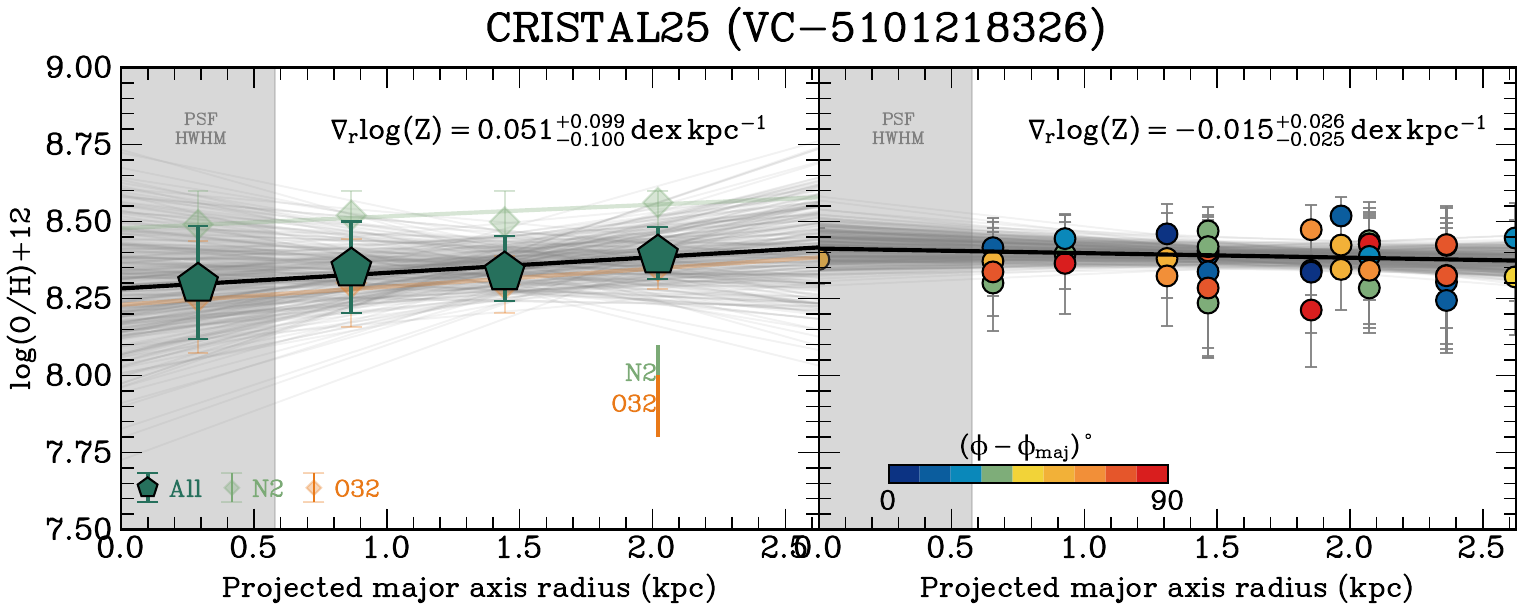}\hfill
  \gradimage{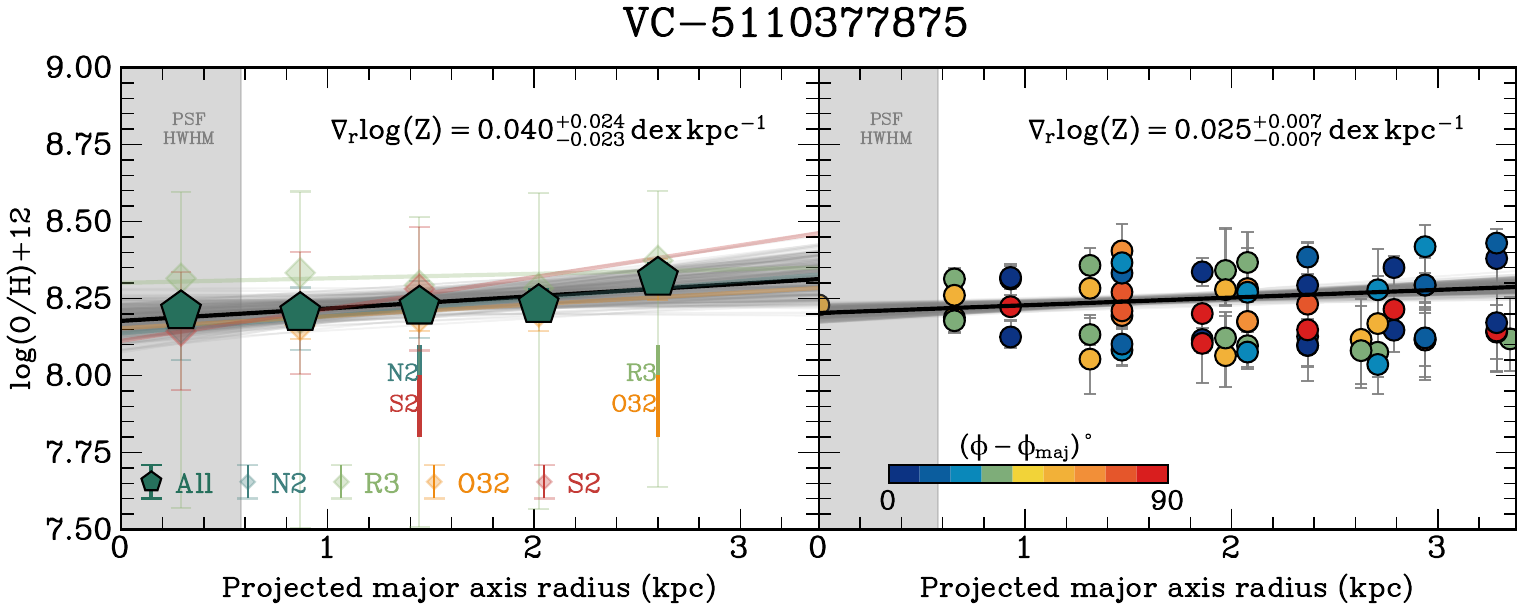}
  \caption{Continued.}
  \label{fig:gradient_fit_2}
\end{figure*}

\section{Literature sample}\label{app:literature_sample}
Table~\ref{tab:literature_sample} lists the references from which we compiled the metallicity gradient measurements, 
as well as kinematics results where available.
As noted in Sect.~\ref{subsec:Z_disks}, the literature compilation is heterogeneous: the data were taken with a variety of instruments, span a wide range of spatial resolutions, and employ different metallicity diagnostics. 
While this diversity is not ideal, applying a stellar mass cut to select a subsample within the mass range of our sample results in a significantly reduced sample size: only \Ngallowz\ at cosmic noon and $35$ galaxies beyond (see Fig.~\ref{fig:DelZ_disknondisk}).
Therefore, to maximise the statistics, we have chosen to retain the literature sample as is. Future work on a more unified sample selection would be highly valuable.

\begin{figure}
\centering
    \includegraphics[width=0.48\textwidth]{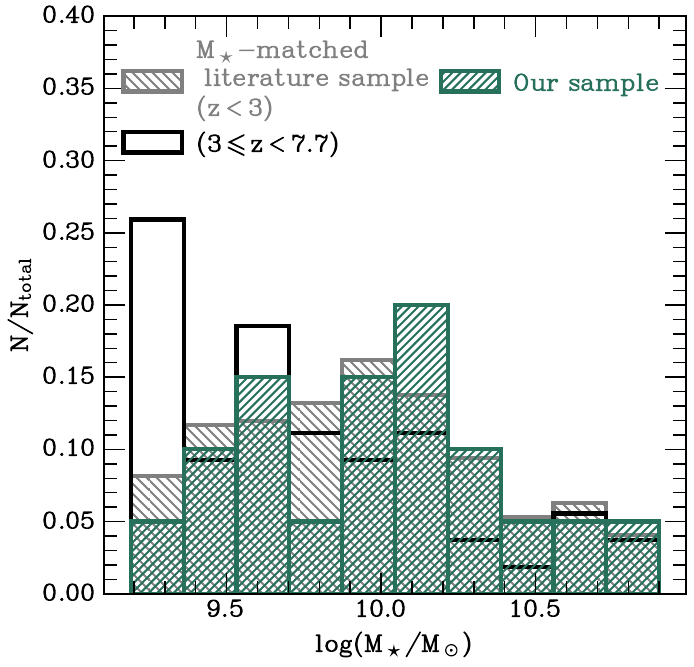}
      \caption{Distributions of stellar mass $M_\star$ for literature samples within the mass range of our sample ($z\geqslant3$ in black, $z<3$ in hatched grey). 
      The $M_\star$ distribution of our sample is overlaid in hatched green. 
      The peak at $\log(M_\star/M_\odot)\lesssim9.4$ for the $z\geqslant3$ sample is primarily contributed by the sources at $z>6$.
      }
      \label{fig:mstar_dist_lit}
\end{figure}

\begin{figure}
\centering
    \includegraphics[width=0.45\textwidth]{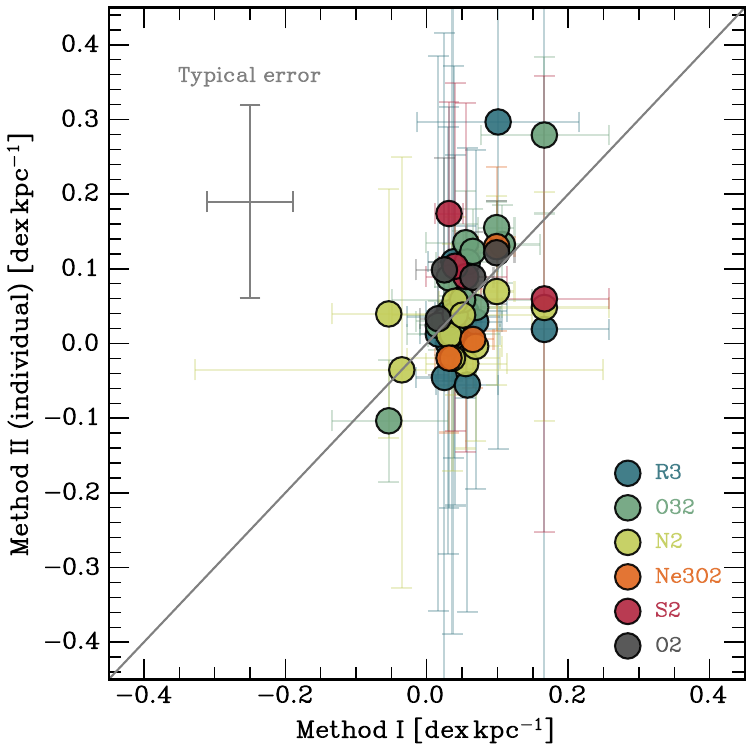}
      \caption{Comparison of the metallicity gradients derived from Method I and individual line diagnostics (Method II). The values are listed in Table~\ref{tab:gradients_eachdiag}. The grey diagonal line is the line of equality. 
      The typical uncertainty is shown at the upper left.
      The uncertainties associated with individual line diagnostics are large, 
      but combining multiple line diagnostics with the Bayesian approach in Method I can tightens the overall constraints.
      }
      \label{fig:m1m2indiv}
\end{figure}

Figure~\ref{fig:mstar_dist_lit} presents the stellar mass distribution of the final mass-matched sample at $z$\,$<$\,$3$ and $z$\,$\geqslant$\,$3$. The stellar mass distribution of the $z$\,$<$\,$3$ sample is generally well-matched with our sample. 
However, the $z$\,$\geqslant$\,$3$ sample skews towards $\log(M_\star/M_\odot)$\,$\lesssim$\,$9.4$, which is contributed mostly from $z$\,$>$\,$6$ galaxies in this redshift range. These high-redshift galaxies are predominantly low-mass sources in the literature.

As pointed out in Sect.~\ref{subsec:Z_disks}, our compilation does not attempt to homogenise the gas tracer nor the modelling methodologies used in the various studies.
Whether the intrinsic velocity dispersion ($\sigma_0$) and the ratio $V_{\rm rot}/\sigma_0$ 
differ systematically between ionised (e.g. \Ha, \oiii) and molecular or neutral gas tracers at $z \gtrsim 1$ is still uncertain, largely due to limited statistics
from studies that have employed multiple tracers for the same object. Future work would benefit from comparing kinematics from multiple tracers for a consistent set of objects.
For \acj, \gomezespinoza\ will present the kinematics of our sample as traced by ionised gas.

\section{Azimuthally binned metallicity gradients}
To examine the azimuthal variation at a fixed radius, as discussed in Sect.~\ref{subsec:azimuthal_variation}, we binned the pixels within every azimuthal angle range of $\Delta\phi=30^\circ$, where the azimuthal angle is defined as the acute angle difference from the morphological major axis. The resulting metallicity gradients for 11 galaxies for each azimuthal bin are summarised in Table~\ref{tab:azibin}.

\section{Beam smearing-corrected metallicity gradients}
Table~\ref{tab:bmc_gradients} lists the metallicity gradients that are corrected for beam smearing, based on the correction curves derived in Sect.~\ref{sec:effects_of_beam_smearing}.
Since the correction factors for a given angular resolution relative to galaxy size depend on the absolute intrinsic unknown gradients, we report the corrected values for a range of assumed intrinsic gradients that we have examined, i.e. $|$\dlOH$|$\,$=$\,$[0.05,0.10,0.20]$.

\begin{table*}
    \caption{Metallicity gradients inferred from individual diagnostics.}
    \label{tab:gradients_eachdiag}
    \centering
\begin{tabular}{>{\small}l>{\small}c>{\small}c>{\small}c>{\small}c>{\small}c>{\small}c}
\hline\hline  
ID  & R3 & O32 & N2 & Ne3O2 & S2 & O2 \\  
  & {\scriptsize (dex\,${\rm kpc^{-1}}$)} & {\scriptsize (dex\,${\rm kpc^{-1}}$)} & {\scriptsize (dex\,${\rm kpc^{-1}}$)} & {\scriptsize (dex\,${\rm kpc^{-1}}$)} & {\scriptsize (dex\,${\rm kpc^{-1}}$)} & {\scriptsize (dex\,${\rm kpc^{-1}}$)} \\ \hline  
01a & $\ldots$ & ${0.133}_{-0.052}^{+0.052}$ & $\ldots$ & $\ldots$ & $\ldots$ & $\ldots$ \\ 
02 & ${0.000}_{-0.282}^{+0.290}$ & ${0.028}_{-0.026}^{+0.027}$ & $\ldots$ & $\ldots$ & $\ldots$ & $\ldots$ \\ 
03 & $\ldots$ & ${0.123}_{-0.058}^{+0.057}$ & $\ldots$ & ${0.005}_{-0.065}^{+0.068}$ & $\ldots$ & ${0.089}_{-0.060}^{+0.061}$ \\ 
04a & ${0.041}_{-0.261}^{+0.276}$ & ${0.088}_{-0.024}^{+0.024}$ & ${0.011}_{-0.068}^{+0.069}$ & ${-0.020}_{-0.100}^{+0.107}$ & ${0.174}_{-0.156}^{+0.149}$ & $\ldots$ \\ 
05a & ${0.043}_{-0.116}^{+0.115}$ & ${0.135}_{-0.073}^{+0.069}$ & ${-0.027}_{-0.114}^{+0.114}$ & $\ldots$ & ${0.090}_{-0.234}^{+0.233}$ & $\ldots$ \\ 
06a & $\ldots$ & ${-0.104}_{-0.082}^{+0.081}$ & ${0.040}_{-0.166}^{+0.167}$ & $\ldots$ & $\ldots$ & $\ldots$ \\ 
06b & $\ldots$ & $\ldots$ & ${-0.035}_{-0.292}^{+0.285}$ & $\ldots$ & $\ldots$ & $\ldots$ \\ 
07a & $\ldots$ & ${0.155}_{-0.037}^{+0.035}$ & ${0.070}_{-0.125}^{+0.127}$ & ${0.130}_{-0.113}^{+0.107}$ & $\ldots$ & ${0.122}_{-0.069}^{+0.070}$ \\ 
07b & ${0.019}_{-0.512}^{+0.512}$ & ${0.279}_{-0.105}^{+0.105}$ & ${0.048}_{-0.152}^{+0.155}$ & $\ldots$ & ${0.060}_{-0.312}^{+0.299}$ & $\ldots$ \\ 
09 & ${0.297}_{-0.439}^{+0.406}$ & $\ldots$ & $\ldots$ & $\ldots$ & $\ldots$ & $\ldots$ \\ 
10a-E & ${0.013}_{-0.370}^{+0.372}$ & ${0.024}_{-0.032}^{+0.033}$ & $\ldots$ & $\ldots$ & $\ldots$ & ${0.033}_{-0.075}^{+0.076}$ \\ 
11 & ${-0.046}_{-0.445}^{+0.462}$ & ${0.030}_{-0.046}^{+0.046}$ & $\ldots$ & $\ldots$ & $\ldots$ & ${0.099}_{-0.148}^{+0.150}$ \\ 
13 & ${0.029}_{-0.224}^{+0.230}$ & ${0.048}_{-0.037}^{+0.036}$ & ${-0.004}_{-0.127}^{+0.129}$ & $\ldots$ & $\ldots$ & $\ldots$ \\ 
14 & ${0.036}_{-0.426}^{+0.429}$ & ${0.013}_{-0.062}^{+0.062}$ & ${-0.020}_{-0.151}^{+0.153}$ & $\ldots$ & $\ldots$ & $\ldots$ \\ 
15 & ${-0.056}_{-0.304}^{+0.317}$ & ${0.109}_{-0.052}^{+0.050}$ & $\ldots$ & $\ldots$ & $\ldots$ & $\ldots$ \\ 
19 & ${0.110}_{-0.263}^{+0.263}$ & ${0.009}_{-0.037}^{+0.037}$ & $\ldots$ & $\ldots$ & $\ldots$ & $\ldots$ \\ 
25 & $\ldots$ & ${0.059}_{-0.093}^{+0.092}$ & ${0.038}_{-0.081}^{+0.081}$ & $\ldots$ & $\ldots$ & $\ldots$ \\ 
VC-7875 & ${0.015}_{-0.232}^{+0.238}$ & ${0.039}_{-0.031}^{+0.030}$ & ${0.057}_{-0.125}^{+0.113}$ & $\ldots$ & ${0.103}_{-0.221}^{+0.245}$ & $\ldots$ \\ 
\hline
\end{tabular}
\tablefoot{
The outermost radius at which each diagnostic line is measured for each galaxy is annotated in Fig.~\ref{fig:gradient_fit}.
We only fit a gradient when there are more than two data points available for the diagnostic. 
Figure~\ref{fig:m1m2indiv} shows a comparison between these gradients with that inferred from Method I.
}
\end{table*}

\begin{table}
\caption{Metallicity gradients of 11 galaxies in three azimuthal bins.}
\label{tab:azibin}
\centering
\begin{tabular}{l c c c c}
\hline\hline  
ID  &  $[0^\circ,30^\circ)$\tablefootmark{a} & $[30^\circ,60^\circ)$\tablefootmark{a} & $[60^\circ,90^\circ)$\tablefootmark{a}  \\  
  &  {\scriptsize (dex\,${\rm kpc^{-1}}$)} & {\scriptsize (dex\,${\rm kpc^{-1}}$)} & {\scriptsize (dex\,${\rm kpc^{-1}}$)} \\ \hline  
02 & ${0.044_{-0.012}^{+0.011}}$  &${0.050_{-0.020}^{+0.020}}$  &${0.008_{-0.016}^{+0.017}}$ \\ 
03 & ${0.048_{-0.018}^{+0.018}}$  &${0.138_{-0.020}^{+0.021}}$  &${0.070_{-0.018}^{+0.017}}$ \\ 
04a & ${0.043_{-0.011}^{+0.010}}$  &${0.073_{-0.020}^{+0.023}}$  &${0.052_{-0.020}^{+0.023}}$ \\ 
11 & ${0.006_{-0.014}^{+0.015}}$  &${0.020_{-0.012}^{+0.012}}$  &${0.051_{-0.022}^{+0.022}}$ \\ 
13 & ${-0.018_{-0.008}^{+0.008}}$  &${-0.011_{-0.011}^{+0.011}}$  &${-0.000_{-0.037}^{+0.038}}$ \\ 
15 & ${0.011_{-0.010}^{+0.010}}$  &${0.032_{-0.019}^{+0.020}}$  &${0.068_{-0.035}^{+0.034}}$ \\ 
19 & ${-0.041_{-0.020}^{+0.020}}$  &${0.019_{-0.038}^{+0.039}}$  &${0.049_{-0.023}^{+0.022}}$ \\ 
20 & ${-0.027_{-0.022}^{+0.021}}$  &${-0.015_{-0.037}^{+0.038}}$  &${-0.031_{-0.053}^{+0.051}}$ \\ 
22a & ${-0.059_{-0.030}^{+0.029}}$  &${-0.040_{-0.018}^{+0.019}}$  &${0.013_{-0.021}^{+0.022}}$ \\ 
25 & ${-0.049_{-0.031}^{+0.032}}$  &${0.003_{-0.032}^{+0.033}}$  &${0.037_{-0.058}^{+0.064}}$ \\ 
VC-7875 & ${0.062_{-0.011}^{+0.011}}$  &${-0.020_{-0.015}^{+0.015}}$  &${-0.027_{-0.019}^{+0.019}}$ \\ 
\hline
\end{tabular}
\tablefoot{
\tablefoottext{a}{The acute azimuthal angle difference from the major axis.}

}
\end{table}

\begin{table}
    \caption{Metallicity gradients corrected for beam smearing.}
    \label{tab:bmc_gradients}
    \centering
\begin{tabular}{>{\small}l>{\small}c>{\small}c>{\small}c}
\hline\hline  
ID  & \dlOH\ &  \dlOH\ &  \dlOH\ \\  
  &  {\scriptsize ($|$\dlOH$|=0.20$)}  & {\scriptsize ($|$\dlOH$|=0.10$)} & {\scriptsize ($|$\dlOH$|=0.05$)} \\ 
  &  {\scriptsize (dex\,${\rm kpc^{-1}}$)} & {\scriptsize (dex\,${\rm kpc^{-1}}$)} & {\scriptsize (dex\,${\rm kpc^{-1}}$)} \\ \hline  
01a& ${0.115}$ & ${0.120}$ & ${0.108}$ \\ 
02& ${0.038}$ & ${0.037}$ & ${0.032}$ \\ 
03& ${0.094}$ & ${0.087}$ & ${0.072}$ \\ 
04a& ${0.049}$ & ${0.044}$ & ${0.035}$ \\ 
05a& ${0.071}$ & ${0.069}$ & ${0.058}$ \\ 
06a& ${-0.072}$ & ${-0.068}$ & ${-0.058}$ \\ 
06b& ${-0.053}$ & ${-0.049}$ & ${-0.039}$ \\ 
07a& ${0.135}$ & ${0.128}$ & ${0.107}$ \\ 
07b& ${0.234}$ & ${0.219}$ & ${0.181}$ \\ 
09& ${0.189}$ & ${0.159}$ & ${0.120}$ \\ 
10a-E& ${0.023}$ & ${0.021}$ & ${0.018}$ \\ 
11& ${0.040}$ & ${0.035}$ & ${0.028}$ \\ 
13& ${0.077}$ & ${0.079}$ & ${0.070}$ \\ 
14& ${0.053}$ & ${0.049}$ & ${0.040}$ \\ 
15& ${0.062}$ & ${0.064}$ & ${0.058}$ \\ 
19& ${0.063}$ & ${0.055}$ & ${0.044}$ \\ 
20& ${-0.035}$ & ${-0.035}$ & ${-0.030}$ \\ 
22a& ${-0.033}$ & ${-0.028}$ & ${-0.022}$ \\ 
25& ${0.105}$ & ${0.085}$ & ${0.062}$ \\ 
VC-7875& ${0.043}$ & ${0.043}$ & ${0.041}$ \\
\hline
\end{tabular}
\tablefoot{Similar to Table~\ref{tab:gradients}, 
but with beam-correction factors applied based on the correction curves in Fig.~\ref{fig:bmsc_grad} derived in Sect.~\ref{sec:effects_of_beam_smearing}, for different assumed intrinsic metallicity gradients $|$\dlOH$|=[0.20,0.10,0.05]$ (from left to right).
}
\end{table}

\begin{table*}
\caption{Literature sample of metallicity gradients and kinematics.}
\label{tab:literature_sample}
\centering
\begin{tabular}{>{\small}c>{\small}l>{\small}l}
\hline \hline
$z$&Reference (Survey) & Kinematics Reference \\ \hline 
$0.1<z<3.5$ 
&\citet{Stark2008}\tablefootmark{*}& \citet{Stark2008}\\   
&\citet{Queyrel2012} (MASSIV) &\citet{Epinat2012}\\ 
&\citet{Swinbank2012} (SHiZELS)&\citet{Swinbank2012}\\
&\citet{Jones2013}\tablefootmark{*} & \citet{Jones2013}\\ 
&\citet{Stott2014} (KMOS-HiZELS)\tablefootmark{b} & \citet{Stott2014}\\
&\citet{Troncoso2014} (AMAZE-LSD)&\citet{Gnerucci2011}\\
&\citep[see also][]{Cresci2010} & \\ 
&\citet{Jones2015} (GLASS-HST)\tablefootmark{*} &$\ldots$ \\
&\citet{Leethochawalit2016}\tablefootmark{a,*} & \citet{Leethochawalit2016}\\
&\citet{EWuyts2016} (KMOS$^{\rm 3D}$)&\citet{Uebler2019}\\
&\citet{Molina2017} (SHiZELS) & \citet{Molina2017} \\ 
&\citet{XWang2017} (GLASS-HST)\tablefootmark{*}&\citet{DiTeodoro2018,Hirtenstein2019,Girard2020}\\ 
&\citep[see also][]{Yuan2011} & \\
&\citet{Carton2018} (MUSE-WIDE) &\citet{Sharda2021}\\
&\citet{nmfs2018} (SINS/zC-SINF) & \citet{nmfs2018} \\ 
&\citet{Girard2018}\tablefootmark{*} & \citet{Girard2018} \\
&\citet{Patricio2019}\tablefootmark{*} &$\ldots$ \\ 
&\citet{Sharon2019} \tablefootmark{*}&\citet{Liu2023} \\
&\citet{XWang2019} (GLASS-HST)&$\ldots$\\Spectroscopy
&\citet{Curti2020a} (KLEVER)\tablefootmark{*} &$\ldots$\\
&\citet{XWang2020}\tablefootmark{*}&\citet{Hirtenstein2019,Girard2020}\\ 
&\citet{Gillman2021} (KROSS-KGES) & \citet{Harrison2017,Johnson2018,Tiley2021} \\
&\citet{Simons2021} (CLEAR) &$\ldots$\\
&\citet{Gillman2022} (KURVS) & \citet{Puglisi2023} \\
&\citet{XWang2022} (GLASS-JWST)\tablefootmark{*} &$\ldots$\\
&\citet{ZLi2022} (MAMMOTH-Grism) &$\ldots$\\
&\citet{Barisic2025} (MSA-3D)& \citet{Barisic2025}\\
&\citet{Ju2025} (MSA-3D)&\citet{Ju2025}\\
\hline
$z>3.5$  & \citet{Troncoso2014} (AMAZE-LSD) & \citet{Troncoso2014}\\
&\citet{Arribas2024} (GA-NIFS) & \citet{Arribas2024}\\
&\citet{Vallini2024} & \citet{Parlanti2023,Posses2023}\\
&\citet{Venturi2024} (JWST GO \#1893\tablefootmark{c}) &$\ldots$\\
&\citet{Li2025}& $\ldots$ \\
\hline
\end{tabular}
\tablefoot{
\tablefoottext{a}{When available, stellar masses are taken from \citet{Mainali2023} and \citet{HWang2024}, respectively.
For other objects, we take the approximate median $\log{(M_\star/M_{\odot})}=9.3$ for the entire sample.}
\tablefoottext{b}{Included the KMOS-HiZELS-SV1 sample from \citet{Sobral2013b}, and we adopt the kinematics results presented therein.}
\tablefoottext{c}{\citet{Carniani2021}}
\tablefoottext{*}{Gravitationally lensed.}
}
\end{table*}

\end{appendix}
\end{document}